\documentclass[twocolumn,usenatbib]{mnras}

\usepackage{newtxtext,newtxmath}
\usepackage{pdfpages}
\usepackage{float}
\usepackage{tabularx}
\usepackage{comment}
\usepackage[normalem]{ulem}
\usepackage[mathlines, pagewise]{lineno}
\usepackage{subcaption}
\usepackage{graphicx}	% Including figure files
\usepackage{amsmath}	% Advanced maths commands
\usepackage{hyperref}
\usepackage{physics}
\usepackage{arydshln}
\usepackage{enumitem}	
\usepackage{eso-pic}% http://ctan.org/pkg/eso-pic

\AddToShipoutPictureBG*{%
  \AtPageUpperLeft{%
    \hspace{0.75\paperwidth}%
    \raisebox{-3.5\baselineskip}{%
      \makebox[0pt][l]{\textnormal{DES-2024-0827}}
}}}%

\AddToShipoutPictureBG*{%
  \AtPageUpperLeft{%
    \hspace{0.75\paperwidth}%
    \raisebox{-4.5\baselineskip}{%
      \makebox[0pt][l]{\textnormal{FERMILAB-PUB-24-0130-PPD}}
}}}%

\newcommand{\msun}{\mbox{$M_{\odot}$}}
\newcommand{\Msun}{\mbox{${\bf M_{\odot} }$}}

\newcommand{\addcite}[1]{\textcolor{red}{\textbf{[cite]}}}

\begin{document}
  
%-------- TITLE  ---------------------

\title[Weak lensing + kSZ]{Weak lensing combined with the kinetic Sunyaev Zel'dovich effect: \\ A study of baryonic feedback}
\author[L.~Bigwood, A.~Amon et al.]{
\parbox{\textwidth}{
\Large{
L.~Bigwood,$^{1,2}$\thanks{E-mail: lmb224@cam.ac.uk}
A.~Amon,$^{3,1,2}$\thanks{E-mail: alexandra.amon@princeton.edu}
A.~Schneider,$^{4}$
J.~Salcido,$^{5}$
I.~G.~McCarthy,$^{5}$
C.~Preston,$^{1}$
D.~Sanchez,$^{6}$
D.~Sijacki,$^{1,2}$
E.~Schaan,$^{7,8}$
S.~Ferraro,$^{9,10}$
% ...
\\
%\centerline{(DES Collaboration)}\\
}
\vspace{-0.5cm}
\parbox{\textwidth}{\small
\hfill \textit{The authors' affiliations are shown at the end of this paper. }}
}}
%\vspace{-0.5cm}

\maketitle
\label{firstpage}

%-------- ABSTRACT  ---------------------
\begin{abstract} 
Extracting precise cosmology from weak lensing surveys requires modelling the non-linear matter power spectrum, which is suppressed at small scales due to baryonic feedback processes. However, hydrodynamical galaxy formation simulations make widely varying predictions for the amplitude and extent of this effect. We use measurements of Dark Energy Survey Year 3 weak lensing (WL) and Atacama Cosmology Telescope DR5 kinematic Sunyaev-Zel’dovich (kSZ) to jointly constrain cosmological and astrophysical baryonic feedback parameters using a flexible analytical model, `baryonification'. First, using WL only, we compare the $S_8$ constraints using baryonification to a simulation-calibrated halo model, a simulation-based emulator model and the approach of discarding WL measurements on small angular scales. We find that model flexibility can shift the value of $S_8$ and degrade the uncertainty. The kSZ provides additional constraints on the astrophysical parameters and shifts $S_8$ to $S_8=0.823^{+0.019}_{-0.020}$, a higher value than attained using the WL-only analysis.  We measure the suppression of the non-linear matter power spectrum using WL + kSZ and constrain a mean feedback scenario that is more extreme than the predictions from most hydrodynamical simulations.
We constrain the baryon fractions and the gas mass fractions and find them to be generally lower than inferred from X-ray observations and simulation predictions. We conclude that the WL + kSZ measurements provide a new and complementary benchmark for building a coherent picture of the impact of gas around galaxies across observations.
\end{abstract}


\section{Introduction}
The standard cosmological model, $\Lambda$CDM, has been very successful when tested against observations of the cosmic microwave background \citep[CMB;][]{Planck2018, WMAP:2003ivt}, the lensing of the CMB at intermediate redshifts \citep{Plensing:2020, ACT_madhavacheril2023} and low-redshift observations of the expansion history as probed by baryon acoustic oscillations \citep{BAO}.  Measurements of weak galaxy lensing provide a strong test of $\Lambda$CDM at relatively small scales in the low-redshift Universe.  In order to extract unbiased cosmological constraints from weak lensing, accurate modelling of the non-linear matter distribution at $k>0.1h$Mpc$^{-1}$ is crucial.  This requires understanding both the non-linear dark matter evolution due to gravity to percent-level accuracy, as well as the impact of the baryons.  In particular, a number of physical processes associated with baryons redistribute gas and impact the non-linear matter power spectrum by up to $\sim$30$\%$ \citep[see e.g.][]{Chisari:2019, vandaalen:2020}. This `baryonic feedback' encapsulates the energetic effect of Active Galactic Nuclei (AGN) heating gas and ejecting it to the outskirts of groups and clusters, as well as the likely sub-dominant effects of stellar winds, supernovae feedback and gas cooling \citep{vanDaalen:2011}. At present, astrophysical model uncertainties, such as those due to baryonic feedback, have been shown to limit the precision of weak lensing surveys \citep{Amon:2021,KiDSDES}. Therefore, to extract maximal information from weak-lensing data demands improved modelling of baryonic effects and their impact on the matter distribution.  Moreover, it has been proposed that the `$S_8$ tension' -- the finding that weak lensing constraints on the clustering amplitude parameter, $S_8=\sigma_8(\Omega_{\rm m}/0.3)^{0.5}$, are lower than predictions from the CMB -- could be explained by a suppression of the non-linear matter power spectrum\citep{AAGPE2022,preston/etal:2023}. This could either be caused by a more extreme baryonic feedback effects than hydrodynamical simulations predict or extensions to the standard model of cosmology. In order to isolate a departure from the standard cosmological model, baryonic effects must be better understood.

Powerful AGN feedback is believed to have the ability to eject baryons beyond the virial radius of galaxies, redistributing the gas to the outskirts of galaxy groups and clusters \citep{McCarthy2011,Springel:2018, Dubois2016, Henden2018}.   Indeed, studies of hydrodynamical simulations have demonstrated that AGN feedback alters the total matter distribution relative to dark matter-only simulations, and that it causes a suppression of the power spectrum at scales  $0.1\lesssim k \lesssim 10$$h$Mpc$^{-1}$, whereas increased star formation can enhance power on the smallest scales \citep[see][for a review]{Chisari:2019}. 
These simulations reproduce many of the observed properties of galaxies, including optical properties, galaxy group/cluster profiles, scaling relations and Sunyaev-Zel’dovich counts.
However, despite these successes, the scale, amplitude and redshift dependence of the larger-scale power suppression remain largely uncertain, with significant variation between simulations. These differences are direct outcomes of the `sub-grid' modelling of astrophysical processes, which take place below the resolution scale of the simulation.

Specifically, sub-grid models are required to follow the formation, growth and feedback of black holes, as well as gas cooling, metal enrichment, star formation and associated stellar feedback. The AGN feedback may operate in either kinetic or thermal modes (generally associated with the radio and quasar modes respectively), or alternate between the two depending on the black hole accretion rate in a `two-mode' feedback scenario \citep[see][]{Sijacki2007}. The feedback model normally has a number of associated ill-constrained parameters encoding the feedback strength, such as the efficiency of thermal/kinetic coupling, the black hole accretion rate, and, in some models, the minimum heating temperature of gas cells before a feedback event occurs. While physical arguments can be used to narrow the plausible range of some of these parameters, this is normally not sufficiently constraining for precision cosmology purposes. Thus, it is often the case that the parameters are calibrated against key observables.  
 
X-ray measurements of the hot gas fractions of groups and clusters within the virial radius are most widely used to benchmark the simulations, along with galaxy stellar mass function (GSMF), star formation history and galaxy sizes \citep{McCarthy2017,Henden2018,Schaye2023,Kugel2023, Nelson:2019}. 
Even with identical subgrid physics, the box size and resolution of the simulation can also have a non-negligible impact on the matter distribution \citep{vanDaalen:2011,Pandey2023}. Indeed, there is a large parameter space of feedback prescriptions, modelling choices and simulation properties that result in significant variation in the suppression of the matter power spectrum \citep{Salcido2023,Schaye2023,Hernandez-Aguayo2022,Dave:2019,Henden2018}.  

Recent weak lensing analyses have devised various approaches to mitigate the impact of baryonic feedback on cosmological constraints. The DES Y3 cosmic shear analysis opted to discard measurements from the analysis on angular scales that are impacted by baryonic effects from the analysis \citetext{\citealp{Krause:2022}; \citealp{amon:2022}; \citealp*{Secco:2022}}. Alternatively, baryon feedback has been modelled using a halo model approach \citep{Asgari:2021, li2023hyper}, and using a halo model that is calibrated to a hydrodynamical simulations \citep{mead:2021, KiDSDES}. More recently, \cite{Salcido2023} have developed an emulator trained using a suite of hydrodynamical simulations with varied feedback efficiencies.
 
Instead of relying on the hydrodynamical simulations, the baryonification model is another approach which shifts particle outputs in gravity-only simulations to attain modified halo profiles, modelling the rearrangement of baryon material caused by feedback effects \citep{Schneider:2015}.  This approach has been used in \citet{Schneider:2022}, \citet{ChenDES:2022} and \citet{arico2023}. Other approaches include a principal component analysis \citep{huang:2019}. 

An alternative approach to using models that are informed by hydrodynamical simulations (and therefore indirectly benchmarked against X-ray data) is to jointly analyse weak lensing data with observations of the gas content in and around galaxy groups and clusters. This has been done to improve cosmological constraints by reducing the model-space of the nuisance parameters \citep{troester:2020} and to constrain the suppression of the matter power spectrum \citep{Schneider:2022}. 
 
A highly complementary observable to X-ray is the kinetic Sunyaev-Zel’dovich (kSZ) effect, caused by the Thomson scattering of the CMB photons by free electrons moving with bulk motion in groups and clusters of galaxies relative to the CMB rest-frame \citep{SZ:1980,SZ:1972}. This causes a shift in the CMB temperature while preserving the frequency dependence.  If the bulk line-of-sight velocity is known, the kSZ effect directly measures the free electron number density, independent of temperature \citep{Hand2012, Planck2016, Soergel2016, Schaan:2021}. The kSZ effect is well-suited to probe low density and low temperature environments like the outskirts of galaxies and clusters, whereas X-ray measurements are more sensitive to the inner regions \citep{Amodeo:2021}.  

\noindent
The goals of this work are threefold:
\begin{enumerate}[label=\Roman*., itemsep=0pt, topsep=0pt]
    \item We test the performance of four baryon feedback mitigation strategies for analysing mock and DES Y3 weak lensing data: the DES Y3 scale cut approach \citep{Krause:2022}, a halo model approach \citep{mead:2021} calibrated to a hydrodynamical simulation, an emulator built using a suite of hydrodynamical simulations \citep{Salcido2023} and the baryonification model \citep{Schneider:2015, Schneider:2019}. 
    \item As an alternative to simulation-driven models, we use the most flexible model, baryonification, to jointly analyse the lensing data with the ACT kSZ measurements for improved constraints on the baryonic feedback parameters and therefore the cosmological parameters.
    \item  We use these data to constrain astrophysical observables for the first time, providing a new avenue to benchmark the hydrodynamical simulations.
\end{enumerate}

The paper is structured as follows.  Section~\ref{sec:data} describes the DES Y3 cosmic shear and ACT DR5 kSZ datasets used in this analysis.  Section~\ref{model} outlines the modelling of the cosmic WL and kSZ measurements, including the four baryon models.
  
In Section~\ref{sec:4} we summarise the findings of a mock analysis.  We compare the cosmological constraints when analysing the shear data with different baryon mitigating models and model complexities in Section~\ref{sec:WLresults}.  In Section~\ref{sec:results} we present our constraints on both cosmological and baryonic parameters obtained from a joint WL + kSZ analysis. Finally, in Section~\ref{sec:XSZ} we consider our constraints on the observables that simulations benchmark against. We summarise key findings and conclude in Section~\ref{conclusions}.

%\newpage
\section{Data}\label{sec:data}

\subsection{Dark Energy Survey cosmic shear}\label{sec:desdata}
The Dark Energy Survey (DES) has completed six years of photometric observations in the \textit{grizY} bands, using the 4-metre Blanco telescope located at the Cerro Tololo Inter-American Observatory. The survey spans $\sim$5000deg$^2$ in the Southern Hemisphere. 

For this analysis, we use data taken during the survey's first three years of operation (DES Y3), between 2013 and 2016 \citep{y3-gold}. The DES Y3 footprint covers 4143 deg$^2$ with a number density of $5.59$ galaxies arcmin$^{-2}$ to a depth of $i$$\sim$23.5. The \textsc{metacalibration} weak lensing catalog has over 100 million galaxies that have passed a raft of validation tests \citealp*{Gatti:2021}. The sample has been divided into four redshift bins and the calibrated redshift distributions and associated uncertainty are defined in \citet*{Myles:2021}. Remaining biases in the shape measurement and redshift distributions, primarily due to blending, are calibrated using image simulations, and the associated corrections for each redshift bin are reported in \citet{MacCrann:2022}. 
The DES Y3 cosmic shear tomographic two-point correlation functions, $\xi_{\pm}$, are measured in twenty angular logarithmic bins spanning 2.5 to 250.0 arcmin \citetext{\citealp{Amon:2021}; \citealp*{Secco:2022}}.

\subsection{Atacama Cosmology Telescope Kinetic Sunyaev Zel'dovich}\label{sec:actdata}
The Atacama Cosmology Telescope (ACT) is a 6-metre millimeter waveband telescope, observing thhe cosmic microwave background (CMB). Since first light in 2007, it has had three generations of receivers, the most recent of which is the polarisation-sensitive Advanced ACTPol (AdvACT), which extended the frequency coverage to five bands spanning 28 to 230~GHz. The fifth data release (herinafter DR5) coadds maps collected from 2008 to 2018 covering approximately 18000 deg$^2$ and utilises data from all three generations of receiver \citep{Naess2020}.

This work uses the kinetic Sunyaev Zel'dovich (kSZ) measurements presented in \citet{Schaan:2021}. These are stacked measurements of the ACT DR5 and \textit{Planck} CMB temperature maps at 98~GHz (hereinafter called f90 for consistency with \citet{Schaan:2021}) and 150~GHz (f150) with the reconstructed velocities of the spectroscopic BOSS CMASS galaxy catalog. The galaxy sample spans the redshift range $0.4<z<0.7$ with a median redshift of $z=0.55$.  It corresponds to a selection of relatively massive galaxies with a mean stellar mass of $\log_{10}(M_{\rm star}/{\rm M}_\odot) \approx 11.3$ and an assumed mean halo mass of $\sim 10^{13} {\rm M}_\odot$ \citep{Schaan:2021, Amodeo:2021}, though the latter is quite uncertain. Given its importance to the modelling of the stacked kSZ sample, we will discuss the mean halo mass of this sample later in the work.

\subsection{Hydrodynamical simulations}

\begin{figure}
	\centering
	\includegraphics[width=0.9\linewidth]{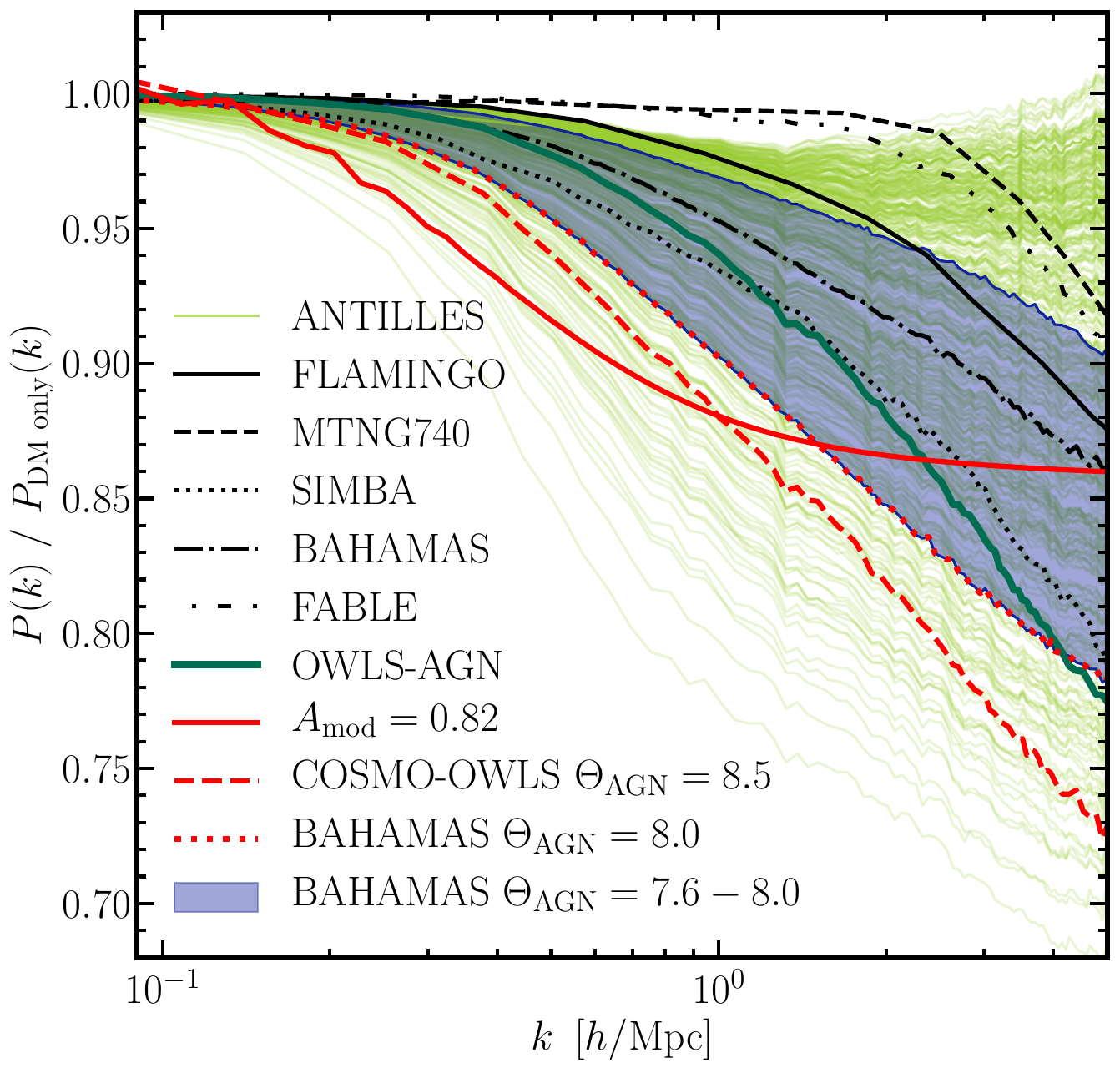} 
	\caption{The suppression of the matter power spectrum due to baryonic effects predicted by hydrodynamical simulations.  We show OWLS-AGN \citep[][dark green solid line]{vanDaalen:2011,Schaye2010}, which informed the scale cuts baryonic feedback mitigation approach.  We also plot the matter power spectrum suppression measured in the 400 simulations of the ANTILLES suite, which calibrated the SP(k) emulator \citep[][light green region]{Salcido2023}.  The range in the suppression predicted by the BAHAMAS suite spanning $\Theta_{\rm AGN}=7.6-8.0$ from which HM20 was calibrated \citep[][dark blue region]{McCarthy2017}.  We further plot FLAMINGO \citep[][black solid line]{Schaye2023}; BAHAMAS \citep[][black dash-dotted line]{McCarthy2017}; SIMBA \citep[][black dotted line]{Dave:2019}, MillenniumTNG \citep[][black dashed line]{Pakmor2023} and FABLE (doubledot-dashed line, \citealt{Henden2018},  Bigwood et al. in prep.).  Finally, we show in red the three baryonic feedback scenarios we test in our mock analysis (Section~\ref{sec:mocks}): BAHAMAS $T_{\mathrm{heat}}=8.0$ (dotted red line), cosmo-OWLS $T_{\mathrm{heat}}=8.5$ (dashed red line) and $A_{\mathrm{mod}}=0.82$ (solid red line) \citep{preston/etal:2023}.}
	\label{fig:sims}
\end{figure}

Throughout the paper, we compare our constraints on the matter power spectrum to predictions from a range of hydrodynamical simulations: the FLAMINGO $(1~\rm{Gpc})^3$ box with baryonic particle masses of $10^9$ $M_{\odot}$ \citep[][solid line]{Schaye2023}; BAHAMAS $(400~\rm{Mpc}/h)^3$ box with $\Theta_{\rm AGN}=7.8$ and $1024^3$ dark matter and baryonic particles \citep[][dash-dotted line]{McCarthy2017}; SIMBA $(100~\rm{Mpc}/h)^3$ box with $1024^3$ gas elements \citep[][dotted line]{Dave:2019}; MillenniumTNG 740~Mpc$^3$ with baryonic mass resolution of $3.1\times 10^7 M_{\odot}$ \citep[][dashed line]{Pakmor2023} and FABLE $(100~\rm{Mpc}/h)^3$ box with $1280^3$ dark matter particles and $~1280^3$ baryonic resultion elements (doubledotted-dashed line, \citealt{Henden2018}, Bigwood et al. in prep.). . These simulations not only span a range of box sizes and resolutions, but also feedback implementations, hydrodynamical schemes and calibration strategies.  Fig.~\ref{fig:sims} shows the suppression of the non-linear matter power spectrum due to baryonic effects, $P(k)/P_{\mathrm{DM only}}$, as predicted by each simulation, demonstrating the spread in the predicted amplitude and scale-dependence.  
Several additional hydro simulations are also used to calibrate the baryonic feedback models tested throughout this work.  We therefore also plot the prediction from the OWLS-AGN  $(100~\rm{Mpc}/h)^3$ box with $512^3$ dark matter and baryonic particles respectively \citep[][dark green solid line]{vanDaalen:2011,Schaye2010}, the span in the suppression predicted by BAHAMAS when $\Theta_{\rm AGN}$ is varied within the range 7.6-8.0 (see Section~\ref{sec:hmcode}), and range modification measured by the ANTILLES suite of 400 simulations \citep[][light green region]{Salcido2023}, each with box size $(100~\rm{Mpc}/h)^3$ and $256^3$ dark matter and baryonic particles respectively.  ANTILLES spans significantly more extreme feedback scenarios than the other simulations considered, such that baryonic effects impact the power spectrum with greater amplitude and at smaller $k$ scales.

\section{Modelling}\label{model}

\subsection{Cosmic shear signal}

The shear two-point correlation functions, $\xi_\pm(\theta)$, for a given angular separation, $\theta$, computed for redshift bins $i,j$,
can be related to the 3D non-linear matter power spectrum. First, it can be expressed as a decomposition of the 2D convergence power spectrum $C^{i,j}_{\kappa}(\ell)$ at an angular wave number, $\ell$, as

\begin{equation}
\centering
\label{eqn:2ptP}
 \xi_{\pm}(\theta) = \sum_{\ell} \frac{2\ell+1}{4\pi} G^\pm_\ell(\cos \theta) [C_{\kappa,\textrm{EE}}^{ij}(\ell) \pm C_{\kappa, \textrm{BB}}^{ij}(\ell)] \, .
\end{equation}
\noindent
We note that weak gravitational lensing does not produce B-modes. However, we show the more general expression here, as a B-mode contribution from intrinsic alignments is possible. The functions $G^{\pm}_{\ell}$ are computed from the Legendre polynomials following \citet{Stebbins1996}.

Under the Limber and flat-sky approximations \citep{Limber1953,LoVerde2008}, we can relate $C^{i,j}_{\kappa}(\ell)$ to the 3D non-linear matter power spectrum, $P$, via
\begin{equation}\label{eq:c_kappa}
    C_{\kappa}^{ij}(\ell) = \int_0^{\chi_{\rm H}} d\chi  \frac{W_i(\chi)W_j(\chi)}{\chi^2}  P\left(k=\frac{\ell+0.5}{\chi(z)}, z\right)\,.
\end{equation}
Here, we assume a spatially flat Universe, $\chi$ is the comoving angular diameter distance and $\chi_{\rm H}$ is the horizon distance. $W_i(\chi)$ are the lensing efficiency kernels, given by: 
\begin{equation}\label{eq:kernel}
W_i(\chi)= \frac{3H_0^2\Omega_{\rm m}}{2c^2}\frac{\chi}{a(\chi)} \int_{\chi}^{\chi_{\rm H}} d\chi' \, n_i(\chi') \frac{\chi'-\chi}{\chi'} \,,
\end{equation}
where $a(\chi)$ is the scale factor at comoving distance $\chi$, and $n_i(\chi) d\chi$ is the effective number density of galaxies in $d\chi$, normalised to unity. 

In this analysis, the linear matter power spectrum is calculated using {\sc CAMB} \citep{CAMB}, and the non-linear correction is determined using {\sc HMCode2020} \citet{mead:2021}. We refer the reader to \citet{KiDSDES} for a detailed comparison of cosmic shear analysis choices, which we use as a guide to formulate the baseline model in this work. We assume three neutrino species with two massless states and one massive state with a mass fixed at the minimum mass allowed by oscillation experiments, $m_\nu=0.06$eV \citep{Patrignani_2016}. 

The intrinsic alignment (IA) of galaxies with their local environment also contributes to the shear correlation function and must be modelled. We chose to do this using the non-linear linear-alignment model (NLA), which describes the linear tidal alignment of galaxies with the density field \citep{hirata/etal:2004}, with a non-linear correction to the linear matter power spectrum \citep{bridle/king:2007}. This approach requires two additional free nuisance parameters: $A_{\rm IA}$, modulating the amplitude of the intrinsic alignment model (see equation~[3-5] in \citealt{bridle/king:2007} for the NLA intrinsic alignment power spectra, $C_{\rm GI}$ and $C_{\rm II}$) and a redshift-dependence parameter, using a power-law with $[(1+z)/(1+0.62)]^{\eta_{\rm IA}}$. 

We model the uncertainty in mean redshift and the shear calibration for each redshift bin $i$ as the free parameters $\Delta z^i$ and $m^i$, respectively, following \citet{Amon:2021} and \citet*{Secco:2022}, and preserve values of the uncertainty determined by \citet*{Myles:2021} and \citet{ MacCrann:2022}. In the cases where the small-scale measurements are analysed, we refer the reader to \citet{ChenDES:2022} for validation that higher-order cosmic shear modelling corrections remain subdominant.

\subsection{Modelling baryonic feedback for cosmic shear}\label{sec:models} 

Strategies have been devised to account for baryonic effects when analysing weak lensing measurements in order to extract unbiased cosmological information. In this section we outline the four approaches we investigate in this work: (1) restrict angular scales, (2) a halo model approach, (3) a hydrodynamical simulation-based emulator and (4) an analytical N-body simulation model. In Table~\ref{tab:modelparameters} we summarise the free parameters of each baryon feedback model and prior choices.

\begin{table*}
\setlength\extrarowheight{3pt}
\caption{Qualitative descriptions of the parameters associated with the HM20, SP(k) and BCEmu methods, which each model the impact of baryonic feedback on the non-linear power spectrum. We also show the priors utilised on each parameter in this analysis, and the conservative prior alternative if applicable. U$[\,]$ brackets indicate flat uniform priors within the range shown.  $\mathcal{N}(\,)$ brakets indicate Gaussian priors described by their mean and 1$\sigma$ width.  The BCEmu1 model has $\log_{10}M_{\rm c}$ as the only free parameter, BCEmu3 varies $\log_{10}M_{\rm c}$,  $\theta_{\rm ej}$ and $\eta_\mathrm{cga}$, and all seven parameters are free in BCEmu7. In the case of using the reduced model complexity of BCEmu1 or BCEmu3 models, we also show the values that the BCEmu parameters are fixed to. Note that the $M_{200}$ mass parameter is only included in the BCEmu joint WL + kSZ analysis and the choice of the fixed value is adopted from \citet{Schaan:2021}.}
\begin{tabular}{ccccc}
\hline
Parameter & Description & Prior & Wide prior & Fixed value\\
\hline
\bf{Halo model: HM20}\\
$\Theta_{\rm AGN}\log_{10}(T_{\mathrm{AGN}}/K)$ & Subgrid heating parameter calibrated to the BAHAMAS & U[7.6, 8.0] & U[7.3, 9.0]  &- \\
&   simulations designed to modulate the amplitude and shape of    \\
& the `one-halo' term in the halo model \\
\hline
\bf{Simulation-based: SP(k)}\\
$\alpha$ & Power-law normalisation for the $f_{\rm b}$ - $M_\mathrm{halo}$ & $\mathcal{N}$(4.16, 0.07) & U[2.85, 4.50] & -\\
& relation (equation~\ref{eq:spksingle}) \\
$\beta$ & Power-law slope for the $f_{\rm b}$ - $M_\mathrm{halo}$ relation (equation~\ref{eq:spksingle}) & $\mathcal{N}$(1.20, 0.05) &  U[0.95, 1.60] & -\\
$\gamma$ & Redshift dependence of the power-law normalisation of  & $\mathcal{N}$(0.39, 0.09) & U[0.12, 0.85] & -\\
& the $f_{\rm b}$ - $M_\mathrm{halo}$ relation (equation~\ref{eq:spksingle})
\\

% $f_{\rm b}$ &  Median total baryon fraction of haloes of mass $M_{200{\rm c}} = 10^{14}\Msun$\\
% &normalised by the universal baryon fraction $(\Omega_{\rm b}/\Omega_{\rm m})$ &  \\
\hline
\bf{Baryonification: BCEmu}\\
$\log_{10}M_{\rm c}$ & The
characteristic mass scale at which the slope of the gas & U[11.0, 15.0] & - & - \\  & profile becomes shallower
than -3 (equation~\ref{eq:rhogas}, , equation~\ref{eq:beta}). \\
\hdashline
$\theta_{\rm ej}$ & Specifies the maximum radius of gas ejection
 & U[2.0, 8.0]&- & 3.5 \\
& relative to the virial radius. \\
$\eta_{\delta}$ & Related to the stellar fraction of the central galaxy:  & U[0.05, 0.4] & - & 0.20 \\ 
&  $\eta_{\delta}= \eta_{\rm{cga}}-\eta$ (equation~\ref{eq:fstar}) \\
\hdashline
$\mu$ & Defines
how fast the slope of the gas profile becomes & U[0.0, 2.0]&- & 1.0 \\
& shallower towards small halo masses  (equation~\ref{eq:rhogas}, equation~\ref{eq:beta}).\\ 
$\gamma$ & Exponent in gas profile parameterisation (equation~\ref{eq:rhogas}) & U[1.0, 4.0]& - & 2.5\\
$\delta$ & Exponent in the gas profile parameterisation (equation~\ref{eq:rhogas}) & U[3.0, 11.0]& - & 7.0\\
$\eta$ & Specifies the total stellar fraction within a halo (equation~\ref{eq:fstar}) & U[0.05, 0.4]&  - &0.20\\
\hdashline
$M_{\rm h, 200}$ ($M_{\odot}$)& Halo mass of the kSZ sample, used in the joint analysis only  & U[$5\times10^{12}$, $7\times10^{13}$] &  - & $3\times10^{13}$ \\
\hline
\end{tabular}\label{tab:modelparameters}
\end{table*}

\subsubsection{Restricting angular scales}\label{sec:sc}
The DES Y3 cosmic shear analysis mitigates the impact of baryonic effects by discarding the measurements at small angular scales \citetext{\citealp{Krause:2022}; \citealp{amon:2022}; \citealp*{Secco:2022}} and analysing the data assuming a dark matter-only model. In this work, we adopt their `$\Lambda$CDM Optimised' scale cuts, which were designed to minimise the bias on $\Lambda$CDM cosmological parameters due to unmodelled baryonic effects to be less than 0.3$\sigma_{\rm{2D}}$ in the $\Omega_{\rm m}-S_8$ parameter space, for a joint lensing and clustering analysis. For a $\Lambda$CDM analysis of cosmic shear alone, this corresponds to up to a 0.14$\sigma_{\rm{2D}}$ potential bias. Note that to determine the angular scales to be used, the OWLS-AGN hydrodynamical simulation \citep{vanDaalen:2011,Schaye2010} was chosen as a representative feedback scenario (shown as the green line, Fig~\ref{fig:sims}). Synthetic cosmic shear data vectors were contaminated according to 
\begin{equation}
    P_{b}(k,z) = \frac{P_{\rm hydro}(k,z)}{P_{\rm DMO}(k,z)}P (k,z)\,,
\end{equation}
where $P_{\rm hydro}(k,z)$ and $P_{\rm DMO}(k,z)$ are the full hydrodynamical and dark matter-only matter power spectra from the OWLS-AGN suite and $P(k,z)$ is the non-linear matter power spectrum. 

A benefit of this approach is that it is agnostic to the exact shape and physics of the matter power spectrum suppression, once the feedback in the Universe is lower in amplitude and scale extent than the simulation chosen (in this case, OWLS-AGN). However, this approach misses the opportunity to extract high-signal-to-noise information about the underlying cosmological model and the astrophysical effects.

\subsubsection{Halo-model approach: HM20}\label{sec:hmcode}

\textsc{HMcode2020}, hereinafter HM20, models the non-linear power spectrum and includes a free parameter to modulate the amount of baryonic feedback, $\Theta_{\mathrm{AGN}}=\log_{10}(T_{\mathrm{AGN}}/K)$ \citep{mead:2021}. This parameter scales the halo concentration and the stellar and gas content, leading to a modification in the overall amplitude and shape of the `one-halo' term in the halo model. The model is calibrated to fit the  power spectrum `response' (the matter-matter power spectrum divided by the same measurement in an equivalent dark matter-only box) of the {\sc BAHAMAS} hydrodynamical simulations \citep{McCarthy2017, vandaalen:2020} in the range $\Theta_{\mathrm{AGN}}=[7.6-8.0]$ (blue shaded region, Fig.~\ref{fig:sims}). We note that $T_{\mathrm{AGN}}$ is related to $\Delta T_{\mathrm{heat}}$, which is the {\sc BAHAMAS} subgrid heating parameter, where an AGN feedback event will only occur after the black hole has stored sufficient energy to heat a fixed number of gas particles by $\Delta T_{\mathrm{heat}}$. 

We define two prior ranges for this case. The first spans the range of $\Theta_{\mathrm{AGN}}$ values that bracket the BAHAMAS $\Theta_{\mathrm{AGN}}=7.6-8.0$ simulations that the model was calibrated against. The `wide prior' chosen here to be $\Theta_{\mathrm{AGN}}=7.3-9.0$ extends beyond the calibration range to span more extreme scenarios and allow for a dark matter-only case. 

This approach has been shown to be accurate at the level of $< 2.5$\% to $k < 10 h$ Mpc$^{-1}$ \citep{mead:2021} when fitting simulated power spectra at $z<2$ with a range of cosmologies 
thus allowing all measured angular scales of the DES Y3 dataset to be utilised (in this case, to $2.5$~arcmin).  A downside of this model is that it relies on the accuracy of the specific feedback implementation and predicted power suppression of a particular simulation and may not be flexible enough to capture the true scenario.

\subsubsection{Hydrodynamical simulation emulator: SP(k)}\label{sec:Spk} 
SP(k) is a flexible empirical model trained on the ANTILLES suite of 400 cosmological hydrodynamical simulations \citep[plotted as the green lines in Fig.~\ref{fig:sims},][]{Salcido2023}. The model predicts the power spectrum suppression given the baryon fraction--halo mass relation of galaxy groups and clusters as input, building upon previous insights from \citet{vandaalen:2020}.  The ANTILLES suite span a range of feedback scenarios, allowing the emulator to achieve a ${\sim}2\%$ level accuracy to describe baryonic effects at scales of up to {$k \, {\lesssim} \, 10 h$ Mpc$^{-1}$} and redshifts up to $z=3$.

In particular, SP(k) casts the suppression in terms of the baryon fraction at the \textit{optimal mass}, $\hat{M}_{k}$, defined as the halo mass that maximizes the strength of the correlation between the suppression of the total matter power spectrum and the total baryon fraction of haloes of different mass. It uses an exponential plateau function to model for the fractional impact of baryons as,
\begin{equation}\label{eq:sup_fit}
    P_\mathrm{hydro}(k)/P_\mathrm{DM}(k) = \lambda(k,z) - \qty[\lambda(k,z) - \mu(k,z)] \exp[{-\nu(k,z)\tilde{f}_{\rm b}}] \ \ ,
\end{equation}
where $\tilde{f}_{\rm b}$ is the baryon fraction at the optimal halo mass normalised by the universal baryon fraction, i.e.: 
\begin{equation}\label{eq:opt_fb}
{\tilde{f}_{\rm b} = f_{\rm b}(\hat{M}_{k,\mathrm{SO}}(k,z))/\qty(\Omega_{\rm b}/\Omega_{\rm m})} \ \ ,
\end{equation}
and $\hat{M}_{k,\mathrm{SO}}$, $\lambda(k,z)$, $\mu(k,z)$ and $\nu(k,z)$ are functions with best-fit parameters given in \cite{Salcido2023}.

For the mass range that can be relatively well probed in current X-ray and Sunyaev-Zel'dovich effect observations ($10^{13} \lesssim M_{200} \,\, [\mathrm{M}_\odot] \lesssim 10^{15} $), the total baryon fraction of haloes can be roughly approximated by a power-law with constant slope \citep[e.g.][]{Mulroy_2019,Akino_2022}. \cite{Salcido2023} find that a modified version of the functional form\footnote{See \href{https://github.com/jemme07/pyspk}{https://github.com/jemme07/pyspk}.} presented in \cite{Akino_2022} provides a reasonable agreement with simulations up to redshift $z=1$,
\begin{equation}\label{eq:spksingle}
    f_{\rm b}/(\Omega_{\rm b}/\Omega_{\rm m})= \left(\frac{e^\alpha}{100}\right) \left(\frac{M_{500c}}{10^{14} \mathrm{M}_ \odot}\right)^{\beta - 1} \left(\frac{E(z)}{E(0.3)}\right)^{\gamma},
\end{equation}
where $\alpha$ sets the power-law normalisation, $\beta$ sets power-law slope, $\gamma$ provides the redshift dependence, $E(z)$ is the dimensionless Hubble parameter and $f_{\rm b}$ is the baryon fraction measured within $R_{500}$.  We use this function to facilitate the marginalization over the uncertainties in the observed baryon fractions, introducing $\alpha$, $\beta$ and $\gamma$ as free parameters in our WL analysis.

SP(k) effectively depends only on a single physically-meaningful parameter, i.e. the baryon fraction. The benefit of this approach is that observational constraints on the baryon fraction could be used to inform the priors used in cosmological analysis. For our study we use two different sets of priors to marginalize over the uncertainties in the observed baryon fractions: wide (conservative) priors consistent with the range of feedback models probed by the ANTILLES simulations used to calibrate the SP(k) model \citep{Salcido2023}, and `observational' priors that encompass current observational constraints on the baryon fraction – halo mass relation from \cite{Akino_2022}. 
Table~\ref{tab:modelparameters} reports the priors on $\alpha$, $\beta$ and $\gamma$ for the two choices.  

As with the scale cuts and HM20 model, we caveat that the SP(k) method is limited by the specific feedback implementation used in the hydrodynamical simulation it was trained on, in this case the ANTILLES suite. While ANTILLES is currently the largest and widest suite of hydro simulations in terms of feedback variations, it still may not cover all possible baryonic responses (see e.g. Appendix ~\ref{sec:vandaalen}).

\subsubsection{Analytical N-body simulation  model: Baryonification}\label{sec:baryonification} 

Baryonification is a method for including the effects of baryonic feedback in dark matter-only $N$-body simulations based on perturbative shifts of particles that mimic the effects of feedback at cosmological scales \citep{Schneider:2015, Schneider:2019, Arico:2020}. The particles are shifted in order to obtain modified halo profiles that include the presence of gas and stars which are shaped by feedback effects. We provide a summary of the method including some important aspects of the parametrisation and refer the reader to \citet{Schneider:2019} and \citet{Giri:2021} for a more detailed explanation. 

In practice, the baryonification (bfc) method relies on a modification of profiles via spherically symmetrical particle shifts around the centres of $N$-body haloes with NFW-like profiles, $\rho_{\rm nfw}$, following
\begin{equation}
\label{eq:bfc_profile}
\rho_{\rm nfw}(r)\,\,\rightarrow\,\,\rho_{\rm bfc}(r)=\rho_{\rm clm}(r) + \rho_{\rm gas}(r) + \rho_{\rm cga}(r) \,.
\end{equation}
The final \emph{baryonified} profiles, $\rho_{\rm bfc}$, consist of three components: the collisionless matter, gas and central galaxy. The collisionless matter ($\rho_{\rm clm}$) profile is dominated by dark matter but also contains satellite galaxies and halo stars. Its shape is modified with respect to the original NFW shape via adiabatic contraction and expansion \citep{Teyssier2011}. The central galaxy profile, $\rho_{\rm cga}$, is parametrised as a power law with an exponential cutoff. This component affects only the innermost part of the halo, rather than cosmological scales. 

At cosmological scales, baryonic effects are primarily caused by feedback-induced changes of the gas distribution around haloes, described by the gas profile, $\rho_{\rm gas}$. Motivated by X-ray observations \citep{Eckert2016, Schneider:2019}, these effects are parametrised with five model parameters 
%($M_{\rm c}$, $\mu$, $\theta_{\rm ej}$, $\delta$, $\gamma$) 
as
\begin{equation}
\label{eq:rhogas}
\rho_{\rm gas}(r) \propto \frac{\Omega_{\rm b}/\Omega_{\rm m}-f_{\rm star}(M)}{\left[1+\left(\frac{r}{r_{\rm core}}\right)\right]^{\beta(M)}\left[1+\left(\frac{r}{r_{\rm ej}}\right)^{\gamma}\right]^{\frac{\delta-\beta(M)}{\gamma}}} \, ,
%\rho_{\rm gas}(r) \propto \frac{1}{\left[1+\left(\frac{r}{r_{\rm core}}\right)\right]^{\beta(M)}\left[1+\left(\frac{r}{\theta_{\rm ej}R_{200}}\right)^{\gamma}\right]^{\frac{\delta-\beta(M)}{\gamma}}},
\end{equation}
which consists of a cored power-law profile with a truncation at the ejection radius, $r_{\rm ej}=\theta_{\rm ej}R_{200}$, where $\theta_{\rm ej}$ is a dimensionless free model parameter\footnote{Note that the gas profile is normalised so that an integral over $r^2/2\pi^2 \times \rho$ gives the total halo mass.}.  The shape of the truncation beyond the ejection radius is controlled by the $\gamma$ and $\delta$ parameters, where the former defines the abruptness and the latter defines the slope of the function. The core of the profile is fixed to $r_{\rm core}=\theta_{\rm core}R_{200}$ with $\theta_{\rm core}=0.1$. The slope of the power law $\beta(M)$ is a function that varies with halo mass \footnote{Here we define the halo mass as the mass enclosed within a radius, centred on the group or cluster, within which the mean density is 200 times the critical density of the Universe.  The notation $M_{200}$ is also used throughout the paper.} and is parameterised as
\begin{equation}\label{eq:beta}
\beta(M)=\frac{3(M/M_{\rm c})^{\mu}}{1+(M/M_{\rm c})^{\mu}} \,.
\end{equation}
The halo mass-dependence of $\beta$ accounts for the fact that AGN feedback is more efficient in removing gas around galaxy groups while large clusters tend to keep most of their gas inside the virial radius. The free model parameters $M_{\rm c}$ and $\mu$ thereby define the scale and the abruptness of the transition when the slope of the profile goes from 3 to 0 for decreasing halo masses.

The total fraction of stars, $f_{\rm star}$, and the fraction of stars that belong to the central galaxy, $f_{\rm cga}$, indirectly affect the available gas that can be pushed out by feedback processes. They are parametrised as
\begin{equation}\label{eq:fstar}
f_i(M)= 0.055\left(\frac{M}{M_{\rm s}}\right)^{\eta_i} \,,
\end{equation}
with $i=\lbrace{\rm star, cga}\rbrace$ and where $M_{\rm s}=2.5\times10^{11}$ M$_{\odot}$/h. The power law is constructed to match the Moster relation \citep{Moster2013}, which is reasonably well-understood for galaxy groups and clusters. Note that following \citet{Giri:2021}, we redefine the parameters as $\eta\equiv\eta_{\rm star}$ and $\delta_\eta\equiv \eta_{\rm cga}-\eta_{\rm, star}$. These two additional bfc parameters, together with  the five gas parameters, are summarised in Table~\ref{tab:modelparameters}.

An efficient cosmological analysis will marginalise over a minimum number of baryonic feedback parameters. How many parameters are required is an open research question that ultimately depends on the unknown baryonic feedback realised in nature. For now, we adhere to the requirement that a given parametrisation needs to be able to fit the matter power spectrum suppression predicted by a range of hydrodynamical simulations. Following \citet{Giri:2021}, in this work we consider the models BCEmu7, BCEmu3, and BCEmu1, referring to the number of model parameters varied in the analysis. While in BCEmu7 all parameters introduced above are kept free, BCEmu3 only allows $\log_{10}M_{\rm c}$, $\theta_{\rm ej}$ and $\eta_{\delta}$ to vary and BCEmu1 only uses $\log_{10}M_{\rm c}$ as a free parameter. The fixed parameters in the BCEmu3 and BCEmu1 models are listed in Table~\ref{tab:modelparameters} and have been selected so that they provide the best fit to a variety of hydrodynamical simulations \citep[see][for more information about the method]{Giri:2021}. Both BCEmu7 and BCEmu3 are able to reproduce the baryonic suppression or the power spectrum predicted by a variety of hydrodynamical simulations to better than one percent. The BCEmu1 model, on the other hand, shows deviations of order five per cent, hinting towards the possibility that one parameter is generally insufficient to describe the variety of existing results from  simulations \citep{Giri:2021}.

A key feature of the baryonification model is that it is based on physically motivated profiles around halo centres. The model is not restricted to the power spectrum but can also be used to obtain the three-dimensional baryonified density field and therefore many corresponding summary statistics. In the following we will take advantage of the connection between power spectrum and halo profiles to obtain simultaneous predictions for the cosmic shear and the stacked kinetic Sunyaev-Zel'dovich signal. Compared to other approaches, the baryonification model BCEmu7 is not calibrated to specific hydrodynamical simulations. It rather depends on the parameters of empirical density profiles that are broadly motivated by observations. The model therefore provides a independent check with very different modelling choices and systematics compared to the subgrid modelling in hydrodynamical simulations.  We note that despite the model's valuable flexibility, it can result in potentially non-physical scenarios.  Other limitations include the fact that the model parameters are currently assumed to have no redshift dependence, as well as the fact that the gas profiles do not separate a hot and cold gas component.

\subsection{Kinetic Sunyaev
Zel’dovich signal}\label{sec:ksz}

The kSZ measurements can be used to constrain the gas density. This effect arises from the bulk motion of the ionised gas in and around galaxies, galaxy groups and clusters, which imparts a Doppler shift on CMB photons.  It preserves the blackbody frequency spectrum of the CMB and instead the thermodynamic temperature fluctuates as
\begin{equation}
    \frac{\Delta T_{\mathrm{kSZ}}}{T_{\mathrm{CMB}}}=\frac{\sigma_{\rm T}}{c} \int_{\rm los} e^{-\tau} n_{\rm e} v_{\rm p} dl\, ,
\end{equation}
where $T_{\mathrm{CMB}}$ is the present-day temperature of the CMB, $\sigma_{\rm T}$ is the Thomson cross-section, $c$ is the speed of light, $n_e$ is the free-electron physical number density, $v_{\rm p}$ is the peculiar velocity and $\tau$ is the optical depth due to Thomson scattering along the line-of-sight distance, $dl$. Following \citet{Schaan:2021} and \citet{Amodeo:2021}, for the redshift range of the kSZ measurements used in this work, the mean optical depth is observed to be below the percent level \citep{Planck2016} and the CMASS galaxy groups are optically thin, therefore we can assume that the integral $e^{-\tau}\approx1$. Furthermore, as the measurements are stacked, the velocity field is independently estimated from the large-scale distribution of galaxies via a reconstruction method, thereby eliminating the dependence on the velocity. The peculiar velocity, $v_{\rm p}$, projected along the line-of-sight can be written as the root mean square (RMS) of the peculiar velocity, $v_{\rm r}$, so that the resulting shift in the CMB temperature can be approximated as
\begin{equation}\label{eq:ksz}
    \frac{\Delta T_{\mathrm{kSZ}}}{T_{\mathrm{CMB}}}=\tau_{\rm gal}(\theta)\frac{v_{\rm r}}{c}\, ,
\end{equation}
where $\tau_{\rm gal}$ refers to the contribution of the optical depth to Thomson scattering of the galaxy group considered. For the median redshift of the CMASS sample in the linear approximation, $z=0.55$, the RMS of the peculiar velocities projected along the line of sight is $v_r=1.06\times 10^{-3} c$ \citep{Schaan:2021}.  The uncertainty on the velocity reconstruction is estimated to be less than a few percent, which we ignore given the statistical precision \citep{Schaan:2021,RiedGuachalla2023,Hadzhiyska2023}.

To model the ACT kSZ measurements it is necessary to convolve equation~\ref{eq:ksz} with the beam profiles utilised at the f90 and f150 frequencies.  We follow \citet{Schaan:2021} and approximate the beam using a Gaussian with $\mathrm{FWHM}=2.1$~arcmin for the former band and $\mathrm{FWHM}=1.3$~arcmin for the latter.  Furthermore, to minimise noise due to degree-scale CMB fluctuations, compensated aperture photometry filters were also used on the observations.  We therefore apply the same filter function that was used in the analysis of the data.  The smoothing function is $+1$ between $\theta<\theta_{\rm d}$, $-1$ between $\theta_{\rm d}<\theta<\sqrt{2}\theta_{\rm d}$ and 0 otherwise, where $\theta_{\rm d}$ is the aperture radius centred around each galaxy (\citealt{Schaan:2021}, equation~9).

To calculate $\tau_{\rm gal}$ measured within a disk of radius $\theta$ centred on the group or cluster, we assume spherical symmetry and integrate the electron number density, $n_{\rm e}$, over the line-of-sight to a cut-off radius of $R_{200}$ as 

\begin{equation}
    \tau_{\rm gal}(\theta)=2\sigma_T\int_{0}^{R_{200}}n_{\rm e}(\sqrt{l^2+d_{\rm A}(z)^2\theta^2})dl \,,
\end{equation}
where $d_{\rm A}$ is the angular diameter distance.  We assume a fully ionised medium with primordial abundances to describe the electron density in terms of the gas density as
\begin{equation}\label{eq:ne}
    n_{\rm e}(r)=\frac{(X_{\rm H}+1)}{2}\frac{\rho_{\mathrm{gas}}(r)}{m_{\mathrm{amu}}} \, ,
\end{equation}
with $X_{\rm H}=0.76$ being the hydrogen mass fraction and $m_{\mathrm{amu}}$ the atomic mass unit.  

In order to calibrate the model for the kSZ profile, the mean halo mass of the CMASS galaxy sample is needed. Note that because the integrated kSZ signal scales with the gas mass and this quantity approximately tracks the halo mass, it is important that the theoretical predictions are for a sample with the same mean halo mass as the CMASS sample.  
Given the significant scatter in the literature, we choose to include an additional model parameter in the analysis, $M_{\rm h, 200}$, corresponding to the mean $M_{200}$ of the CMASS sample, with a prior range provided in Table~\ref{tab:modelparameters}. The justification for this prior choice, and an investigation of its impact are given in Appendix~\ref{app:mhalo}.

\begin{figure*}
\begin{minipage}{\linewidth}
\centering
\subfloat{\includegraphics[width=1\linewidth]{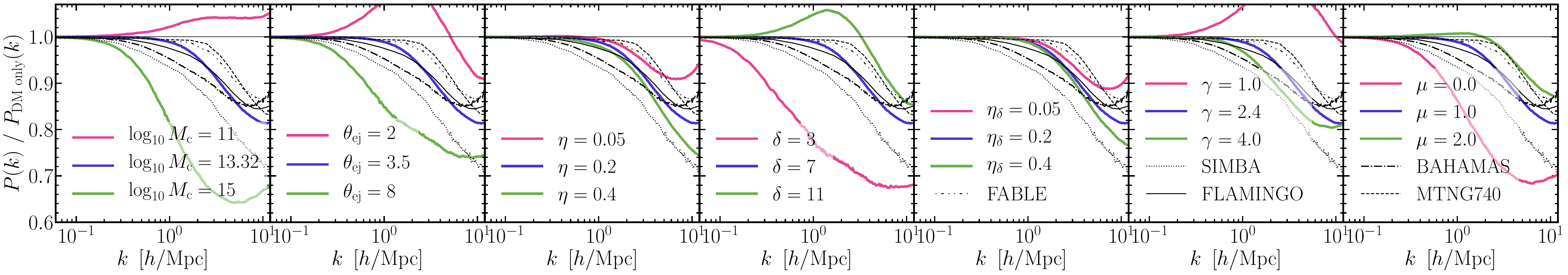}}
\end{minipage}
\begin{minipage}{\linewidth}
\centering
\subfloat{\includegraphics[width=1\linewidth]{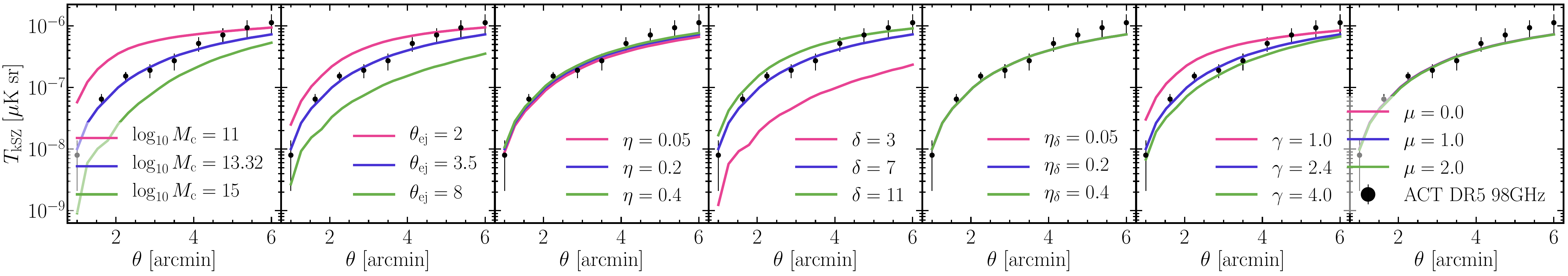}}
\end{minipage}
\caption{Top: The impact of varying parameters of the baryonification model, BCEmu, on the ratio of the total matter power spectrum compared to a dark-matter only power spectrum, $P(k)/P_{\rm DM only}(k)$. Each panel varies one baryonification parameter at a time within the prior bounds (reported in Table~\ref{tab:priors}), whilst keeping the remaining six parameters at their fiducial fixed value, corresponding to the fit to a range of hydrodynamical simulations \citep{Giri:2021}.  For reference, various predictions for the suppression of the matter power spectrum from simulations are over-plotted in black: FLAMINGO \citep[][solid line]{Schaye2023}; BAHAMAS \citep[][dash-dotted line]{McCarthy2017}; SIMBA \citep[][dotted line]{Dave:2019}, MillenniumTNG \citep[][dashed line]{Pakmor2023} and FABLE (doubledotted-dashed line, \citealt{Henden2018}, Bigwood et al. in prep.). Bottom: The stacked kSZ temperature profile at 98GHz as a function of angular radius, $\theta$, centred on the group or cluster (bottom) for the same baryonification parameters as above.  The ACT CMASS measurements at 98GHz are shown as the black data points in the bottom panels and the model profiles are convolved with the f90 beam profile for comparison. }
\label{fig:bfcintuition}
\end{figure*}

\subsection{Modelling kSZ with baryonification}

The baryonification method provides a model for the gas density (equation~\ref{eq:rhogas}) and describes the gas content as measured by kSZ (equations~\ref{eq:ksz}-\ref{eq:ne}). We select this model to jointly analyse the kSZ measurements with the lensing data given that it is agnostic to any choice of hydrodynamical simulation.

In Fig.~\ref{fig:bfcintuition}, we aim to build a better intuition on the BCEmu model by exploring the impact of each parameter on the suppression of the matter power spectrum, $P(k)/P_{\mathrm{DM only}}(k)$ (top row) and the kSZ radial temperature profile $T_{\mathrm{kSZ}}(\theta)$ (bottom row).  At a fixed cosmology\footnote{We chose the parameters $h=0.742$, $\Omega_{\rm m}=0.255$, $f_{\rm b}=\Omega_{\rm b}/\Omega_{\rm m}=0.166$.}, we test the dependence on each of the seven BCEmu parameters from left to right within the defined prior range, whilst fixing the remaining six model parameters at their fiducial fixed value (see Table~\ref{tab:modelparameters}).  

The $M_{\rm c}$ parameter controls the mass or proportion of groups and clusters that have had feedback-induced gas removal and has been previously identified as the most important in the model \citep{Schneider:2015, Giri:2021}. Indeed, we see that within its prior range, this parameter modulates the amplitude, slope and extent of the suppression of the matter power spectrum at scales $k\gtrsim 0.1$ and the amplitude and slope of the kSZ profile. For clusters of mass greater than $M_{\rm c}$, the slope of the gas profile (equation~\ref{eq:beta}) tends to $\beta=3$ and the gas profile approaches a truncated NFW profile.  However, for groups of mass smaller than $M_{\rm c}$, the slope of the power law decreases and the gas profile more closely resembles one that has experienced AGN feedback and had gas ejected from the halo.  Hence, a larger value of $M_{\rm c}$ results in a greater proportion of groups and clusters having gas profiles that mimic the effect of baryonic feedback, leading to a greater suppression of the matter power spectrum and simultaneously, a decrease in the amount of gas within an aperture centred on the galaxy, which gives a smaller kSZ signal. 

In the reduced complexity BCEmu1 model, the remaining parameters are kept fixed, although it is clear that their value choice can have a significant impact on the predictions for the matter power spectrum and the kSZ signal. We observe that even at fixed $M_{\rm c}$, a larger value of $\theta_{\mathrm{ej}}$ increases the radius that gas is `ejected to' in the gas profile and effectively causes matter to be redistributed on smaller $k$-scales. This corresponds to a decrease in the amount of gas as measured by the amplitude of the kSZ signal.  Similarly, we find that increasing the exponent in the gas profile parameterisation $\gamma$ results in a greater suppression of the power spectrum and lower kSZ amplitude, although the impact is on smaller scales ($k\gtrsim 1$) than the effect of $\log_{10} M_{\rm c}$ and $\theta_{\mathrm{ej}}$.  
The remaining gas parameters, $\delta$ and $\mu$, have the reverse effect, i.e., decreasing them results in an enhanced power spectrum suppression and a lower kSZ signal.  

Decreasing the values of stellar parameters, $\eta$ and $\eta_{\delta}$, makes a less steep stellar-halo mass relation, so that a larger amount of stars condense out of the gas and form galaxies.  This causes a boost in the matter power spectrum at large $k$, as we observe in Fig.~\ref{fig:bfcintuition}.  The kSZ radial temperature profile is only dependent on the gas density profile of groups and clusters (equation~\ref{eq:ne}) and so not directly impacted by the stellar parameters $\eta$ and $\eta_{\delta}$.  However, $\eta$ can indirectly have a small impact on the kSZ signal as it alters the number of stars in groups and clusters, and therefore the the reservoir of available gas.

We note two limitations in our modelling of the kSZ signal. While we vary the halo mass of the kSZ sample in our analysis, we assume that the mass dependence of the model is valid beyond the range probed by the kSZ measurement and at varying redshift.  The kSZ measurements span a mass range $M_{200}\approx[0.5-7]\times10^{13}\msun$, which is similar to the halo mass range that cosmic shear is most sensitive to ($M_{200}\approx10^{13.5}\msun$). Nevertheless, cosmic shear still has contributions from higher and lower mass halos \citep[e.g.][]{Salcido2023}. 

Future kSZ measurements that span multiple mass and redshift bins will provide a better understanding of the dependence of the kSZ signal on galaxy properties and the suitability of this assumption. 
Another issue that we do not address in the present study is the role of centrals vs.~satellites in the observed stacked kSZ profiles.  Our theoretical predictions correspond to central galaxies that are assumed to be perfectly centred within their host haloes and we may reasonably expect some degree of bias (with respect to theoretical predictions) to be introduced by the inclusion of satellites.  Without forward modelling the BOSS CMASS selection function, which is beyond the scope of this work, it is difficult to predict the magnitude or sign of this effect.  On the one hand, satellites will obviously be mis-centred with respect to their host haloes and one may expect this to lead to a lower kSZ signal.  On the other hand, a stellar mass-based selection implies that the satellites will typically be in hosts that are more massive than a host which has a central of similar stellar mass.  This will tend to boost the kSZ signal.  For the present study we neglect these uncertainties, leaving their careful consideration for a future study.

\section{Model pipeline and validation}\label{sec:4}
In this section, we briefly describe the set up of the cosmological inference pipeline (Section~\ref{sec:pipe}).  We validate the robustness of this pipeline using each of the four baryon models with a mock analysis, described in Appendix~\ref{app:mock}.  Here, we briefly motivate the choices made in the construction of the mock data (Section~\ref{sec:mocks}) and summarise the findings (Section~\ref{sec:mockresults}).
\begin{table}
    \caption{Summary of cosmological, observational and astrophysical parameters and priors used in the analysis. In the case of flat priors, the prior is bound to the range indicated in the `value' column, while Gaussian priors are described by their mean and 1$\sigma$ width.}   
    \label{tab:priors}
\begin{center}
%\begin{ruledtabular}
\begin{tabular}{ccc}
\hline
\hline
Parameter & Type & Value \tabularnewline
\hline 
\bf{Cosmological} \tabularnewline
$\Omega_{\rm m}$, Total matter density & Flat & [0.1, 0.9] \tabularnewline
 $\Omega_{\rm b}$, Baryon density & Flat & [0.03, 0.07] \tabularnewline
$10^{-9}A_{\rm s}$, Scalar spectrum amplitude  & Flat & [0.5, 5.0]\tabularnewline
$h$, Hubble parameter  & Flat & [0.55, 0.91]  \tabularnewline
$n_{\rm s}$, Spectral index  & Flat & [0.87,1.07] \tabularnewline
$\Omega_{\nu}h^2$, Neutrino mass density &  Flat & 0.06 \tabularnewline
\hline 
\bf{Observational} \tabularnewline
$\Delta z^1$, Source redshift 1 & Gaussian & ( 0.0, 0.018 ) \tabularnewline
$\Delta z^2$, Source redshift 2 & Gaussian  & ( 0.0, 0.015 ) \tabularnewline
$\Delta z^3$, Source redshift 3 & Gaussian  & ( 0.0, 0.011 ) \tabularnewline
$\Delta z^4$, Source redshift 4  & Gaussian  & ( 0.0, 0.017 ) \tabularnewline
$m^1$, Shear calibration 1 & Gaussian & ( -0.006, 0.009 )\tabularnewline
$m^2$, Shear calibration 2 & Gaussian  & ( -0.020, 0.008 )\tabularnewline
 $m^3$, Shear calibration 3 & Gaussian  & ( -0.024, 0.008 )\tabularnewline
$m^4$, Shear calibration 4 & Gaussian  & ( -0.037, 0.008 )\tabularnewline
\hline 
\bf{Intrinsic alignment} \tabularnewline
$a_1$, Tidal alignment amplitude        & Flat & $[-5,5]$ \tabularnewline
$\eta_1$, Tidal alignment redshift index     & Flat & $[-5,5]$  \tabularnewline
\hline
\end{tabular}
%\end{ruledtabular}
\end{center}
\end{table}

\subsection{Inference pipeline}\label{sec:pipe}

To analyse the cosmic shear data we build upon the public DES Y3 cosmological inference pipeline.  We utilise one parameter for redshift and shear calibration per tomographic bin, with prior ranges set to those used in DES Y3.  Without feedback models, we have 15 parameters that we marginalise over in the analysis. 
Table~\ref{tab:priors} summarises the cosmological, observational and astrophysical priors used in this work. 

Parameters are estimated via nesting sampling using {\sc Multinest}\footnote{{\sc\,Multinest}:\,\url{https://github.com/farhanferoz/MultiNest}} \citep{Feroz_2009} within the COSMOSIS framework \citep{Zuntz:2015}, with the sampler settings listed in Appendix~\ref{app:sampler}.  We note however that \citet{KiDSDES} demonstrates that {\sc Multinest} can underestimate the 68\% confidence levels for $S_8$, at the level of $\sim 10-20\%$, while \texttt{Polychord} \citep{Handley:2015} is more accurate. In agreement with this finding, Appendix~\ref{app:sampler} reports that sampling with \texttt{Multinest} instead of \texttt{Polychord} in a WL-only analysis with BCEmu7 results in a 18\% smaller 68\% confidence level for $S_8$, and a 9\% smaller 68\% confidence level for $S_8$ in a joint WL and kSZ analysis with BCEmu7.  We follow \citet{Amon:2021} and \citet*{Secco:2022} when reporting the parameter constraints and quote the mean of the 1D marginal distribution, along with the 68\% confidence limit, which defines the area around the peak of the posterior within which 68\% of the probability lies{\footnote{Formally, these are credible intervals; however, we choose to use the term `confidence interval' in this paper to retain consistency with the language used in \citet{Amon:2021} and \citet*{Secco:2022}. }}.

\subsection{Mock data}\label{sec:mocks}

In order to assess the robustness of our lensing inference pipeline, non-linear power spectrum model and baryon feedback models, we perform analyses using synthetic data and test the ability to recover unbiased cosmology. The mock data was created using the best fit cosmological parameters obtained from the DES Y3 joint lensing and clustering analysis \citep{DES-3x2}\footnote{That is, with $S_8=0.7805$, $\Omega_{\rm m}=0.3380$ and $\sigma_8=0.7353$ (see Appendix~\ref{app:mock} for more detail).}. 

We create two dark matter-only mock data vectors, using different models for the non-linear matter power spectrum. The first uses HM20, the same model used to analyse the data throughout this work. This mock is important for testing that the analysis pipeline can accurately recover cosmological parameters before considering feedback effects. The second mock uses the EuclidEmulator2 (EE2), which has been shown to be accurate to 1$\%$ for $k\leq10h$Mpc$^{-1}$ \citep{EE1,EE2, adamek:2022}. This mock is used to ensure that HM20 provides a sufficiently accurate description for the non-linear matter power spectrum when compared to EE2, for the full angular scale range of the DES Y3 cosmic shear data. 

To test the four baryon mitigation strategies, we chose three baryon feedback scenarios to build the mock data and their predictions for the suppression of the matter power spectrum are shown as the red lines in Fig.~\ref{fig:sims}.
 These choices include extreme scenarios as they are designed to test the flexibility and limits of the baryon models, rather than an attempt to chose the most accurate prescription. First, we consider the upper limit of the BAHAMAS hydrodynamic simulation suite, with $T_{\mathrm{heat}}=8.0$
\citep{McCarthy2017}.  Next, we consider a mock universe with a more extreme power spectrum prescribed by cosmo-OWLS $T_{\mathrm{heat}}=8.5$ \citep{LeBrun:2014}, although this simulation does not replicate the local gas fractions in groups and clusters.  As it is possible that hydrodynamical simulations do not capture the complexities of feedback, we want to test the ability of the baryon models to accurately capture a scenario that modulates the matter power spectrum with a different shape than that in typical simulations. We consider a mock with a suppression described by the $A_{\rm{mod}}$ parameter \citep{preston/etal:2023, AAGPE2022}, where $A_{\rm{mod}}\approx0.82$ is that required to reconcile the DES Y3 cosmic shear with \textit{Planck} $\Lambda$CDM cosmology. 

\subsection{Mock results}\label{sec:mockresults}
The results of these mock tests are detailed in Appendix~\ref{app:mock}. Here, we summarise the findings, which are shown in Fig.~\ref{fig:mocksum}:
\begin{itemize}
\item  When analysing a EE2-generated dark-matter only mock with the HM20 dark matter-only model, $S_8$ is over-estimated by $\sim$0.4$\sigma$. (This is reduced to $\sim$0.2$\sigma$ when analysing restricted angular scale measurements.) Furthermore, we find that $\Omega_{\rm m}$ is underestimated by $\sim$0.9$\sigma$. However, we identify projection effects in this parameter of $\sim$0.5$\sigma$ by analysing an HM20 mock with an HM20 model. 
In this work, we focus on the $S_8$ parameter, and we note that further testing is needed to assess the reliability of the $\Omega_{\rm m}$ constraints. 
\item The DES Y3 ‘$\Lambda$CDM-optimised’ scale cuts underestimate $S_8$ in all three mock baryonic feedback scenarios. This is as expected, as these scale cuts were defined with the OWLS-AGN scenario, which predicts less power suppression than BAHAMAS~8.0, cosmo-OWLS~8.5 and $A_{\rm mod}$. While the scale cuts, by design, remove the sensitivity of the analysis to the impact of baryon feedback, this method's success relies on the true feedback scenario to be less extreme than the simulation used to define the cuts.  
\item HM20, as used with their fiducial BAHAMAS-based prior, underestimates $S_8$ by $\sim$0.7$\sigma$ for the BAHAMAS~8.0 mock, and by more than $1\sigma$ when analyzing a mock with a more extreme feedback scenario. Using a wide prior alleviates this, with the model recovering the true cosmology within $\sim$0.5$\sigma$ for all mock scenarios, although with a cost of almost a factor of two in the precision of the $S_8$ constraint. 
\item The SP(k) model, used with both a wide and X-ray prior, can recover the input cosmology to within $\sim$0.2$\sigma$ for a BAHAMAS~8.0 mock. With the more extreme cosmo-OWLS~8.5 and $A_{\rm mod}$ mocks, this model underestimates the value of $S_8$ by up to $\sim$0.5$\sigma$. This is as expected, as both cosmo-OWLS~8.5 and $A_{\rm{mod}}$ are outside the expected baryon fractions as compared to observations by \citet{Akino_2022}. 
\item When allowing only one parameter to vary in BCEmu emulator, BCEmu1, we recover $S_8$ to within $\sim$0.2$\sigma$ for the BAHAMAS~8.0 mock, and $\sim$0.5$\sigma$ for cosmo-OWLS~8.5. When we use the more flexible BCEmu7, all mock scenarios recover the true cosmology within $\sim$0.5$\sigma$, and the error bar on $S_8$ is up to 1.5 times wider. 
\end{itemize}
Overall, we find that when more restrictive modelling choices are used,  we tend to underestimate $S_8$.  The bias is worsened when restrictive choices are used to analyse mocks with the more extreme baryonic feedback scenarios, i.e. cosmo-OWLS~8.5 and $A_{\rm{mod}}$. We note that despite the greater accuracy of using more conservative priors or marginalising over a greater number of baryonic feedback parameters, it is at the expense of the precision.  We find that the uncertainty on the constraint of $S_8$ can degrade by up to a factor of two when switching to more flexible modelling choices.

\section{Results: Assessing models for baryonic effects}\label{sec:WLresults}

\begin{figure}
	\centering
	\includegraphics[width=\columnwidth]{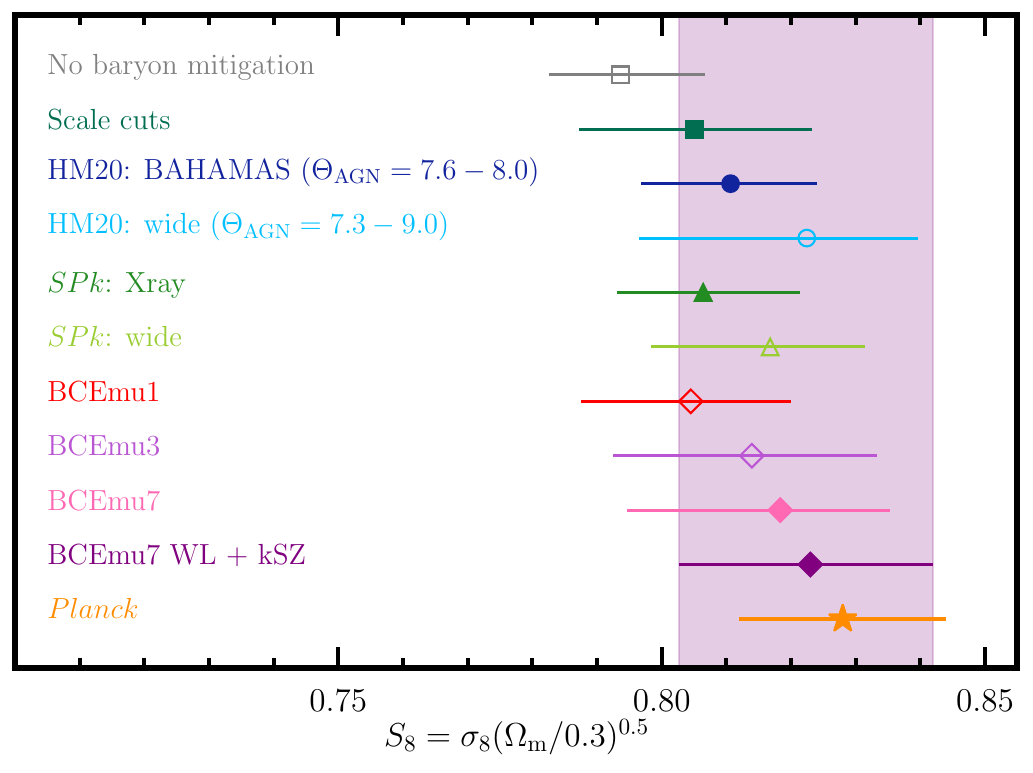} 
	\caption{Summary of the 1D marginalised constraints on $S_8$ from analysing the DES Y3 cosmic shear data with different baryonic feedback mitigation strategies. The mean of the $S_8$ marginalized posterior is indicated by the symbol and 68\% confidence levels are shown as horizontal bars. The primary result from the joint analysis of WL + kSZ using BCEmu7 is represented as the purple shaded region. We compare to the \textit{Planck} TTTEEE result presented in \citet{EfstathiouGratton:2021}. }
	\label{fig:1ds8data}
\end{figure}

The results of the WL-only DES Year 3 analysis are divided into three sections. In Section~\ref{sec:shearcosmo}, we present the headline cosmological constraints using the four baryon feedback strategies outlined in Section~\ref{sec:models}.  In Section~\ref{sec:shearpk}, we show the constraints on the suppression of the power spectrum.  Finally, in \ref{sec:shearcosmomodels}, we explore the dependence of our results on the model complexity and prior choices within each strategy.  

\subsection{Cosmological parameter constraints}\label{sec:shearcosmo}

\begin{figure*}
	\centering
	\includegraphics[width=0.48\textwidth]{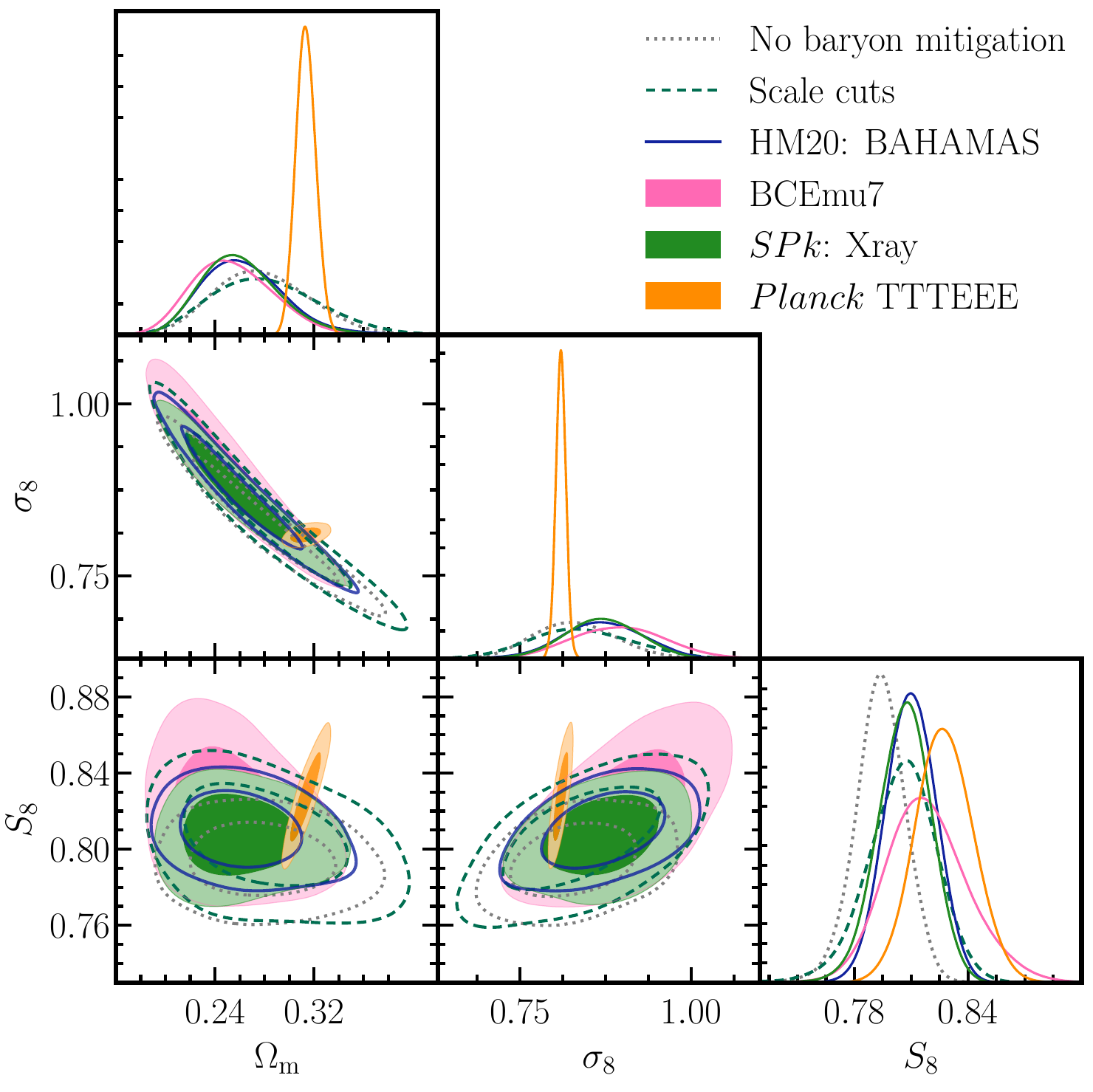} 
    \includegraphics[width=0.5\textwidth]{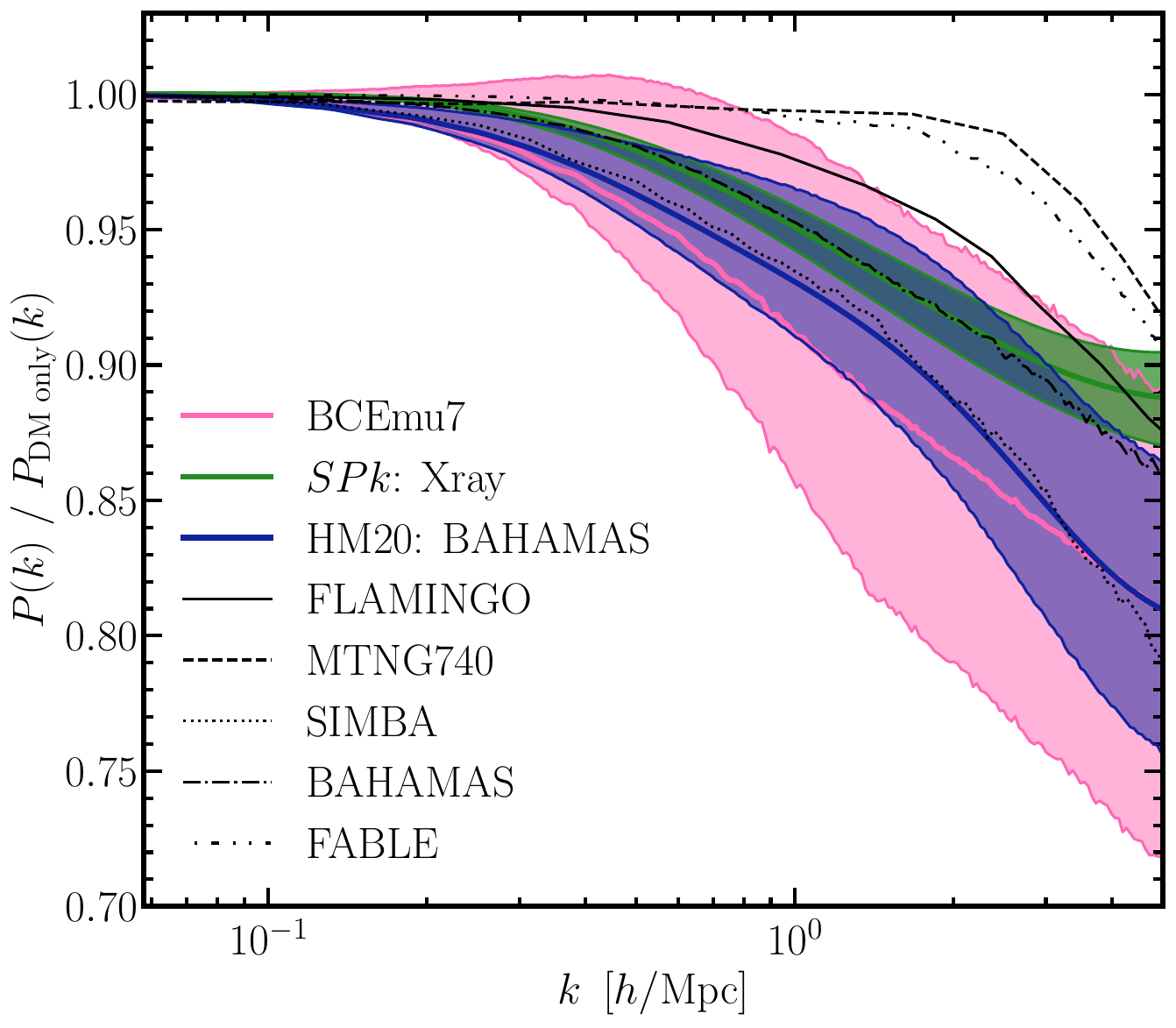}
	\caption{\textit{Left:} The marginalised posteriors for $\Omega_{\rm m}$, $\sigma_8$ and $S_8$ using the DES Y3 cosmic shear data and different baryon feedback models. We compare the DES Y3 optimised scale cut approach (green dashed) to scenarios where all angular scales of the DES Y3 lensing data are modelled: \textsc{HM20}, using their recommended BAHAMAS-based prior (HM20: BAHAMAS, navy), the seven-parameter baryonification model (BCEmu7, pink) and the SP(k) emulator, using their 'X-ray observational prior' (SPk: Xray, solid green).  For reference, we show the case where all scales of the DES data vector are used but baryons are not modelled (gray dotted) and the \textit{Planck} TTTEEE likelihood \citep[orange,][]{EfstathiouGratton:2021}. The inner and outer contours show the 68\% and 95\% confidence levels, respectively.  \textit{Right:} The corresponding constraints on the suppression of the total matter power spectrum compared to a dark matter-only scenario, $P(k)/P_{\rm DM only}(k)$, at $z=0$ using the DES Y3 cosmic shear and again, HM20 (HM20: BAHAMAS, navy), BCEmu (BCEmu7, pink) and SP(k) (SPk: Xray, green). The solid lines show the mean suppression and the shaded regions indicate the 68\% confidence levels.  For reference, various predictions from hydrodynamical simulations are over-plotted in black: FLAMINGO \citep[][solid line]{Schaye2023}; BAHAMAS \citep[][dash-dotted line]{McCarthy2017}; SIMBA \citep[][dotted line]{Dave:2019}, MillenniumTNG \citep[][dashed line]{Pakmor2023} and FABLE (doubledot-dashed line, \citealt{Henden2018},  Bigwood et al. in prep.).}
\label{fig:BCM}
\end{figure*}

\begin{table*}
\setlength\extrarowheight{3pt}
    \caption{The constraints on $S_8$, $\Omega_{\rm m}$ attained for each method of baryonic feedback mitigation.  We report the mean value of each parameter with errors given by the 68\% confidence levels.  We also demonstrate the quality of the fit by reporting $\chi_{\rm{min}}^2$, the minimum value of $\chi^2$, for each analysis variant.  We report $\chi_{\rm{red}}^2=\chi_{\rm{min}}^2/N_{\rm dof}$, with $N_{\rm dof}$ being the number of degrees of freedom $N_{\rm dof}=N_{\rm dp}-N_{\rm param}$, where $N_{\rm dp}$ and $N_{\rm param}$ are the number of data points and model parameters utilised in the analysis.  With the exception of the scale cuts method, angular scales down to 2.5~arcmin of the DES Y3 data were used.}   
\label{tab:cosmologyresults}
\centering
\begin{tabular}{cccccccc}

\hline
\hline
\ Model & $S_{8, \rm {mean}}$ &$\Omega_{8, \rm{mean}}$   & $\chi_{\rm{min}}^2$ & $N_{\rm dp}$ & $N_{\rm param}$& $N_{\rm dof}$ & $\chi^2_{\rm red}$ \\

\hline
No baryon mitigation & $ 0.794^{+0.013}_{0.011} $ & $0.278^{+0.033}_{0.044} $& 418.37 & 400 & 15 & 385 & 1.09 \\ 
 Scale cuts & $ 0.805^{+0.018}_{-0.018} $ & $0.281^{+0.035}_{-0.051} $ & 284.86 & 273 & 15 & 258 & 1.10\\ 
  HM20 $\Theta_{\mathrm{AGN}}=7.6-8.0$ & $ 0.811^{+0.013}_{-0.014} $ & $0.261^{+0.026}_{-0.034} $& 415.25 & 400 & 16 & 384 & 1.08\\ 
  HM20 $\Theta_{\mathrm{AGN}}=7.3-9.0$ & $ 0.822^{+0.017}_{-0.026} $ & $0.252^{+0.027}_{-0.039} $ & 415.82 & 400 & 16 & 384 & 1.08 \\ 
  BCEmu1 & $ 0.804^{+0.016}_{-0.017} $ & $0.274^{+0.033}_{-0.042} $ & 414.49 & 400 & 16 & 384  & 1.08\\ 
  BCEmu3 & $ 0.814^{+0.019}_{-0.021} $ & $0.261^{+0.029}_{-0.044} $ & 414.97 & 400 & 18 & 382 & 1.09\\ 
  BCEmu7 WL & $ 0.818^{+0.017}_{-0.024} $ & $0.255^{+0.027}_{-0.038} $ & 414.21 & 400  &22 & 378 & 1.10 \\ 
  BCEmu7 WL + kSZ & $ 0.823^{+0.019}_{0.020} $ & $0.250^{+0.025}_{0.036} $ &439.33  & 418 & 23 & 395 & 1.11 \\ 

  BCEmu7, $\Sigma m_{\nu}: [0.06,0.6]$ & $ 0.813^{+0.019}_{-0.023} $ & $0.269^{+0.028}_{-0.043} $  & 414.64& 400 & 23 & 377 & 1.10 \\ 
  BCEmu7, TATT & $ 0.802^{+0.028}_{-0.024} $ & $0.239^{+0.021}_{-0.041} $ & 408.54 & 400 &25  & 375 & 1.09 \\

   SP(k) conservative prior & $ 0.817^{+0.015}_{0.019} $ & $0.255^{+0.025}_{0.040} $ &415.44&400&18&382 &1.09\\ 
   SP(k) Xray et al. (2022) prior & $ 0.806^{+0.015}_{0.013} $ & $0.261^{+0.025}_{0.036} $ &415.02 &400&18&382 &1.09\\ 
 
\hline
\end{tabular}
\end{table*}

The 1D marginalised constraints obtained for $S_8$ are summarised in Fig.~\ref{fig:1ds8data} for analyses using all model variants. Here, we compare those from the DES Y3 `$\Lambda$CDM optimised' scale cuts, HM20: BAHAMAS, SP(k): Xray and BCEmu7. For these approaches, the mean marginal values of $S_8$ are found with 68\% confidence levels to be
\begin{align}
\label{eqn:S8results}
\begin{aligned}
{\rm BCEmu7:}\,\,\, &S_8 &=&\,\,\,  0.818^{+0.017}_{-0.024} \\  
{\rm Spk:Xray:}\,\,\, &S_8 &=&\,\,\, 0.806^{+0.015}_{-0.013} \\  
{\rm HM20:BAHAMAS:}\,\,\, &S_8 &=&\,\,\,  0.811^{+0.013}_{-0.014} \\  
{\rm Scale \,\, cuts:}\,\,\, &S_8 &=&\,\,\, 0.805^{+0.018}_{-0.018} \,.\\  
\end{aligned}
\end{align}
\noindent 
For reference, we show the result when all angular scales are analysed without any model for baryonic effects and the \textit{Planck} TTTEEE\footnote{TTTEEE refers to the high multipole likelihood attained from combining the temperature power spectra (TT),
temperature-polarization E-mode cross spectra (TE) and polarization E-mode power spectra (EE).} $\Lambda$CDM result \citep[orange,][]{EfstathiouGratton:2021}.  The 2D marginalised posteriors for $S_8$, $\Omega_{\rm m}$ and $\sigma_8$ using DES Y3 weak lensing data are plotted in the left panel of Fig.~\ref{fig:BCM}, also showing no baryon mitigation (dotted grey), DES Y3 `$\Lambda$CDM optimised' scale cuts (dashed green), HM20: BAHAMAS (navy), SP(k): Xray (solid green), BCEmu7 (pink) and \textit{Planck} TTTEEE (orange).
Table~\ref{tab:cosmologyresults} lists the mean constraints on $S_8$ and $\Omega_{\rm m}$ and quantifies the goodness of fit to the data of each modelling variant by quoting the minimum and reduced $\chi^2$, $\chi_{\mathrm{min}}^2$, $\chi_{\mathrm{red}}^2$. We find that each baryon feedback analysis variant demonstrates a suitable fit to the measurements.

When analysing all angular scales without modelling baryonic feedback, we attain a low value of $S_8=0.794^{+0.013}_{0.011}$.  This is up to $1\sigma$ lower than constraints attained with modelled baryonic effects, highlighting the importance of mitigating feedback to avoid biased cosmology.  When accounting for baryonic effects, we find, in agreement with the mock analysis for the models tested, that the measured $S_8$ is consistent at the level of 0.6$\sigma$ ($\sim$2$\%$)\footnote{Note that throughout the following sections we quantify the shift in the measured value of $S_8$ by two analyses using the metric $\Delta S_8/[(\sigma_{S_8}^1)^2+(\sigma_{S_8}^2)^2]^{1/2}$, where $\Delta S_8$ is the difference between the respective mean values and $\sigma_{S_8}^1$,$\sigma_{S_8}^2$ are the $1-\sigma$ errors on $S_8$.}. However, the errorbar on the $S_8$ constraint varies by a factor of 1.5.  In more detail, we see that HM20: BAHAMAS and SP(k):Xray give the tightest constraints on the $S_8$ parameter, which are in excellent agreement with each other. This is expected, as these models are calibrated on hydrodynamical simulations informed by X-ray constraints. The BCEmu7 analysis gives the highest value of $S_8$. As in the case of the mock analysis, this supports our findings that restrictive modelling choices for baryonic feedback leads to lower value of $S_8$ when compared to more flexible models. The flexibility of BCEmu7 comes at a cost, as the uncertainty on $S_8$ is a factor of 1.5 larger than that in the HM20: BAHAMAS analysis.  This is to be expected given the degeneracy between the extremity of feedback and $S_8$; greater flexibility in the modelling of baryons inevitably results in a larger errorbar on the $S_8$ constraint. The analysis using the DES Y3 scale cuts does not suffer as substantial a loss in constraining power as the BCEmu7 case, but results in the lowest value for $S_8$\footnote{Note that the DES Y3 `$\Lambda$CDM optimised' cosmic shear analysis \citetext{\citealp{amon:2022}; \citealp*{Secco:2022}} obtains a lower value of $S_8$ than that obtained here ($S_8=0.772^{+0.018}_{-0.017}$). Based on the study of the impact of analysis choices in \citep{KiDSDES}, we attribute the difference in our results primarily to the use of HM20 to model the dark matter non-linear matter power spectrum, which was shown to be more accurate than Halofit, as well as the intrinsic alignment model that we chose for this analysis, and the choice to fix the neutrino mass in the analysis. In Appendix~\ref{app:IA} we investigate the impact of these choices further.}.

\subsection{Power suppression constraints}\label{sec:shearpk}

Baryon feedback processes modify the gravitational evolution of the cosmic density field and suppress the matter power spectrum compared to a dark matter-only scenario on non-linear scales, as seen in hydrodynamic simulations.  This effect has been previously observed by analyses of weak lensing data using variations of the baryonification model \citep{Schneider:2022, ChenDES:2022, arico2023}.  In this section, we constrain the amplitude and scale dependence of the suppression of the matter power spectrum due to baryonic effects, $P(k)/P_{\mathrm{DM only}}(k)$, using the DES Y3 cosmic shear. For the first time, we show the model-dependence of the constraints by considering the model complexity of the baryonification model and the comparison to the SP(k) and HM20 models. For each baryonic feedback model, we record the power spectrum suppression at each step in the chain.

In the right panel of Fig.~\ref{fig:BCM} we plot the mean suppression and the and 68\% confidence levels inferred from analyses with our three baseline models: BCEmu7, HM20: BAHAMAS ($\Theta_{\mathrm{AGN}}=7.6-8.0$) and SP(k):Xray. We find that the suppression inferred by the three models are consistent within the 68\% confidence limits up to $k$$\approx$3 $h$/Mpc.  However, we find that BCEmu7 allows more extreme suppression of the power spectrum at all non-linear scales.  There are substantial differences in the size of the uncertainties, correlated with the flexibility of the model. BCEmu7, the most flexible model, has the largest uncertainty. SP(k):Xray provides the tightest constraints on the power spectrum suppression and constrains a less extreme feedback scenario in terms of the amplitude and the scale extent of the suppression. 

We compare our constraints to five hydrodynamical simulations: FLAMINGO, \citep[][solid line]{Schaye2023}; BAHAMAS \citep[][dash-dotted line]{McCarthy2017}; SIMBA \citep[][dotted line]{Dave:2019}, MillenniumTNG \citep[][dashed line]{Pakmor2023} and FABLE (doubledot-dashed line, \citealt{Henden2018},  Bigwood et al. in prep.). HM20 predicts a feedback strength that encompasses BAHAMAS $T_{\rm{heat}}=7.8$, which is unsurprising, given that the model is calibrated to span these simulations, but it is notable that the mean constraint is more extreme on all scales. Interestingly, all three models find the mean suppression to be more extreme than FLAMINGO on scales $k$$\sim$0.2-4$h^{-1}$Mpc, at the level of  $1.1\sigma$ (BCEmu7), $3.0\sigma$ (SP(k):Xray) and $1.4\sigma$ (HM20:BAHAMAS) at $k=2h^{-1}$Mpc.

The power suppression constraints are broadly consistent with those from previous weak lensing analyses \citep{Schneider:2022} and slightly more extreme than the constraints of \citet{ChenDES:2022, arico2023, garciagarcia2024, Terasawa2024}. Here, we make note of some details. Owing to the enhanced statistical power of the DES Y3 data over that of the Kilo-Degree Survey, we find substantially improved constraints from the WL-only analysis compared to \citet{Schneider:2022}, even though we include an additional parameter for the intrinsic alignment model and eight additional nuisance parameters to account for the uncertainty in the shear and redshift calibration. Furthermore, we use a different baryonification model to that of \citet{ChenDES:2022} and \citet{arico2023} and we do not impose X-ray priors on any of the baryon parameters. One way in which this model is different from BCEmu is that only particles within the virial radius, $R_{200}$, are displaced \citep[see][for a more detailed discussion of the model comparison]{Grandis2023}. One implication of the model differences is that it is not straightforward to compare constraints on the $M_{\rm c}$ parameter, though we note that in our BCEmu scenario, the posterior on the $M_{\rm c}$ parameter is not limited by the upper value of the prior.

\begin{figure*}
\centering
\begin{subfigure}[b]{\textwidth}
\includegraphics[width=.33\textwidth]{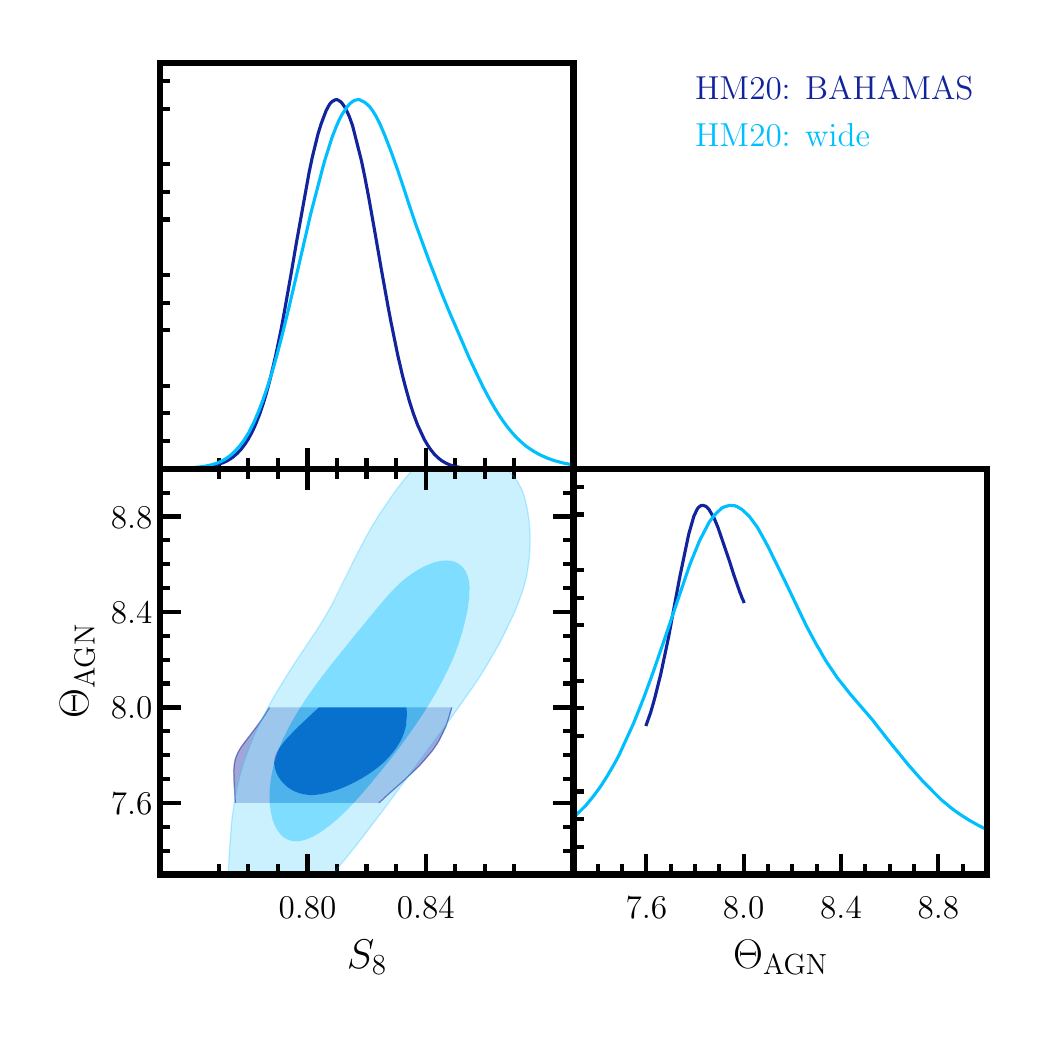}\hfill
\includegraphics[width=.33\textwidth]{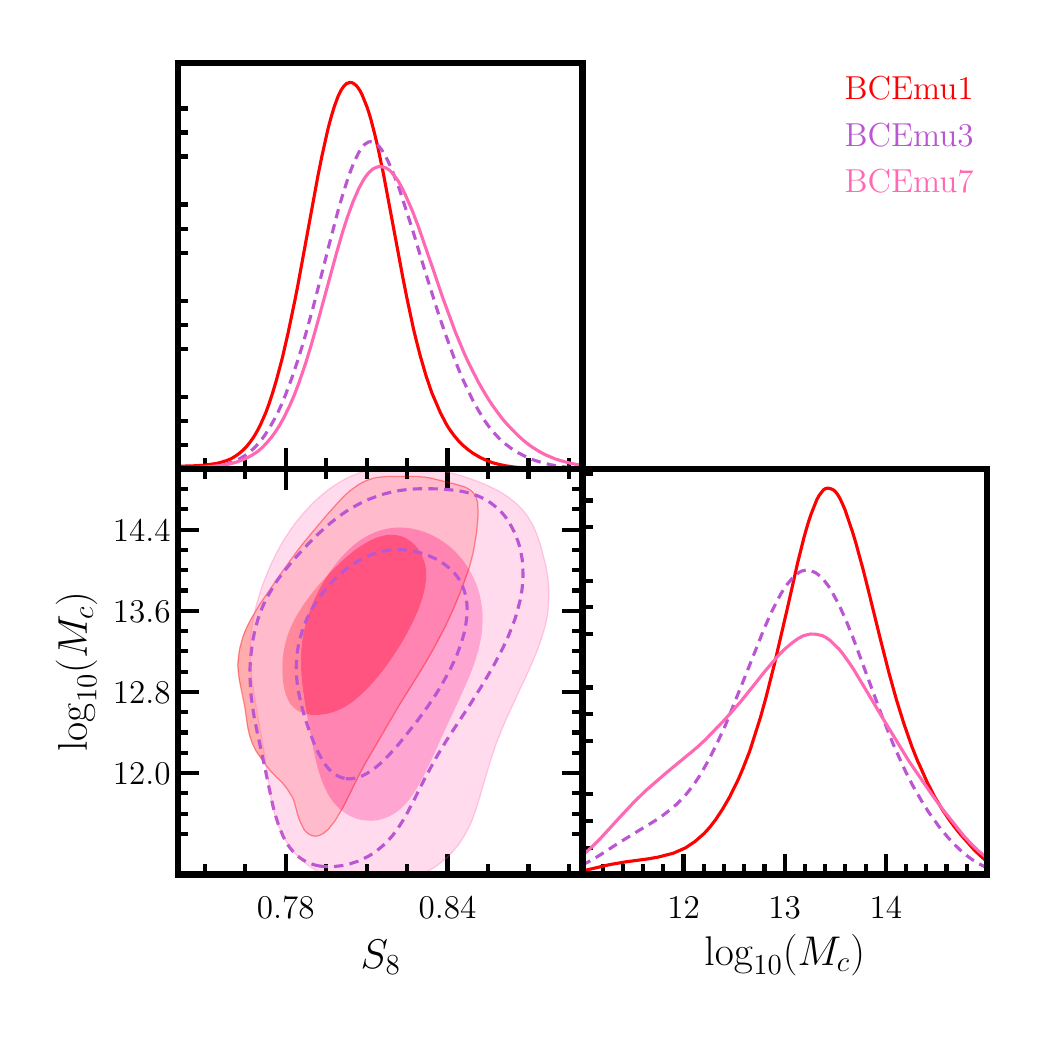}\hfill
\includegraphics[width=.33\textwidth]{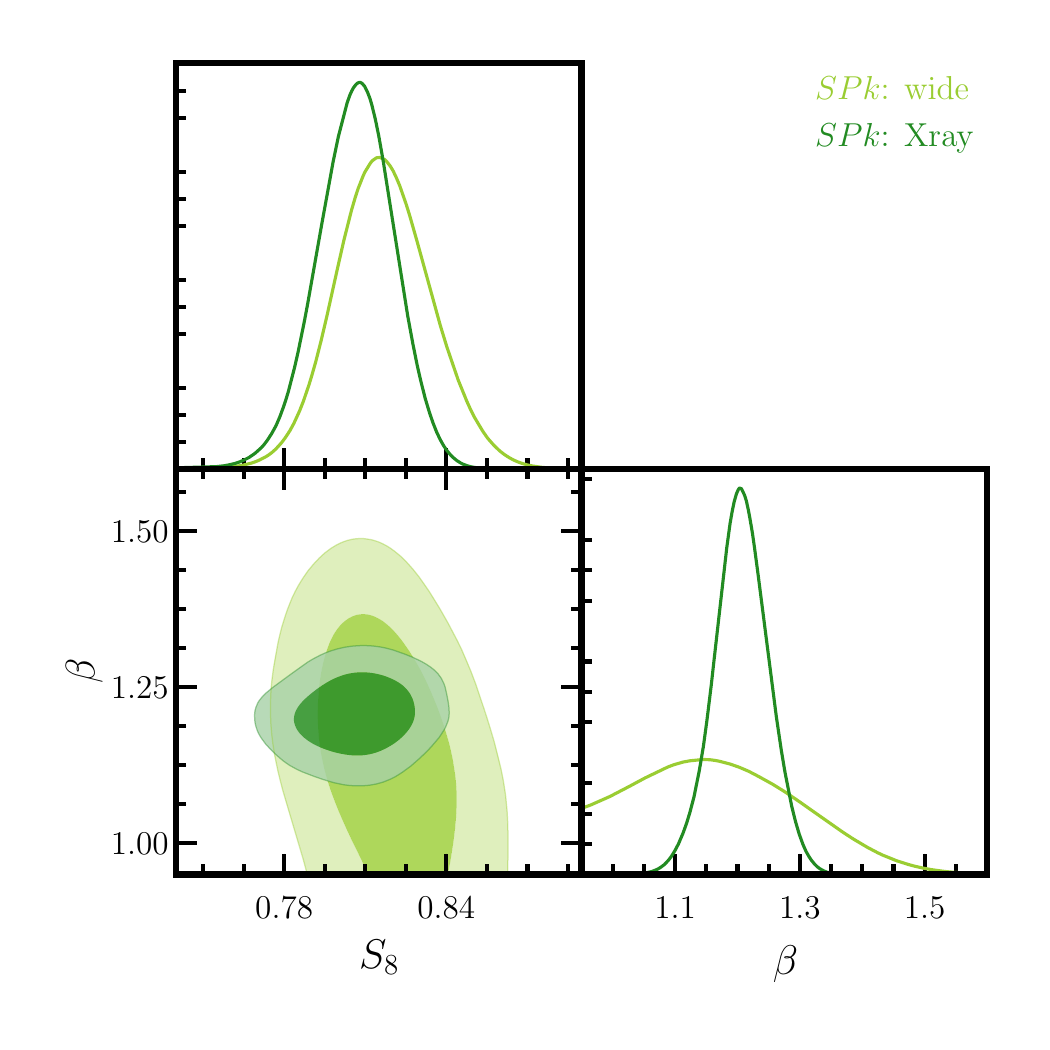}
\end{subfigure}
\begin{subfigure}[b]{\textwidth}
\includegraphics[width=.325\textwidth]{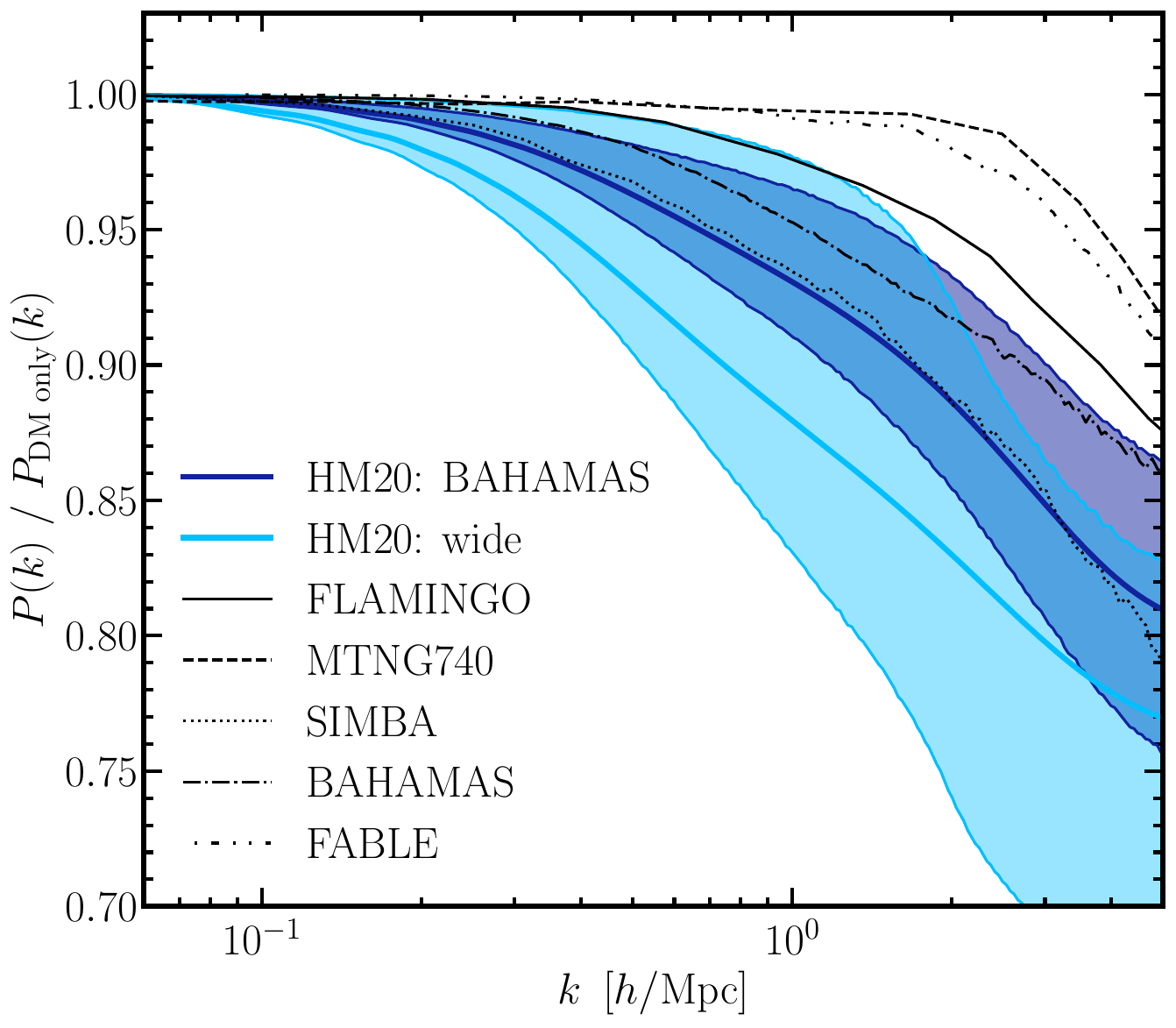}\hfill
\includegraphics[width=.325\textwidth]{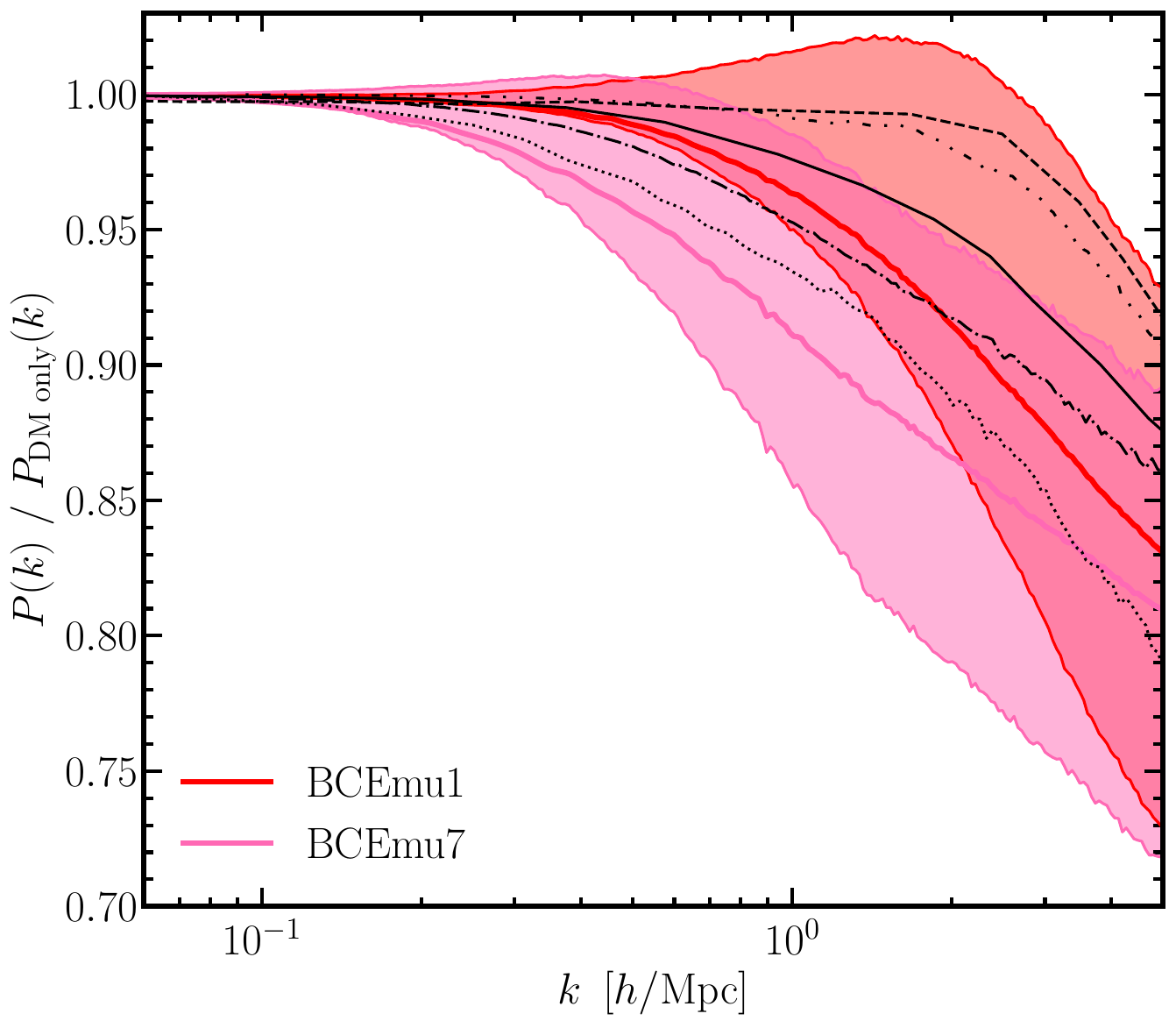}\hfill
\includegraphics[width=.325\textwidth]{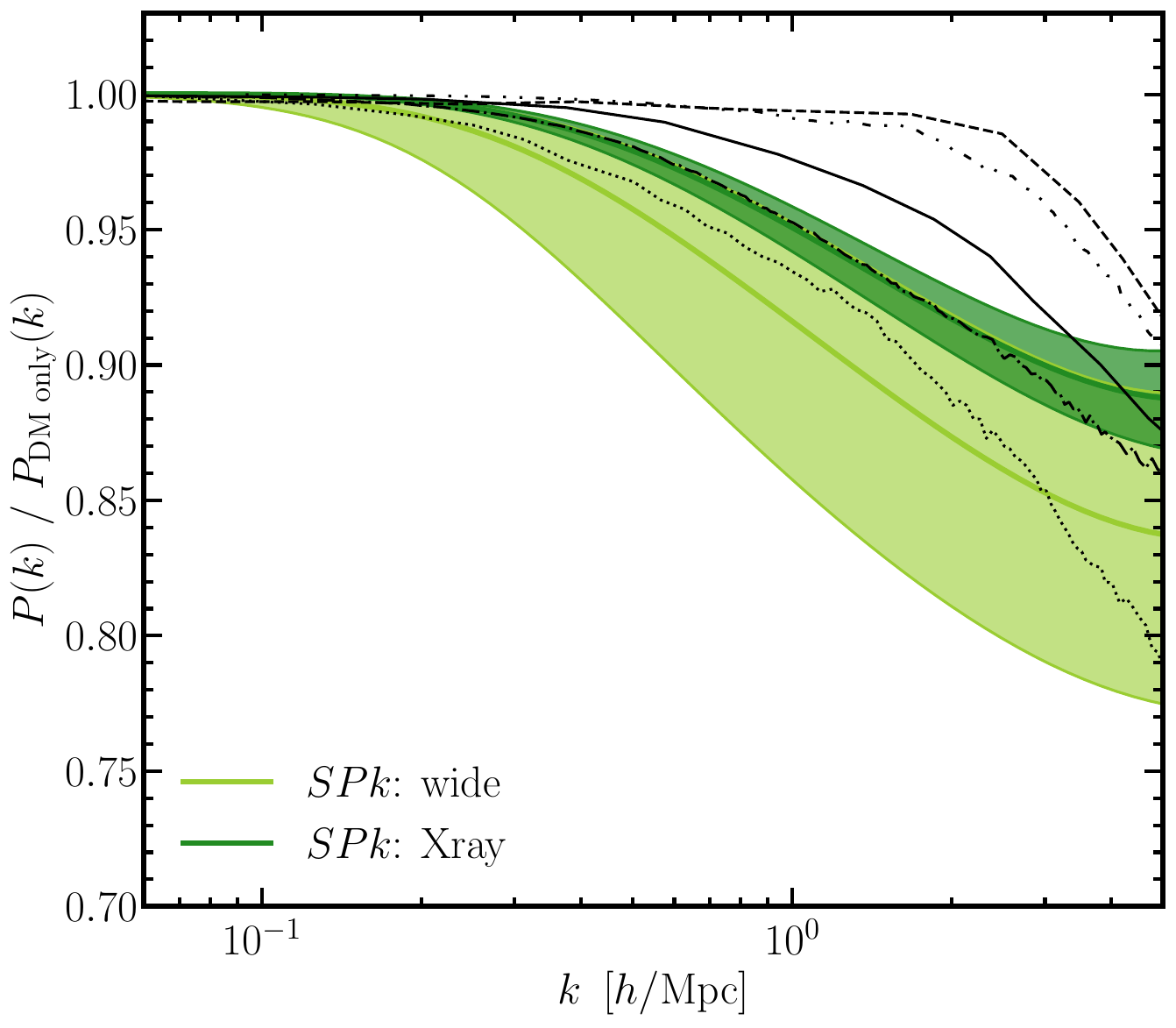}
\end{subfigure}
\caption{\textit{Top:} The degeneracy of the marginalised $S_8$ posterior with varying the baryon model complexity or prior choices on baryon model parameters.  For each panel the inner and outer contours show the 68\% and 95\% confidence levels, respectively. \textit{Left:} HM20 - $S_8$ and $\Theta_{\rm AGN}$ attained when using a prior bracketing the BAHAMAS simulations (dark blue, $\theta_{\rm AGN}=7.6-8.0$) and a less-informative prior choice, (light blue, $\theta_{\rm AGN}=7.3-9.0$).  Centre: BCEMu - $S_8$ and $\log_{10}(M_{\rm c})$ attained for the full seven parameter (pink, BCEmu7), three parameter (purple, BCemu3) and one parameter (blue, BCEmu1) models.  \textit{Right:}  SP(k)-$S_8$ and $\beta$ attained for the X-ray informed model (dark green) and a less-informative prior (light green). 
\textit{Bottom:} The corresponding constraints on the suppression of the total matter power spectrum compared to a dark matter-only scenario, $P(k)/P_{\rm DM only}(k)$, at $z=0$ for each of the models.  In general, when allowing for a more flexible model, the constraints indicate more extreme suppression of power, although with degraded constraining power.  
%\textit{Left:} The scenario using the HM20 model with a prior bracketing the BAHAMAS simulation (dark blue, $\Theta_{\rm AGN}=7.6-8.0$) and more conservative, wide prior choice (light blue, $\Theta_{\rm AGN}=7.3-9.0$).  Centre: the constraints attained using the complete seven parameter (pink, BCEmu7) and one parameter (red, BCEmu1) baryonification models.  \textit{Right:} the results using the SP(k) model with the recommended X-ray prior (dark green) and a less-informed prior (light green). 
For each panel the solid lines show the mean suppression predicted, and the shaded regions the 68\% confidence levels.  For reference, various predictions for the suppression of the matter power spectrum from simulations are over-plotted in black: FLAMINGO \citep[][solid line]{Schaye2023}; BAHAMAS \citep[][dash-dotted line]{McCarthy2017}; SIMBA \citep[][dotted line]{Dave:2019}, MillenniumTNG \citep[][dashed line]{Pakmor2023} and FABLE (doubledot-dashed line, \citealt{Henden2018},  Bigwood et al. in prep.).}
\label{fig:complexitychoice}
\end{figure*}

\subsection{Impact of model complexity}\label{sec:shearcosmomodels}

The mock analysis revealed that marginalising over a greater number of baryonic nuisance parameters, or utilising wider priors on these parameters, generally improved the accuracy of the cosmological constraints.  However, we saw that this was at the expense of inflated errors on the cosmological parameters, which is clearly sub-optimal for an effective cosmological analysis.  In this section, we explore the impact of altering the complexity and prior choices of each baryon feedback model on the measured cosmological and baryonic constraints when analysing the DES Y3 $\xi_{\pm}$ measurements.

The upper left panel of Fig.~\ref{fig:complexitychoice} shows the impact of the BAHAMAS-informed prior on the HM20 feedback parameter $\Theta_{\mathrm{AGN}}$ on the marginalised $S_8$ posterior. In the light blue constraint, we extend the prior range for $\Theta_{\mathrm{AGN}}$ outside of the calibration range to encompass more extreme feedback scenarios, as well as a dark-matter only scenario. The parameters are degenerate and opening up the $\Theta_{\mathrm{AGN}}$ prior leads to long tails that extend to higher values of $S_8$, such that the mean constraint is $>$0.5$\sigma$ higher. This illustrates how high values of $S_8$ are disfavoured by the restricted prior on the baryonic feedback model, and suggests that weak lensing data may favour a higher value of $\Theta_{\mathrm{AGN}}$ than the BAHAMAS simulations span. (Although we note that the HM20 model was only calibrated within the BAHAMAS range, so the mapping between the power spectrum suppression and the baryon fraction outside this range is uncertain.) Similarly, for the case shown in the right-hand panel of  Fig~\ref{fig:complexitychoice}, when the X-ray informed prior is lifted, the data constrains higher values of $S_8$ by $\sim$0.5$\sigma$ (the posteriors on the SP(k) parameters are shown in Appendix~\ref{app:spk}). 
 These shifts are consistent with those we determine in the mock analysis (Section~\ref{sec:mocks}, Appendix~\ref{app:mock}). In Section~\ref{sec:XSZ}, we discuss the implications of these results on our understanding of the gas models and observations.

We test the impact of limiting the baryonification model complexity to the one (BCEmu1) and three-parameter (BCEmu3) case, compared to the fiducial BCEmu7.  The central panel of Fig.~\ref{fig:complexitychoice} shows the marginalised posteriors on $\Omega_{\rm m}$, $S_8$ and the baryonification parameter $\log_{10}M_{\rm c}$.  The posteriors on all of the baryonic feedback parameters are shown in Appendix~\ref{app:bcemu}.  BCEmu7 and BCEmu3 produce comparable constraints on cosmological and feedback parameters, generating a shift in $S_8$ of only $\sim$0.2$\sigma$.  We also do not see any significant improvement of the precision on these constraints when marginalising over four less baryonic parameters. 
However, switching to BCEmu1 from BCEmu7 results in a substantially lower values of $S_8$ by $>$0.5$\sigma$, consistent with the mock analysis.  This reduction in the flexibility of the model forces $\log_{10}M_{\rm c}$ to a larger value, since the full extremity of feedback has to be captured by only the one parameter.  We attribute this shift in the cosmology and feedback parameters to the values at which the remaining six `under-the-hood' baryonification parameters are fixed to in the BCEmu1 model, quoted in Table~\ref{tab:modelparameters}\footnote{The fixed parameters were determined by fitting to the baryonic suppression of the matter power spectrum of a number of hydrodynamical simulations, then summing the likelihoods to find the best-fit parameters to all of the simulations \citep{Giri:2021}.}.  Given that the analysis using BCEmu1 gives a larger value of $\log_{10}M_{\rm c}$ than that attained using BCEmu7, and that for example, the posterior on $\theta_{\rm ej}$ is toward higher values than the fixed 3.5, this implies that the values for the simulation-informed fixed parameters of BCEmu1 represent a less-extreme feedback scenario than those constrained by an analysis of DES cosmic shear.

The lower panels of Fig.~\ref{fig:complexitychoice} illustrate how the corresponding constraints on the power spectrum suppression are sensitive to the restrictiveness of the choices within each model.  Analogously to the upper panels, the left and right-most panels test switching to the more conservative prior choices of the HM20 and SP(k) models, i.e. $\Theta_{\mathrm{AGN}}=7.3-9.0$ and a prior spanning the feedback landscape of the ANTILLES suite, respectively.  The central panel shows the impact of restricting the model complexity of BCEmu, by allowing only one (BCEmu1) out of the full seven (BCEmu7) baryonification parameters to vary in the analysis. As before, we compare the data constraints to predictions from hydrodynamical simulations. 

For each model, when we opt for the more flexible modelling choices, the mean constraint on the power spectrum suppression tends to more extreme scenarios at all non-linear $k$ with respect to their more restrictive counterpart.  In particular, each of the flexible models is most consistent with more extreme feedback scenarios (e.g., SIMBA).  This could suggest that the restrictive baryonic modelling choices do not have the flexibility to capture the full extremity of feedback that the data prefers and therefore the higher values of $S_8$. It is clear, however, that the weak lensing data cannot place strong constraints on feedback on its own.
The use of the more conservative model choices comes at the expense of reduced constraining power.  For example, SP(k) displays over a factor of two increase in the size of the 68\% confidence level at all scales when switching from the restrictive to conservative modelling choice. A complementary avenue to simulation-based models that still maximises cosmological constraints is to use the flexible model framework and jointly analyse the lensing with observations of the gas.

\section{Results: Joint weak lensing + kinetic SZ}
\label{sec:results}

\begin{figure*}
	\centering
	\includegraphics[width=0.48\textwidth]{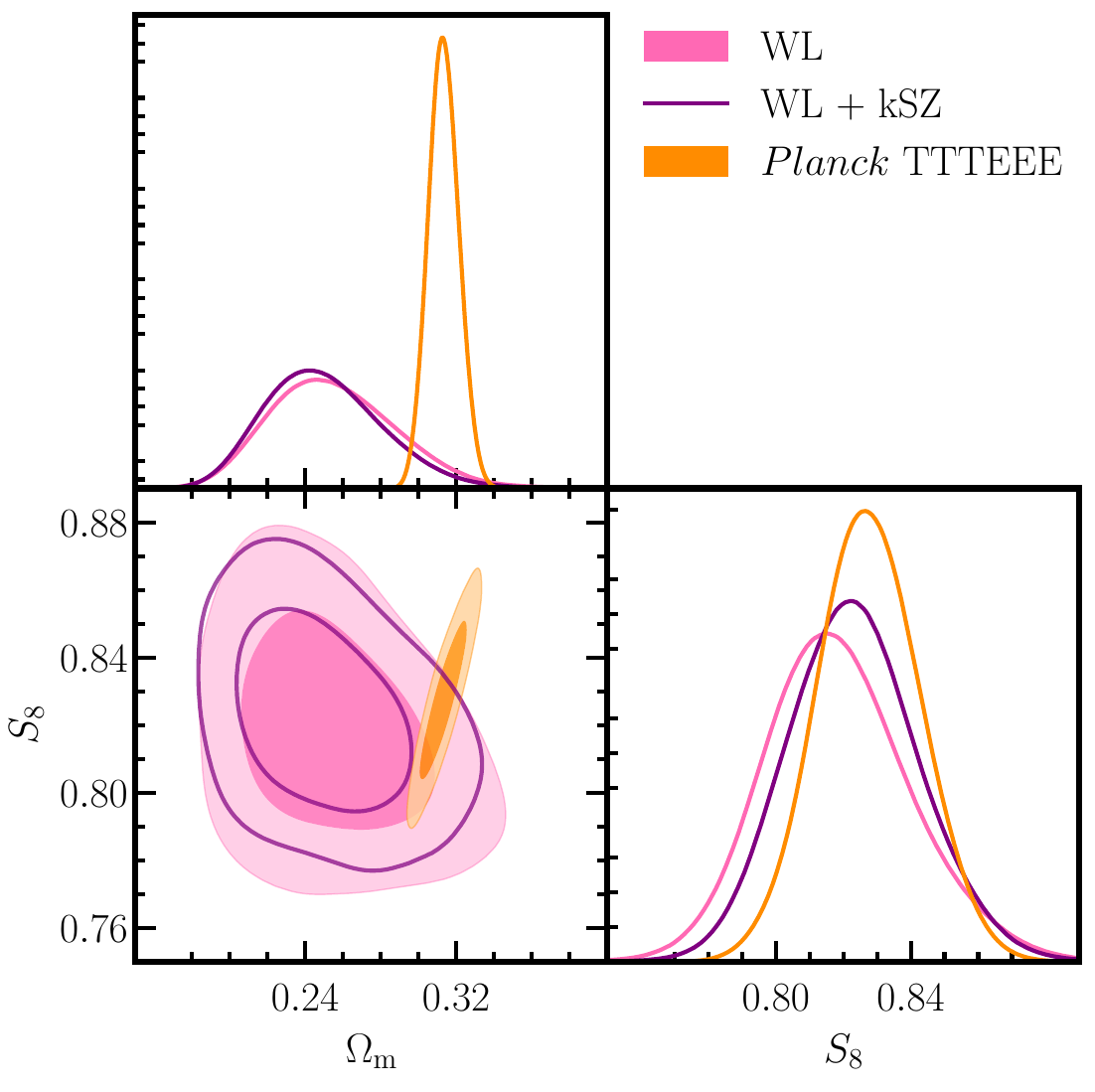} 
 \includegraphics[width=0.5\textwidth]{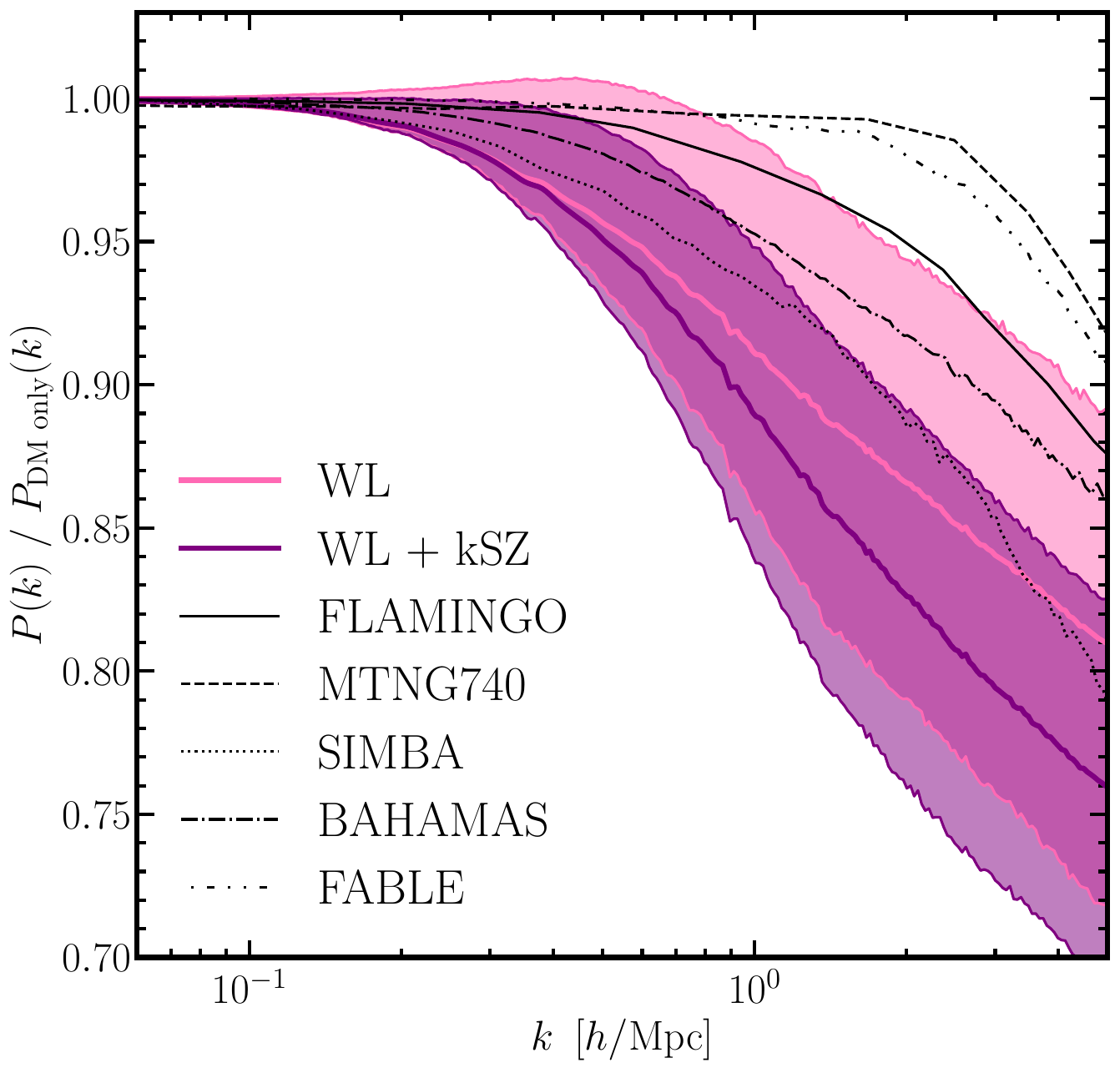}
	\caption{\textit{Left:} The marginalised posteriors for $\Omega_{\rm m}$ and $S_8$ attained modelling baryonic feedback with the BCEmu7 model using the DES Y3 cosmic shear dataset only (pink), or a combined analysis of DES Y3 cosmic shear and ACT DR5 kSZ measurements (purple).  The inner and outer contours show the 68\% and 95\% confidence levels, respectively.   We compare to the CMB $\Lambda$CDM constraint measured by \textit{Planck} TTTEEE likelihood \citep{EfstathiouGratton:2021}.   \textit{Right:} The constraints on the suppression of the total matter power spectrum compared to a dark matter-only scenario, $P(k)/P_{\rm DM only}(k)$, at $z=0$ when modelling baryonic feedback with the BCEmu7 model using the DES Y3 cosmic shear dataset only (pink), or a combined analysis of DES Y3 cosmic shear and ACT DR5 kSZ measurements (purple). The solid lines show the mean suppression and the shaded regions indicate the 68\% confidence levels.  For reference, various predictions for the suppression of the matter power spectrum from simulations are over-plotted in black: FLAMINGO \citep[][solid line]{Schaye2023}; BAHAMAS \citep[][dash-dotted line]{McCarthy2017}; SIMBA \citep[][dotted line]{Dave:2019}, MillenniumTNG \citep[][dashed line]{Pakmor2023} and FABLE (doubledot-dashed line, \citealt{Henden2018},  Bigwood et al. in prep.).}
	\label{fig:BCMjoint}
\end{figure*}

\begin{figure*}
	\centering
	\includegraphics[width=1\linewidth]{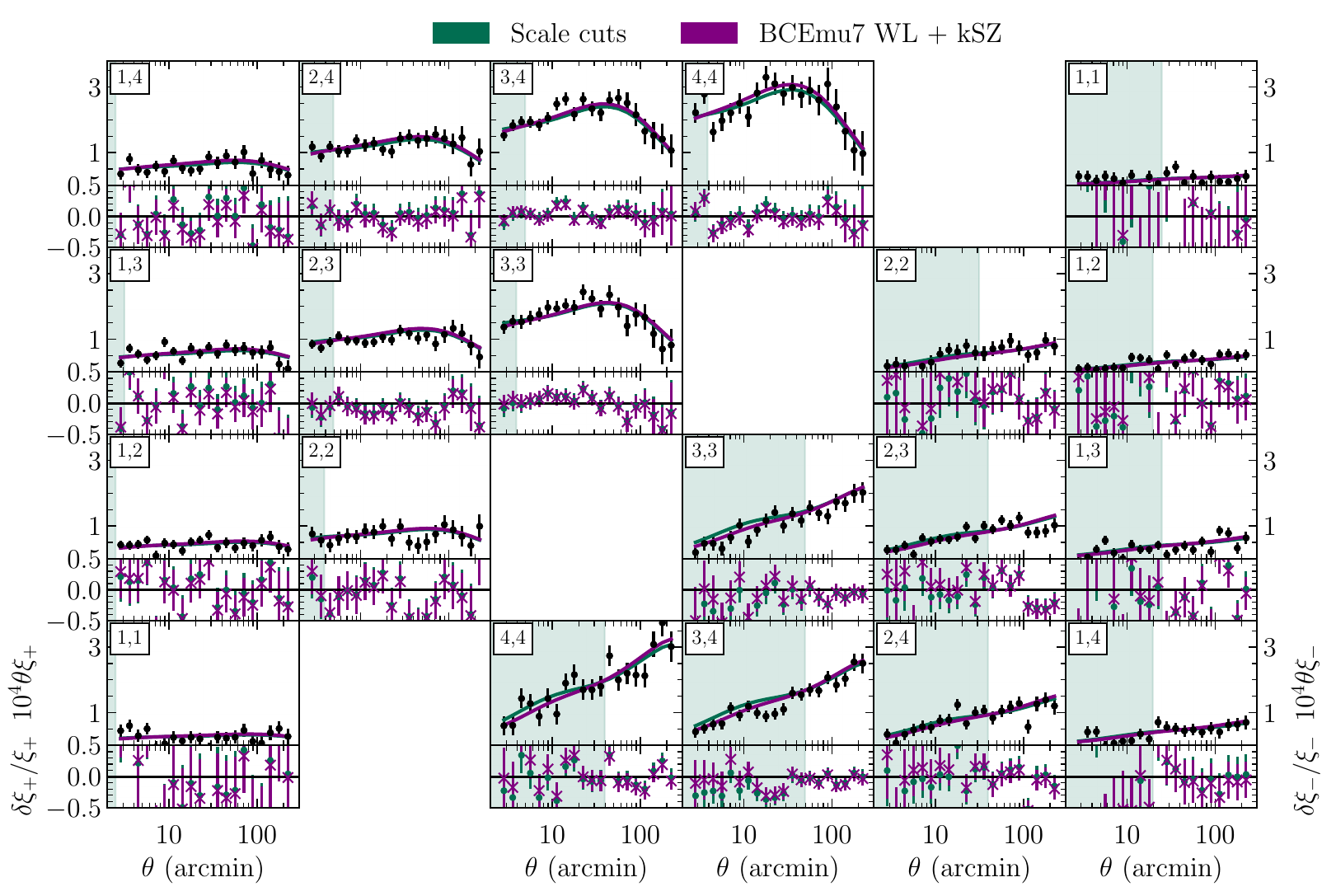} 
	\caption{The DES Y3 cosmic shear two-point correlation function measurements, $\xi_{\pm}$, as a function of angular separation, $\theta$, (black data points, upper panels) from \citet{Amon:2021} and \citet*{Secco:2022} for each pair of redshift bins, indicated by the panel label. The measurements are scaled by $\theta$ for visual aid.  The error bars are calculated as the square root of the diagonal of the analytic covariance matrix.   We show the best fit $\Lambda$CDM theoretical predictions to the large angular scale measurements (indicated by the shaded regions) without a baryon model (green line) and the best fit obtained from jointly analysing all angular scales of the DES shear measurements with ACT kSZ data using the BCEmu7 model (purple line).  The residuals between the measurement and best-fit model predictions $\delta\xi_{\pm}/\xi_{\pm}$ are shown in the lower panels.  }
\label{fig:DV}
\end{figure*}

\begin{figure}
\centering
\includegraphics[width=0.95\columnwidth]{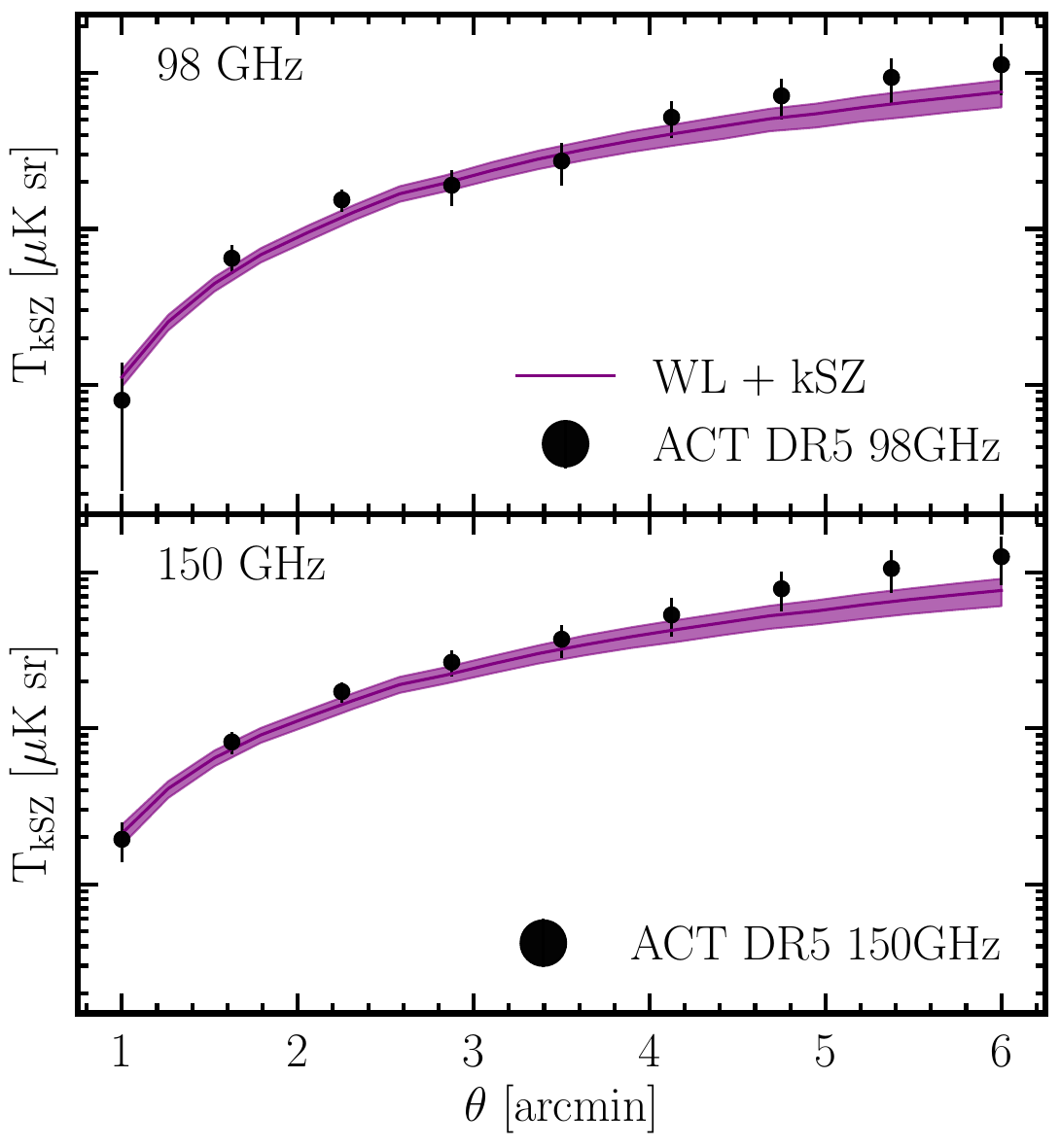}\hfill
\caption{Stacked ACT DR5 kSZ temperature profile measurements (black points) as a function of angular radius, $\theta$, at 98GHz and 150GHz \citep{Schaan:2021} and the best fit from the joint analysis with the DES Y3 cosmic shear using the BCEmu7 model (purple).  The model profiles are convolved with the f90 and f150 beam profile for comparison to the 98 GHz and 150 GHz data, respectively. 
}
\label{fig:kszresults}
\end{figure}

In the previous section we demonstrated that weak lensing $S_8$ constraints are degenerate with the amount of baryonic feedback. A complementary approach to simulation-informed baryon models is to use a  flexible model to jointly analyse cosmic shear with probes of the gas distribution in order to better constrain the model parameters.  In this section, we report the results of a joint analysis of the DES Y3 cosmic shear and ACT kSZ measurements, using the BCEmu7 baryon mitigation model, described in Section~\ref{sec:ksz}.

\subsection{Cosmological parameter constraints}

The left panel of Fig.~\ref{fig:BCMjoint} compares the marginalised posteriors on $S_8$ and $\Omega_{\rm m}$ attained from the WL-only analysis using BCEmu7 (pink) to those obtained from a joint analysis with kSZ (purple).  The \textit{Planck} TTTEEE $\Lambda$CDM posteriors are shown for reference \citep{EfstathiouGratton:2021}.  
The mean marginal value of $S_8$ for the joint analysis is found with 68\% credible levels to be

\begin{align}
\label{eqn:S8kSZ}
\begin{aligned}
%{\rm BCEmu7:}\,\,\, &S_8 &=&\,\,\,  0.818^{+0.017}_{-0.024} \\  
{\rm Lensing+kSZ}\,\,\, &S_8 &=&\,\,\, 0.823^{+0.019}_{-0.020} \,,\\  
\end{aligned}
\end{align}
\noindent which corresponds to a $\sim$0.2$\sigma$ shift towards higher values with respect to the result of the WL-only BCEmu7 analysis of $S_8=0.818^{+0.017}_{-0.024}$.  

\citet{AAGPE2022} and \citet{preston/etal:2023} have proposed that the $S_8$ tension could be resolved if the non-linear matter power spectrum is suppressed more strongly than is currently assumed in weak lensing analyses, either due to unmodelled baryonic feedback effects or non-standard dark matter. For baryonic feedback to be the source, its effects on the matter power spectrum would be more extreme than is currently predicted by the hydrodynamical simulations.  
The impact of the kSZ on the $S_8$ constraint to reduce the uncertainty and shift the value toward \textit{Planck} is minor given the low signal-to-noise of the kSZ measurements. However, it motivates us to investigate the impact of the kSZ on the $P(k)$ constraints and discuss the possibility of a more extreme suppression.

We find that incorporating a joint analysis with kSZ results in a significant improvement in the constraint on $\log_{10}M_{\rm c}$, reducing the uncertainty by a factor of $\sim$3 with respect to the WL-only analysis.  The joint kSZ and WL data prefer larger values of $\theta_{\rm ej}$ and $\gamma$, and lower values of $\mu$ (see Fig.~\ref{fig:bcemuparams} and Table~\ref{tab:bcemuparamconstraints}), suggesting that 
gas is ejected to larger radii, redistributing matter on larger scales (Fig.~\ref{fig:bfcintuition}). This supports the idea of a more extreme feedback scenario, resulting in higher values of $S_8$.  The improvement on the WL constraint on $S_8$ with the inclusion of kSZ is $\sim$10$\%$. Although this is modest, it is clear that the parameter space of the baryonification model is better constrained, even in this case of a kSZ measurement with signal-to-noise of $\sim$7.

We compare the best-fit models to the measured DES Y3 cosmic shear two-point correlation functions, $\xi_{\pm}$ in Fig.~\ref{fig:DV}. The dark green line indicates the best fit for our reanalysis of DES Y3 using their scale cuts, which are indicated by the greeb shaded region. The purple line shows that for the BCEmu joint analysis of all angular scales of the lensing measurement and the kSZ. The lower panels highlight the fractional residuals between the measurements and the model, $(\xi_{\pm}-\xi_{\pm}^{\rm model})/\xi_{\pm}^{\rm model}$, following the same colour scheme. While the fits are indistinguishable at large scales, at small scales, particularly for $\xi_-$, the predictions differ and the best-fit line for the joint analysis has a lower amplitude. Both of these model choices provide a good fit to the data, although their $S_8$ values differ by $\sim$1$\sigma$ and their non-linear matter spectrum predictions differ substantially. This highlights the degeneracy between a low-$S_8$ cosmology and a higher-$S_8$ cosmology with baryonic effects modelled on non-linear scales. Fig.~\ref{fig:kszresults} displays the joint WL + kSZ BCEmu7 constraints on the stacked kSZ radial temperature profile at 98GHz and 150GHz (purple).  We verify that like the DES data vector, the WL + kSZ model provides a good fit to the data.  The $\chi^2_{\rm red}$ values reported in Table~\ref{tab:cosmologyresults} further show that the best-fit models attained from the shear and WL + kSZ analyses are an equally good fit to the datasets. 

It is important to note that the constraints on $S_8$ that we obtain are dependent on the choice of NLA as the IA model.  Appendix~\ref{app:IA} tests the impact of using the Tidal Alignment and Tidal Torquing \citep[`TATT';][]{blazek2019} superspace IA model.  In a WL-only analysis with BCEmu7 and using TATT, we find a value of $S_8$ $\sim0.5\sigma$ lower than that we obtain in our fiducial analysis with NLA. This shift is found to be consistent across the various baryon models that we consider, as well as with previous findings \citep*[e.g.][]{Secco:2022,KiDSDES}. Therefore, when considering the results of this work in the context of the $S_8$ tension, it is important to bear in mind the existing uncertainty in IA modelling and the shifts in cosmological parameters that can occur as a result.

\subsection{Power suppression constraints}\label{sec:WLkSZpk}

We investigate how the constraint on the suppression of the matter power spectrum changes with the addition of the kSZ data.  The right panel of Fig.~\ref{fig:BCMjoint} shows the mean constraint and 68\% confidence level for the BCEmu7 cosmic shear analysis (pink), as shown previously, compared to the joint WL and kSZ analysis (purple).  The joint analysis results in a mean suppression that is more extreme at all displayed $k$-scales. For example, at $k=2h/$Mpc, the 1$\sigma$ bounds on the power suppression  from the WL-only analysis range from 5-20$\%$, and with the inclusion of the kSZ, the suppression ranges from 10-25$\%$ . Consistent with \citet{Schneider:2022}, we find that the inclusion of kSZ data in the analysis improves the constraint from weak lensing only to favour more extreme scenarios. 
Compared to the hydrodynamical simulations, the mean suppression from the joint analysis is more extreme than all of MillenniumTNG, BAHAMAS, FLAMINGO, FABLE and SIMBA at $k=2h/$Mpc by more than $\sim2.5\sigma$ , $\sim1.4\sigma$, $\sim1.9\sigma$, $\sim2.4\sigma$ and $\sim0.9\sigma$ respectively.  Of the simulations, SIMBA shows the best agreement with the WL and kSZ constraint.  We note, however, that for BAHAMAS and FLAMINGO we compare to only their fiducial feedback variants here, but both suites have more extreme feedback variations which are in better agreement with our measurements.

Finally, we consider our findings in the context of \citet{AAGPE2022, preston/etal:2023}, which proposed that the $S_8$ tension could be resolved if a more extreme baryon feedback scenario than that predicted by the state-of-the-art hydrodynamical situations existed. These works analysed the KiDS and DES WL data assuming the \textit{Planck} cosmology on linear scales, and modulating the non-linear power spectrum suppression via the $A_{\rm mod}$ parameter. 
\citet{preston/etal:2023} therefore quantified the small-scale suppression required to resolve the suppression, which corresponds to $A_{\rm mod}\approx0.82$. In this scenario, the matter power suppression is surpressed enough to  reconcile the difference in $S_8$ between DES Y3 cosmic shear and the Planck $\Lambda$CDM model. 
The joint WL + kSZ constraints on the suppression of the matter spectrum that we obtain are consistent with the $A_{\rm mod}=0.82$ prediction.

\section{Discussion: Consistency of X-ray and kinetic SZ data}\label{sec:XSZ}

Our WL + kSZ approach is complementary to existing efforts toward the goal of a complete model of baryonic feedback that is consistent with a wide range of observables. X-ray measurements of cluster gas mass fractions are the primary observable used to calibrate or benchmark many hydrodynamical simulations \citep[e.g.][]{McCarthy2017,Henden2018, Kugel2023}.  Observations of the tSZ power spectrum and tSZ flux–halo mass relation have also been used to assess the simulations' realism \citep{McCarthy:2023, Henden2018,Schaye2023,Pakmor2023}.  In general, while tSZ and X-ray measurements are more sensitive to the inner regions of galaxy groups and clusters,  the kSZ effect is well-suited to probe the outskirts of halos, through its sensitivity to low density and low temperature environments \citep{Schaan:2021}.  Furthermore, X-ray and tSZ observations are typically derived from massive cluster halos ($M_{500}>10^{14}$), while our kSZ measurements represent halos of mass $M_{200}\sim10^{13}$, closer to the mass range that WL is most sensitive to.
Since the kSZ effect probes the gas in halos of a different mass regime and on different scales to that which is currently used to calibrate feedback effects in simulations, it may allow new insights to be gained.

Our weak lensing constraints on the matter power spectrum suppression point to a feedback scenario that is more extreme than most simulations predict. This observation holds in all three flexible model scenarios tested (Fig.~\ref{fig:complexitychoice}).  The addition of the kSZ data pushes the mean constraint towards an even more disruptive feedback scenario: at $k=2$$h$/Mpc, the fiducial FLAMINGO simulation is disfavoured at $\sim$2$\sigma$ and MillenniumTNG at $\sim$2.5$\sigma$.  Beyond the comparison to the simulations, our findings point to a more disruptive feedback scenario than inferred from the predicted $P(k)$ suppression using X-ray gas and stellar fraction observations \citep{Grandis2023}, as well as tSZ-mass relation of clusters \citep{to2024}, which could point to interesting differences between the X-ray view of baryon feedback, compared to that from weak lensing and kSZ. We note that a strong feedback scenario was also found using an analysis of the cross-correlation of diffuse X-ray and weak lensing \citep{Ferreira2023}, and hints of a stronger scenario with the cross-correlation of tSZ with galaxies \citep{Pandey2023}.

\subsection{Mass dependence of the total baryon fraction}

\begin{figure*}
\centering
\includegraphics[width=.492\textwidth]{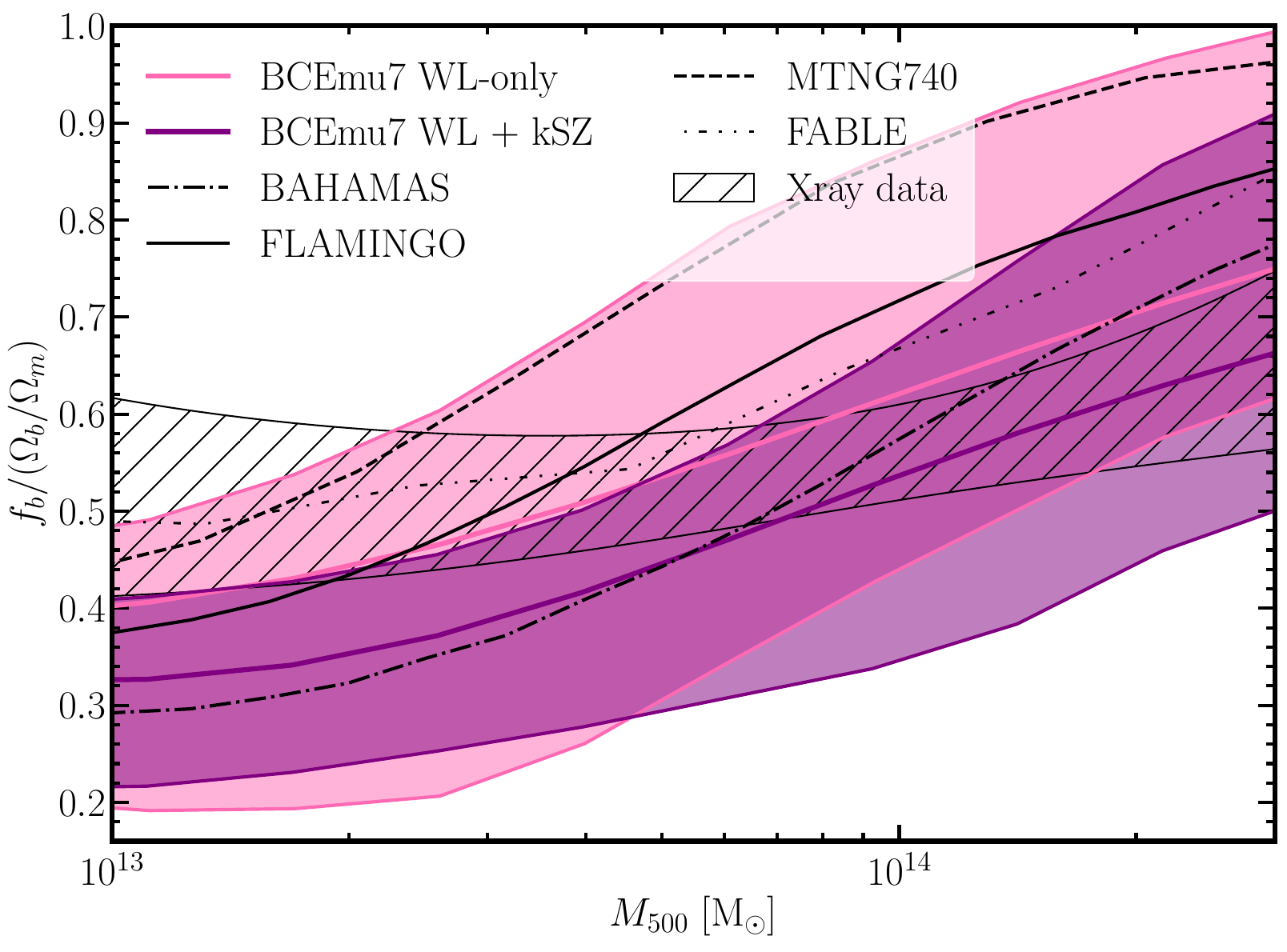}\hfill
\includegraphics[width=.5\textwidth]{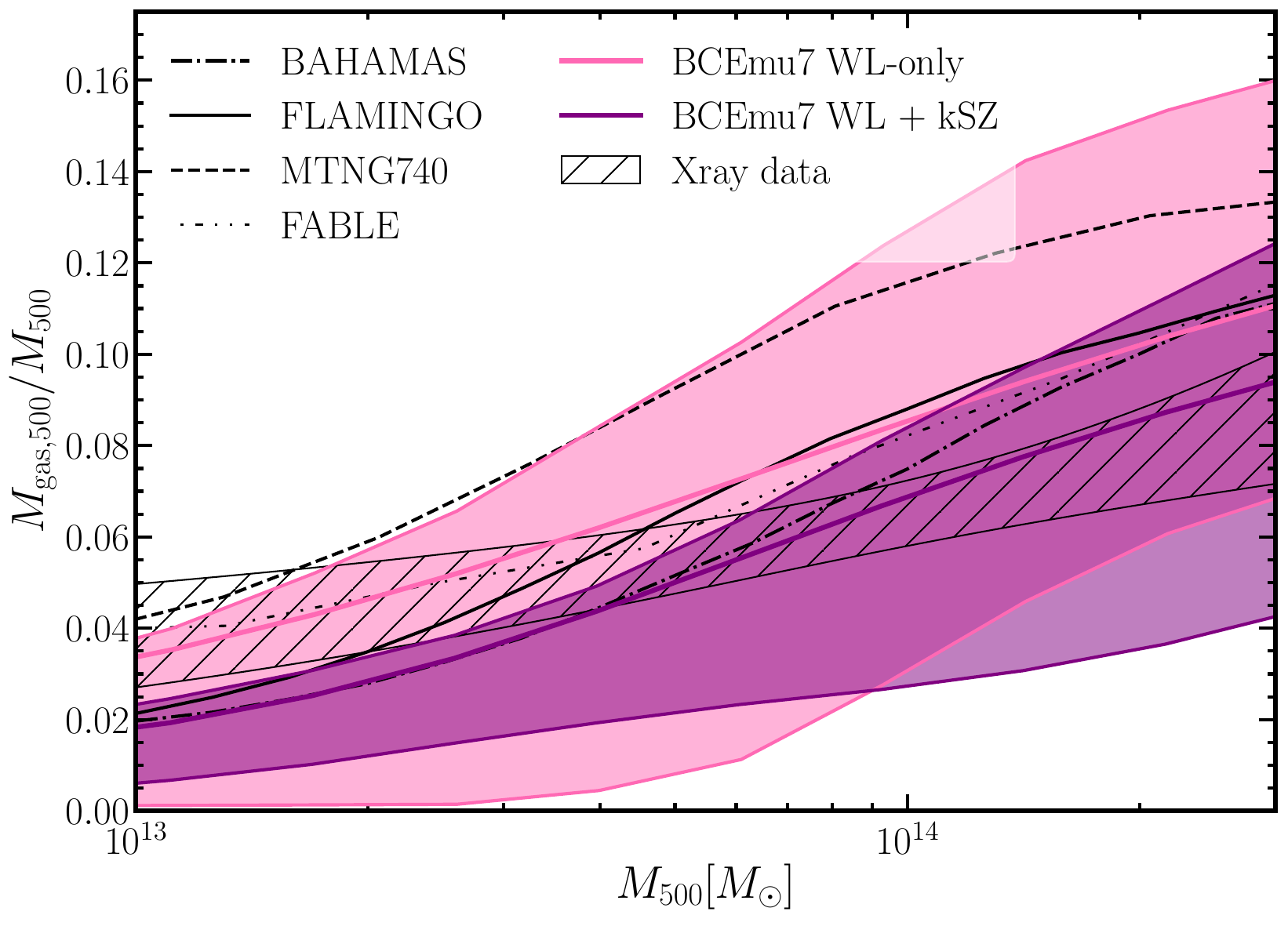}
\caption{\textit{Left:} The total baryon fraction, $f_{\rm b}/\Omega_{\rm b}/\Omega_{\rm m}$,  as a function of halo mass, $M_{500}$, where $\Omega_{\rm b}/\Omega_{\rm m}$ is the cosmic baryon fraction and $f_{\rm b}$ is the fraction of mass in baryons to the total halo mass in groups and clusters.  We plot the constraint attained from analysing the DES Y3 cosmic shear data with BCEmu7 (pink) in addition to the constraint attained from the joint WL + kSZ analysis (purple).   Solid lines show the mean baryon fraction halo mass relation, and the shaded regions enclose the $1\sigma$ constraints.  \textit{Right:}  The fraction of mass in gas to the total halo mass in groups and clusters, $M_{\rm gas, 500}/M_{500}$, as a function of halo mass, $M_{500}$.  We plot the constraint attained from analysing the DES Y3 cosmic shear data with BCEmu7 (pink) in addition to the constraint attained from the joint WL + kSZ analysis (purple).  Solid lines show the mean gas fraction halo mass relation, and the shaded regions enclose the $1\sigma$ constraints.  In both panels we plot the X-ray constraints from HSC-XXL $1\sigma$ \citep{Akino_2022} as the black hatched region, scaled to the mean cosmology obtained from the WL + kSZ analysis.  For reference, we also plot the measurements from BAHAMAS \citep[][dash-dotted line]{McCarthy2017}; FLAMINGO \citep[][solid line]{Schaye2023}; MillenniumTNG \citep[][dashed line]{Pakmor2023} and FABLE (doubledot-dashed line, \citealt{Henden2018},  Bigwood et al. in prep.).  }
\label{fig:kszgas}
\end{figure*}

As discussed in Section~\ref{sec:Spk}, the baryon fraction, $f_{\rm b}$, measured in halos of mass $M_{500}\approx10^{14}M_{\odot}$ can be related directly to the matter power spectrum suppression in a manner which is robust to a number of baryonic feedback prescriptions \citep{vandaalen:2020}. In this section, we scrutinise the weak lensing constraints on the halo mass dependence of the total baryon fraction in halos with respect to the cosmic baryon fraction, $f_{\rm b}/(\Omega_{\rm b}/\Omega_{\rm m})$. For a given set of parameters, we can use the BCEmu model to compute the baryon fraction through the summation of the integrated gas and stellar profiles (see equation~\ref{eq:rhogas} and equation~\ref{eq:fstar}) out to $R_{500}$.  For both the WL-only and WL + kSZ BCEmu7 analyses, we compute the $f_{\rm b}/(\Omega_{\rm b}/\Omega_{\rm m})$-$M_{500}$ relation for the cosmology and baryonic feedback parameters sampled at each step of the chain.

Fig.~\ref{fig:kszgas} shows the constraints for the WL-only analysis in pink and the WL + kSZ in purple, with the mean relation indicated by a solid line and 68\% confidence level as a shaded region.   Measurements of halo masses are generally reliant on the assumption of hydrostatic equilibrium and hence may suffer a bias due to non-thermal support in the halo \citep[see, e.g. ][]{Rasia2006}.  However, estimates have also been obtained from weak lensing data which do not require this assumption \citep{Akino_2022,Mulroy_2019,Hoekstra2015,Eckert2016}.  As a result we compare to the $1\sigma$ region of the X-ray HSC-XXL constraint \citep{Akino_2022} in hatched black.  The X-ray data has a dependence on cosmology through $E(z)$, hence we scale the data to the mean cosmology constrained by the WL + kSZ analysis.  We also show the baryon fractions measured in BAHAMAS \citep{McCarthy2017}, FLAMINGO \citep{Schaye2023}, MillenniumTNG \citep{Pakmor2023} and FABLE (\citealt{Henden2018},  Bigwood et al. in prep.).

Overall, the BCEmu7 WL-only and WL + kSZ constraints are in good agreement. The addition of the kSZ reduces the uncertainty by a factor of $\sim 1.5$ at $M_{500}\approx 10^{13} M_{\odot}$ and prefers lower baryon fractions for all masses. For large groups with $M_{500}\gtrapprox 5\times 10^{13} M_{\odot}$, we find good consistency between the 68\% confidence levels on $f_{\rm b}/(\Omega_{\rm b}/\Omega_{\rm m})$ as predicted by BCEmu7 WL-only and WL + kSZ analyses compared to the X-ray data.
However, for lower mass groups with masses $M_{500}\approx 10^{13} M_{\odot}$, the WL + kSZ data prefers a total baryon fraction that is lower than the X-ray data by $\sim 1.4\sigma$. 
With the exception of BAHAMAS which is in best agreement, all of the simulations predict higher values of $f_{\rm b}/(\Omega_{\rm b}/\Omega_{\rm m})$ for all masses compared to the mean of the WL + kSZ constraint. At $M_{500}\sim 4\times 10^{13} M_{\odot}$, FLAMINGO predicts $\sim1.5\sigma$ higher values, MilleniumTNG $\sim3.1\sigma$ and FABLE $\sim1.4\sigma$.

In Appendix~\ref{sec:vandaalen}, we compare the relationship between the baryon fraction and power suppression that our WL + kSZ constrains to that proposed by \citet{vandaalen:2020}, which is a good fit to many of the hydrodynamical simulations. We note that the BCEmu model does not impose a prior on this relationship and therefore provides a route to place constraints using data. Given that our constraints differ from the simulation-based relationship, we can conclude that either there are unaccounted for systematics in the WL and kSZ data, the BCEmu model allows non-physical scenarios, or  simulations do not currently capture the full possible range of feedback effects, thereby overestimating the relationship \citep[see e.g.][]{Debackere2020}. Future work is needed to further understand this relationship using data. 
 
The SP(k) baryon model also provides a direct mapping from the matter power spectrum suppression to the baryon fraction. In Appendix~\ref{app:spk} we show the mean and 68\% confidence levels for the baryon fraction attained from the WL-only SP(k) analyses.
Similar to the BCEmu case, there is good agreement of the lensing analysis with SP(k):wide and X-ray data at high masses $M_{500}\sim 1\times 10^{14} M_{\odot}$.   In agreement with the findings with BCEmu, we find that lensing prefers slightly lower baryon fractions for halos $M_{500}\sim 10^{13} M_{\odot}$, with the SP(k):wide analysis lying lower by $\sim 1.3\sigma$.  This highlights that the lensing data alone, when analysed with flexible modelling choices, prefers a lower baryon content in halos to that measured by X-ray observations and that predicted by simulations.

\subsection{Mass dependence of the gas mass fraction}
The majority of baryonic mass in galaxy groups and clusters exists as diffuse gas and measurements of the fraction of the halo mass in gas, $M_{\mathrm{gas}}/M_{500}$, are also sensitive to the matter power spectrum suppression \citep{Schneider:2015,vandaalen:2020,arico2023}.  The hot intracluster medium is X-ray luminous, hence studies of the X-ray emissivity of groups and clusters allow measurements of the gas mass to be derived. 

In this section, we compare X-ray derived  measurements of the $M_{\mathrm{gas}}/M_{500}$-$M_{500}$ relation to those constrained by the BCEmu7 WL-only and joint WL + kSZ analyses in this work.  As with the baryon fractions, we calculate the $M_{\mathrm{gas}}/M_{500}$-$M_{500}$ relation at each step in the chain by integrating equation~\ref{eq:rhogas} to $r_{500}$.  The right panel of Fig.~\ref{fig:kszgas} shows the mean and 68\% confidence levels on the gas fraction-halo mass relation for both the WL-only (pink) and WL + kSZ chains (purple).  We compare to the $1\sigma$ region of the X-ray HSC-XXL constraint \citep{Akino_2022} in hatched black (also scaled to the mean cosmology obtained in the WL + kSZ analysis) and the baryon fractions measured in BAHAMAS \citep{McCarthy2017}, FLAMINGO \citep{Schaye2023}, MillenniumTNG \citep{Pakmor2023} and FABLE (\citealt{Henden2018},  Bigwood et al. in prep.).

In consistency with the baryon fraction result, we find that the BCEmu7 WL-only and WL + kSZ constraints are in good agreement with each other and the X-ray data for massive groups of $M_{500}\gtrapprox 5\times 10^{13} M_{\odot}$.  For groups of lower masses we find that the joint WL + kSZ analysis significantly improves the constraining power on the gas fractions, with the 68\% confidence levels on $M_{\mathrm{gas}}/M_{500}$ at $M_{500}\approx 10^{13} M_{\odot}$ shrinking by a factor of $\sim$2 between the WL-only and WL + kSZ analyses.  At $M_{500}\approx 10^{13} M_{\odot}$, the WL + kSZ data constrains a total baryon fraction that is lower than the X-ray data by $\sim$1.6$\sigma$.  The fiducial BAHAMAS simulation is in best agreement with the WL + kSZ constraint, with the remaining simulations predicting higher values of $M_{\rm gas, 500}/M_{500}$ at all halo masses.  At $M_{500}\sim 4\times 10^{13} M_{\odot}$, the fiducial FLAMINGO simulation predicts $\sim$2.3$\sigma$ higher values, MilleniumTNG $\sim$7.2$\sigma$ and FABLE $\sim$2.1$\sigma$.

The baryon and gas fraction constraints by the WL + kSZ data imply that lower mass groups are expelling a greater amount of baryonic matter due to feedback than predicted by X-ray measurements.  When combined with the findings of Section~\ref{sec:WLkSZpk} that WL + kSZ predicts a greater mean matter power spectrum than all of the simulations, we build a consistent picture: that X-ray observations constrain a weaker feedback scenario than that preferred by WL + kSZ, subject to the uncertainties we have discussed. 

\section{Summary and Conclusions}\label{conclusions}

Weak lensing measurements at small angular scales are statistically powerful. Not only do they offer a window to test $\Lambda$CDM in the non-linear regime, but also to constrain baryonic feedback effects and benchmark hydrodynamical simulations.  Extracting accurate cosmological information relies on accurate modelling of physical processes associated with baryons which can re-distribute matter on small scales. Otherwise, unmodelled baryonic effects can bias the cosmological constraints from lensing analyses \citep{AAGPE2022, preston/etal:2023} or severely limit their precision \citep{Amon:2021,KiDSDES}. 
The aim of this work has been two-fold.  Firstly, we test four baryon model approaches, comparing their ability to constrain cosmological parameters and the suppression of the matter power spectrum.  Secondly, we perform a joint analysis of DES cosmic shear with ACT kSZ measurements.  We demonstrate that a combined weak lensing and kSZ analysis provides an exciting opportunity to not only improve constraints on cosmological parameters, but also on astrophysical effects.  
The main results of this study are:
\begin{itemize}
    \item We perform \textbf{the first mock baryon model comparison for cosmic shear} to assess the robustness of the models to recover the underlying cosmology in a DES Y3-like analysis.  We consider three different mock `feedback scenarios' to test four baryon strategies; a halo
    model approach, a simulation-based emulator, an analytical N-body simulation model and the approach of discarding small angular scales. In general, using restrictive modelling choices which do not capture the input matter power spectrum suppression can underestimate the recovery of $S_8$. Given the spread in the true suppression of the matter power spectrum as predicted by simulations and the lack of observational constraints on this quantity, model flexibility is crucial to ensure accurate cosmology, but it comes at a cost: the uncertainty on $S_8$ can degrade by up to a factor of $\sim2$ with different model choices. 
    \item  We analyse the DES Y3 cosmic shear data with the three baryon feedback models, with the scale cut approach, and with no feedback model. \textbf{We find that each baryon model provides a good fit to the DES Y3 data, but the measured value of $S_8$ varies by $\sim$0.5-2$\sigma$ and the errorbar by a factor of $\sim$1.5.}
    \item We output the posterior for the suppression of the matter power spectrum using the three models, with and without their informative priors. \textbf{For all three of the models, without their informative priors, the mean suppression of the matter power spectrum constrained using DES Y3 WL is more extreme than the prediction from the hydrodynamical simulations considered}.
    \item We \textbf{jointly analyse the DES Y3 cosmic shear with kSZ} measurements from ACT DR5 and CMASS, in order to constrain the BCEmu7 model parameters.  We find a slightly higher value of $S_8= 0.823^{+0.019}_{-0.020}$, compared to the value attained by the WL-only analysis. If instead we analyse the DES Y3 cosmic shear jointly with X-ray baryon fraction constraints using SP(k), we find a lower value of $S_8= 0.806^{+0.015}_{-0.013}$.
    \item The kSZ significantly improves the constraint on the suppression of the matter power spectrum from WL. The joint \textbf{WL+kSZ predicts a more extreme suppression of the matter power spectrum than the WL scenario}, with a mean constraint predicting a greater suppression than the fiducial BAHAMAS, MillenniumTNG, fiducial FLAMINGO and FABLE simulations. At $k\gtrsim 1h/$Mpc, only SIMBA falls within the 1$\sigma$ bound.  
    \item  We \textbf{constrain the baryon fraction-halo mass and the gas fraction-halo mass relations using WL + kSZ data}.  Both the baryon fraction and gas fraction is consistent with that from \citet{Akino_2022} X-ray data within the 68\% confidence level for groups $M_{500}\approx 10^{14} M_{\odot}$.  However for lower mass groups $M_{500}\approx 10^{13} M_{\odot}$ the baryon fraction lies $\sim$1.4$\sigma$ lower than the X-ray data and the gas fraction lies $\sim$1.6$\sigma$ lower. 
    \item Our constraints on the matter power spectrum suppression, baryon fractions and gas fractions all point towards a tension between the feedback of groups and clusters predicted by weak lensing + kSZ and X-rays, or X-ray calibrated models. 
\end{itemize}

The next generation of shear surveys, such as Vera C. Rubin Observatory’s Legacy Survey of Space and Time \citep[LSST;][]{LSST}, Euclid \citep{Euclid} and the
Nancy Grace Roman Space Telescope \citep{Roman} will deliver unprecedented statistical power to test the cosmological model in the non-linear regime ($k>0.1 h {\rm Mpc}^{-1}$) using weak lensing. This work highlights the importance of determining an accurate model for baryonic effects, which is flexible enough to not bias cosmological constraints. It will be crucial to test models of feedback with thorough mock analyses. In order for the model uncertainty in weak lensing analyses to not limit the statistical power of the survey, it is crucial to either incorporate external probes of the gas content to constrain the additional baryon parameters, or to reduce the uncertainty in simulation-based priors. The latter requires a consistent picture for baryonic feedback effects, supported by a range of observations. 

We demonstrate that joint analyses of gas measurements with weak lensing data not only improve cosmological constraints, but provide valuable constraints on astrophysical feedback models and benchmark hydrodynamical simulations. We find a consistent picture that could imply that the WL and kSZ data is in tension with the X-ray measurements, and as a result, the predictions from simulations calibrated to X-ray data. This is not particularly surprising as  X-ray measurements are generally sensitive to the hot gas content in the inner regions of clusters, compared to the outer regions and lower mass halos that kSZ measurements probe.

Looking ahead, kSZ measurements as a function of mass and redshift will provide a handle for improved baryonic feedback models. Spectroscopic galaxy surveys, such as the Dark Energy Spectroscopic Instrument \citep[DESI;][]{DESI:2016} and the Prime Focus Spectrograph \citep[PSF;][]{PFS}, will greatly increase the sample size of galaxy catalogues, in combination with the state-of-the-art CMB observations, for example, from Simons Observatory \citep{Simons}. 

\section*{Acknowledgements}
%The authors would like to thank Catherine Heymans and George Efstathiou for insightful feedback on this manuscript. 

Leah Bigwood is supported by a Science and Technology Facilities Council studentship.  Alexandra Amon was supported by a Kavli Fellowship for part of the duration of this work.  Jaime Salcido and Ian McCarthy were supported by an European Research Council (ERC) Consolidator Grant under the European Union’s Horizon 2020 research and innovation programme (grant agreement No 769130). 

Author contributions are as follows.  L. Bigwood and A. Amon performed the analysis and wrote the bulk of the manuscript.  The additional core authors of A. Schneider, J. Salcido and I. McCarthy contributed BCEmu and SP(k) code and continuous insight.  C. Preston and D. Sanchez assisted in the early stages, including the mock analysis.  D. Sijacki provided useful discussions and detailed comments on the manuscript.  E. Schaan, S. Ferraro and N. Battaglia provided the ACT kSZ dataset.  The paper has gone through internal review by the DES collaboration, with A. Chen and S. Dodelson serving as excellent internal reviewers.  The remaining authors have made contributions to the DES instrumentation, data collection, data processing and calibration, in addition to other code and validation tests.

This project has used public archival data from the Dark Energy Survey. Funding for the DES Projects has been provided by the U.S. Department of Energy, the U.S. National Science Foundation, the Ministry of Science and Education of Spain, 
the Science and Technology Facilities Council of the United Kingdom, the Higher Education Funding Council for England, the National Center for Supercomputing 
Applications at the University of Illinois at Urbana-Champaign, the Kavli Institute of Cosmological Physics at the University of Chicago, 
the Center for Cosmology and Astro-Particle Physics at the Ohio State University,
the Mitchell Institute for Fundamental Physics and Astronomy at Texas A\&M University, Financiadora de Estudos e Projetos, 
Funda{\c c}{\~a}o Carlos Chagas Filho de Amparo {\`a} Pesquisa do Estado do Rio de Janeiro, Conselho Nacional de Desenvolvimento Cient{\'i}fico e Tecnol{\'o}gico and 
the Minist{\'e}rio da Ci{\^e}ncia, Tecnologia e Inova{\c c}{\~a}o, the Deutsche Forschungsgemeinschaft and the Collaborating Institutions in the Dark Energy Survey. 

The Collaborating Institutions are Argonne National Laboratory, the University of California at Santa Cruz, the University of Cambridge, Centro de Investigaciones Energ{\'e}ticas, 
Medioambientales y Tecnol{\'o}gicas-Madrid, the University of Chicago, University College London, the DES-Brazil Consortium, the University of Edinburgh, 
the Eidgen{\"o}ssische Technische Hochschule (ETH) Z{\"u}rich, 
Fermi National Accelerator Laboratory, the University of Illinois at Urbana-Champaign, the Institut de Ci{\`e}ncies de l'Espai (IEEC/CSIC), 
the Institut de F{\'i}sica d'Altes Energies, Lawrence Berkeley National Laboratory, the Ludwig-Maximilians Universit{\"a}t M{\"u}nchen and the associated Excellence Cluster Universe, 
the University of Michigan, NSF NOIRLab, the University of Nottingham, The Ohio State University, the University of Pennsylvania, the University of Portsmouth, 
SLAC National Accelerator Laboratory, Stanford University, the University of Sussex, Texas A\&M University, and the OzDES Membership Consortium.

Based in part on observations at NSF Cerro Tololo Inter-American Observatory at NSF NOIRLab (NOIRLab Prop. ID 2012B-0001; PI: J. Frieman), which is managed by the Association of Universities for Research in Astronomy (AURA) under a cooperative agreement with the National Science Foundation.

The DES data management system is supported by the National Science Foundation under Grant Numbers AST-1138766 and AST-1536171.
The DES participants from Spanish institutions are partially supported by MICINN under grants ESP2017-89838, PGC2018-094773, PGC2018-102021, SEV-2016-0588, SEV-2016-0597, and MDM-2015-0509, some of which include ERDF funds from the European Union. IFAE is partially funded by the CERCA program of the Generalitat de Catalunya.
Research leading to these results has received funding from the European Research
Council under the European Union's Seventh Framework Program (FP7/2007-2013) including ERC grant agreements 240672, 291329, and 306478.
We  acknowledge support from the Brazilian Instituto Nacional de Ci\^encia
e Tecnologia (INCT) do e-Universo (CNPq grant 465376/2014-2).

This manuscript has been authored by Fermi Research Alliance, LLC under Contract No. DE-AC02-07CH11359 with the U.S. Department of Energy, Office of Science, Office of High Energy Physics.

\section*{Data availability}
The data underlying this article may be shared on reasonable request to the corresponding authors.

 %-------------- BIBLIO -------------------------------------------------------

\bibliographystyle{mnras_2author}

\bibliography{bib}
\label{lastpage}												

 %-----------------------------------------------------------------------------
  
\appendix

\section{Mock analysis}\label{app:mock}

\begin{figure}
\centering
\begin{subfigure}[b]{\linewidth}
\includegraphics[width=0.85\linewidth]{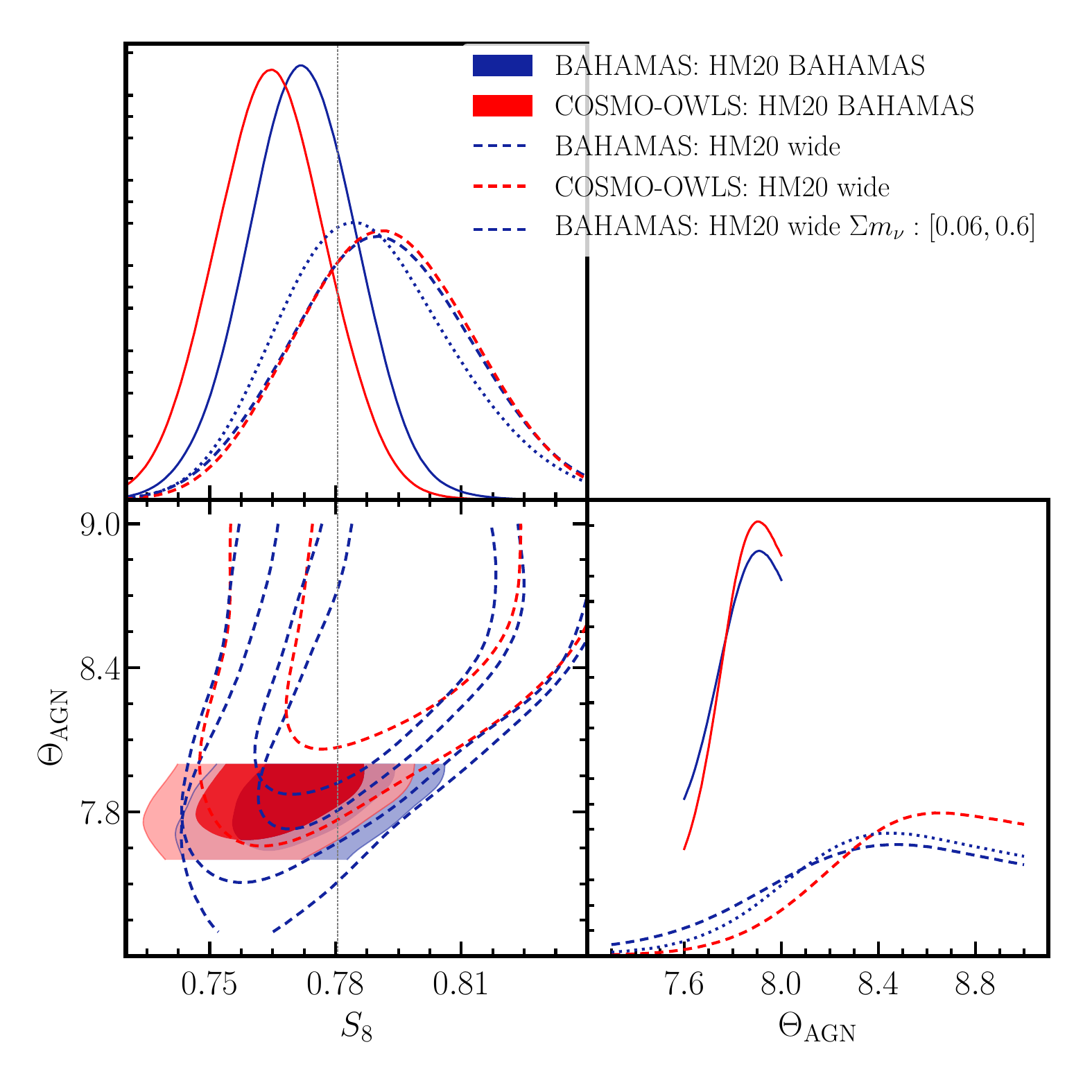}\hfill
\end{subfigure}
\begin{subfigure}[b]{\linewidth}
\includegraphics[width=0.85\linewidth]{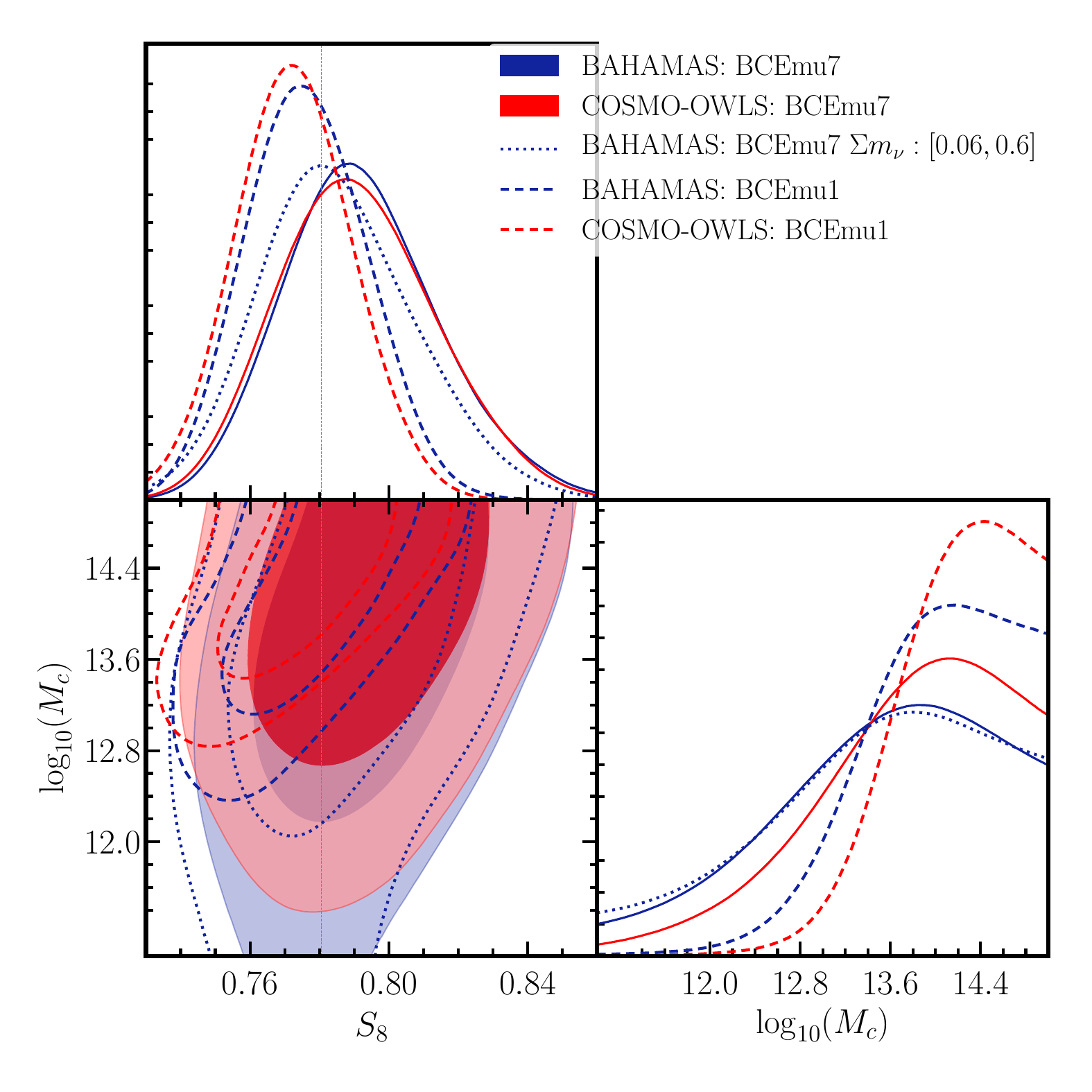}\hfill
\end{subfigure}
\begin{subfigure}[b]{\linewidth}
\includegraphics[width=0.85\linewidth]{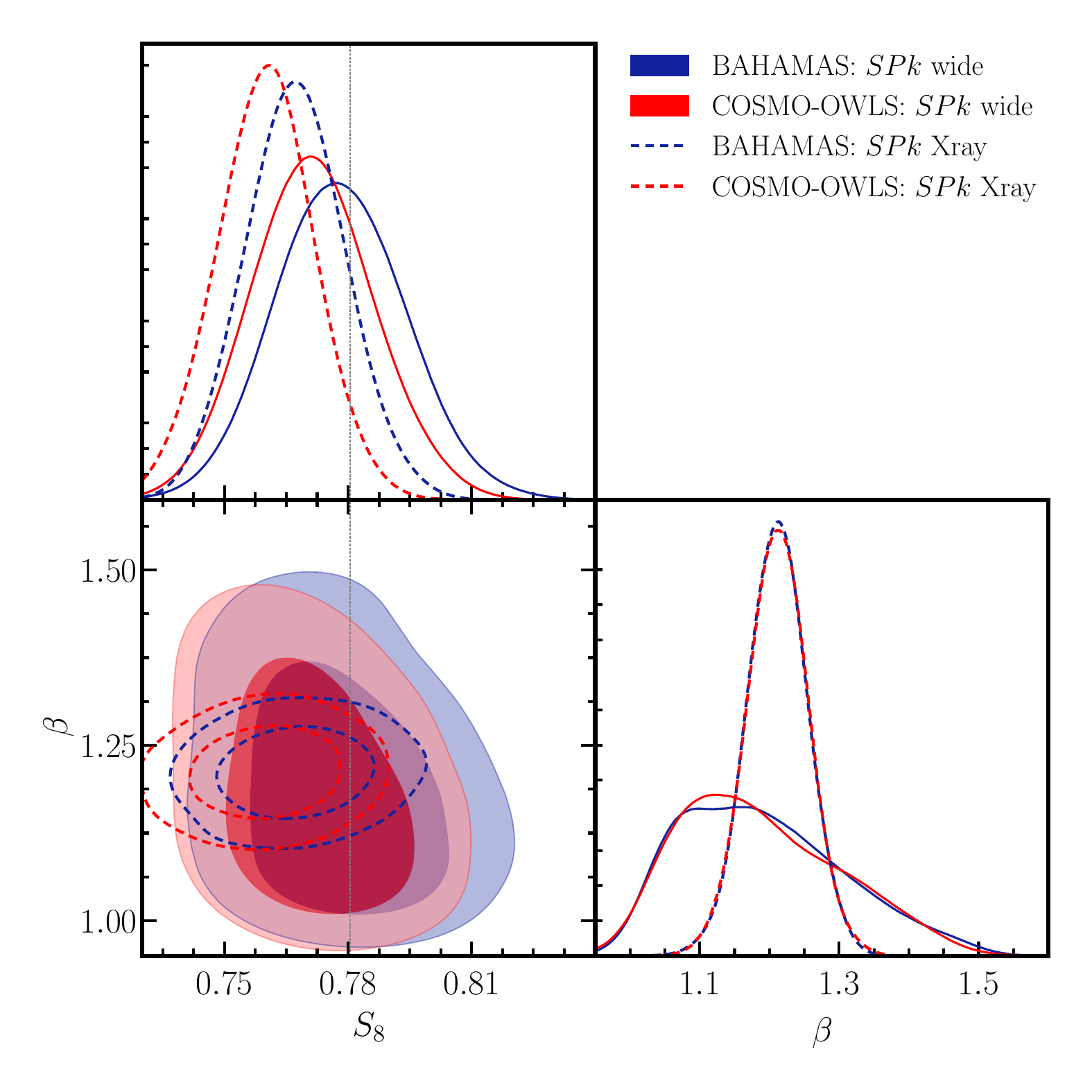}
\end{subfigure}
\caption{Marginalised posteriors for $S_8$ and the baryon model parameter using each mock for analyses using the HM20 (upper), BCEmu (middle) and SP(k) (lower) models.  The mocks are labelled in the format 'model used to create the mock: model/analysis choices used to analyse the mock'.   The inner and outer contours show the 68\% and 95\% confidence levels, respectively and the dashed line indicates the input cosmology.}
\label{fig:mockchoices}
\end{figure}

\begin{figure*}
    \centering
    \includegraphics[width=\linewidth]{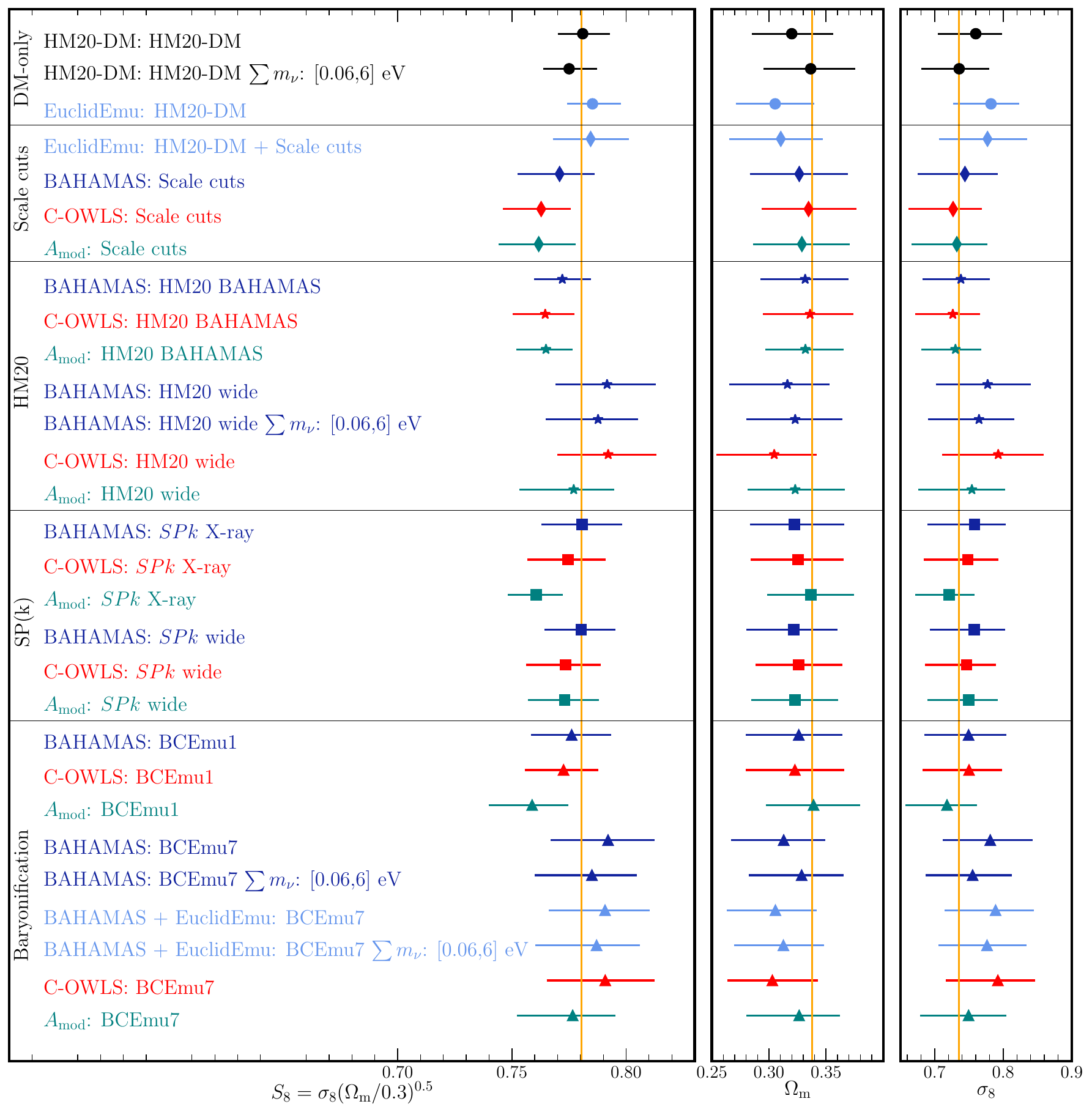}
    \caption{Summary of the 1D marginalised constraints on $S_8$, $\Omega_{\rm m}$ and $\sigma_8$ attained by our mock analysis with respect to the input cosmology.  We plots the mean and 68\% confidence levels as listed in Table~\ref{tab:mockfull}, with crosses showing the {\sc multinest} best-fit result.  The mocks were created with DES-Y3 covariance, with the input cosmology indicated by the vertical yellow line.  The top panel shows the results of validating the non-linear correction to the matter power spectrum due to dark matter only.  The mocks are labelled in the format 'model used to create the mock: model/analysis choices used to analyse the mock'.  For example, 'EuclidEmu: HM20-DM' denotes a mock created with Euclid Emulator for the non-linear correction to the matter power spectrum, and analysed with HM20. 
    The remaining panels validate the four methods of mitigating for baryonic feedback we test in this work; optimized scale cuts, HM20, BCEmu and SP(k).  If not otherwise specified, the mocks were created with HM20 as the dark matter-only non-linear correction to the matter power spectrum, and three 'baryonic feedback scenarios'; BAHAMAS, cosmo-OWLS and $A_{\rm{mod}}=0.82$.  We test analysing the mocks with various prior and analysis choices for each of the four baryon model approaches. 
    Each mock was analysed with HM20 as the dark matter-only non-linear correction to the matter power spectrum.}
    \label{fig:mocksum}
\end{figure*}

In this section, we provide more detail on the mock analysis presented in Sec.~\ref{sec:mocks}. We create the synthetic cosmic shear data using the best-fit cosmological parameters attained from the DES Y3 joint lensing and clustering analysis \citep{DES-3x2}\footnote{$\Omega_{\rm m}=0.3380$, $\Omega_{\rm b}=0.0450$, $10^{-9}A_{\rm s}=1.8418$, $h=0.615$, $n_{\rm s}=0.949$, $S_8= 0.7805$, $\sigma_8=0.7353$.}.  As described in Sec.~\ref{sec:mocks}, the mocks were created with HM20 as the dark matter-only non-linear correction to the matter power spectrum, unless otherwise stated.  We create data vectors with three variants for the impact on the non-linear power spectrum of baryonic feedback; we use the power suppression predicted by BAHAMAS 8.0, cosmo-OWLS 8.5 and $A_{\rm{mod}}=0.82$.  In this section we present a more detailed summary of the dark matter and baryonic feedback model validation tests ran using synthetic data vectors.  We refer to Table~\ref{tab:mockfull} which presents the {\sc multinest} mean values of $S_8$ attained for the full mock suite, reporting the relative shifts from the input cosmology.  The results are also summarised in in Fig.~\ref{fig:mocksum}, where we additionally plot the mean and best-fit $\Omega_{\rm m}$ and $\sigma_8$ for each mock test.  

\begin{table*}
\setlength\extrarowheight{3pt}
\caption{Summary of the mock tests we performed to validate the modelling of the non-linear matter power spectrum.
Mock tests 1-3 validate the non-linear correction to the matter power spectrum due to dark matter only.  The remaining tests validate the baryon feedback mitigation strategies we test in this work; scale cuts (4-7), HM20 (8-14), SP(k) (15-20) and BCEmu (21-29). We test analysing the mocks with various prior and analysis choices for each of the four baryon model approaches.  If not otherwise specified, the mocks were created with HM20 as the dark matter-only non-linear correction to the matter power spectrum, and three 'baryonic feedback scenarios'; BAHAMAS, cosmo-OWLS and $A_{\rm{mod}}=0.82$.  The 'mock' column therefore states the model/baryonic feedback scenario used to create the mock and the 'model' column labels the model/analysis choices used to analyse the mock'.  For example, a mock of 'EuclidEmu' and model of HM20-DM' denotes a mock created with Euclid Emulator for the non-linear correction to the matter power spectrum, and analysed with HM20.  We report the $S_8$ constraints and 68\% confidence level using the mean-marginal approach  $S_{8}$, with $\Delta S_8$ quantifying the offset from the true $S_8= 0.7805$, as a fraction of the $1\sigma$ error.} 

\begin{center}
\begin{tabular}{ccccc}
\hline
\hline
\ No. & Mock & Model & $S_{8}$ &$\Delta S_{8}$ \\

\hline
1 & HM20-DM & HM20-DM & $ 0.781^{+0.012}_{0.011} $ & $0.044\sigma$ \\ 
2 & HM20-DM & HM20-DM $\sum m_{\nu}$: [0.06,6] eV & $ 0.775^{+0.012}_{0.011} $ & $-0.461\sigma$\\ 
3 & EuclidEmu-DM & HM20-DM& $ 0.785^{+0.012}_{0.011} $ & $0.403\sigma$  \\ 
\hline
 4 &EuclidEmu-DM  & HM20-DM + Scale cuts & $ 0.784^{+0.017}_{0.017} $ & $0.239\sigma$  \\ 

5 & BAHAMAS 8.0 &Scale cuts & $ 0.771^{+0.015}_{0.018} $ & $-0.566\sigma$  \\
6 &cosmo-OWLS 8.5 & Scale cuts & $ 0.763^{+0.013}_{0.017} $ & $-1.191\sigma$ \\ 
7& $A_{\rm mod}= 0.820$ & Scale cuts & $ 0.762^{+0.016}_{0.018} $ & $-1.114\sigma$ \\
\hline
8 & BAHAMAS  8.0 & HM20 BAHAMAS & $ 0.772^{+0.013}_{0.012} $ & $-0.675\sigma$ \\
9& cosmo-OWLS 8.5  & HM20 BAHAMAS & $ 0.765^{+0.013}_{0.014} $ & $-1.182\sigma$\\
10& $A_{\rm mod}= 0.820$ & HM20 BAHAMAS & $ 0.765^{+0.012}_{0.013} $ & $-1.270\sigma$ \\
11& BAHAMAS  8.0& HM20 wide & $ 0.792^{+0.021}_{0.023} $ & $0.509\sigma$ \\ 
12 & BAHAMAS  8.0& HM20 wide $\sum m_{\nu}$: [0.06,6] eV &  $ 0.788^{+0.017}_{0.023} $ & $0.358\sigma$ \\ 
13 & cosmo-OWLS 8.5& HM20 wide & $ 0.792^{+0.021}_{0.022} $ & $0.537\sigma$\\ 
14& $A_{\rm mod}= 0.820$ & HM20 wide & $ 0.777^{+0.019}_{0.024} $ & $-0.180\sigma$\\
\hline
15& BAHAMAS  8.0 & SP(k): X-ray & $ 0.781^{+0.017}_{0.018} $ & $0.017\sigma$ \\
16& cosmo-OWLS 8.5 & SP(k): X-ray & $ 0.775^{+0.016}_{0.018} $ & $-0.347\sigma$ \\
17& $A_{\rm mod}= 0.820$ & SP(k): X-ray &$0.761^{+0.012}_{0.012} $ & $-1.643\sigma$ \\
18& BAHAMAS  8.0 & SP(k) wide & $ 0.780^{+0.015}_{0.016} $ & $-0.010\sigma$ \\
19& cosmo-OWLS 8.5 & SP(k) wide & $ 0.774^{+0.015}_{0.017} $ & $-0.424\sigma$ \\
20& $A_{\rm mod}= 0.820$ & SP(k) wide & $ 0.773^{+0.015}_{0.016} $ & $-0.479\sigma$\\
\hline
21 & BAHAMAS  8.0& BCEmu1 & $ 0.776^{+0.017}_{0.018} $ & $-0.248\sigma$  \\ 
22 & cosmo-OWLS 8.5 & BCEmu1 & $ 0.773^{+0.015}_{0.017} $ & $-0.489\sigma$ \\ 
23 &  $A_{\rm mod}= 0.820$ & BCEmu1 & $ 0.759^{+0.016}_{0.019} $ & $-1.239\sigma$ \\
24 &  BAHAMAS  8.0& BCEmu7 & $ 0.792^{+0.020}_{0.025} $ & $0.510\sigma$ \\ 
25 &  BAHAMAS  8.0& BCEmu7 $\sum m_{\nu}$: [0.06,6] eV & $ 0.785^{+0.020}_{0.025} $ & $0.202\sigma$  \\ 
26 & BAHAMAS  8.0 + EuclidEmu-DM  &BCEmu7 & $ 0.791^{+0.020}_{0.025} $ & $0.466\sigma$\\ 
27 & BAHAMAS  8.0 + EuclidEmu-DM   &BCEmu7 $\sum m_{\nu}$: [0.06,6] eV & $ 0.787^{+0.019}_{0.027} $ & $0.287\sigma$\\ 
28 & cosmo-OWLS 8.5 & BCEmu7 & $ 0.791^{+0.022}_{0.026} $ & $0.439\sigma$ \\
29& $A_{\rm mod}= 0.820$ & BCEmu7 &$ 0.777^{+0.018}_{0.024} $ & $-0.165\sigma$ \\
\hline

\end{tabular}\label{tab:mockfull}

\end{center}
\end{table*}

We validate the choice to model the dark matter-only power spectrum with HM20 using dark matter-only mocks created with HM20 and EE2.  Here we summarise the findings of mock tests 1-4.  
\begin{itemize}
    \item This test analyses a mock with the same model choices used to create it and is useful for identifying projection effects. We find these to be present for $\Omega_{\rm m}$ and $\sigma_8$, which are under- and over-estimated, respectively, by $\sim$0.5$\sigma$. 
    %HM20 models the non-linear dark matter-only power spectrum accurately enough to give unbiased $S_8$, but we find that  $\Omega_{\rm m}$ and $\sigma_8$ are under and over estimated, respectively. This effect is exacerbated due to projection effects, which we identify 
    \item When modelling dark-matter with HM20 in a HM20 generated mock, allowing the neutrino mass to vary with the prior $\sum m_{\nu}$: [0.06,6] eV improves the recovery of $\Omega_{\rm m}$ and $\sigma_8$.  It however decreases the accuracy in the recovery of $S_8$, resulting in an underestimation of $0.5\sigma$. 
    \item Modelling the non-linear dark matter-only power spectrum with HM20 in a EE2 generated mock over estimates $S_8$ by $\sim 0.4\sigma$, which is reduced marginally to $\sim 0.2\sigma$ when scale cuts are applied. We note that \citet{KiDSDES} find a smaller bias in the mean of $\sim 0.1\sigma$ when scale cuts are used. In this mock analysis there are two set-up differences that could explain this: we use a higher value of $\Omega_{\rm m}$ to create the mock and we marginalise over $A_{\rm s}$, instead of $S_8$.
\end{itemize}

Next we validate each of our four baryon models (scale cuts, HM20 with free $\Theta_{\rm AGN}$, BCEmu and SP(k)) and their respective prior and analysis variants.  We test the modelling choices on three `baryonic feedback scenarios' of increasing `extremity' in terms of their impact on the matter power spectrum suppression: BAHAMAS 8.0, cosmo-OWLS 8.5 and $A_{\rm{mod}}=0.82$.  Here we summarise the findings of the mock tests 5-29 and Fig.~\ref{fig:mockchoices}.  Note that in the figure, mocks are labelled in the format 'model used to create the mock: model/analysis choices used to analyse the mock'.  For example, 'BAHAMAS: HM20 BAHAMAS' denotes a mock created with BAHAMAS-like baryon feedback scenario and analysed with HM20 with the BAHAMAS $\Theta_{\rm AGN}=7.6-8.0$ prior.

\begin{itemize}
    \item Optimised scale cuts underestimates $S_8$ by $0.5-1.2\sigma$ in the three feedback scenarios we test, with the tension worsening with more extreme feedback. As noted in Sec.~\ref{sec:mocks}, this is as expected given our choice of feedback scenarios used to build the synthetic data. The OWLS-AGN scenario, which was used to decide which angular scales of the measurement would be discarded in the DES Y3 analysis predicts a less extreme feedback scenario than the three scenarios used here.
    \item The left panel of Fig.~\ref{fig:mockchoices} compares prior choices of the HM20 model with $\Theta_{\rm AGN}$ as a free parameter.  We find that HM20 with the BAHAMAS calibrated prior choice of $\Theta_{\rm AGN}=7.6-8.0$ underestimates $S_8$ by $\sim0.7-1.2\sigma$, with the tension worsening with a more extreme feedback scenario.  Using the wide prior choice of $\Theta_{\rm AGN}=7.3-9.0$ overestimates $S_8$ in BAHAMAS 8.0 and cosmo-OWLS 8.5 by $\sim0.5\sigma$, but still underestimates $S_8$ in an $A_{\rm{mod}}=0.82$-like feedback scenario by $\sim0.2\sigma$.
    \item The right panel of Fig.~\ref{fig:mockchoices} shows that SP(k) with the X-ray informed prior demonstrates the best recovery of $S_8$ in a BAHAMAS 8.0-like feedback scenario of all the baryonic feedback modelling choices.  In more extreme feedback scenarios, it however underestimates $S_8$ by $\sim0.3-1.6\sigma$.  Using the wide prior choice improves the recovery of $S_8$ to be within $\sim0.4\sigma$ for the three feedback scenarios.   
    \item The central panel of Fig.~\ref{fig:mockchoices} demonstrates that BCEmu7 overestimates $S_8$ in both BAHAMAS 8.0 and cosmo-OWLS 8.5 baryonic feedback scenarios by $\sim 0.4\sigma$, but underestimates $S_8$ in an $A_{\rm{mod}}=0.82$-like feedback scenario by $\sim0.2\sigma$.  BCEmu1 underestimates  $S_8$ by $0.3-1.2\sigma$, again with the tension worsening when testing more extreme feedback scenarios.  
    \item In the mock tests which over-estimate $S_8$, freeing the neutrino mass with the prior $\sum m_{\nu}$: [0.06,6] eV appears to improve recovery of the input cosmology.  The shifts we see in $S_8$ are however consistent with the $\sim 0.2\sigma$ shift to lower values we find when testing the dark matter-only modelling of HM20 with and without free neutrinos, i.e. tests 1 and 2.  

\end{itemize}

Ultimately we find that choosing the restrictive analysis variant of a model tends to result in an under-estimation of $S_8$.  
This motivates the use of more flexible modelling choices in cosmic shear analyses, since these tend to result in improved recovery of $S_8$, however this is at the consequence of a loss in precision.  

\section{Baryon feedback constraints}\label{app:bary}

\subsection{SP(k)}\label{app:spk}
\begin{figure}
\centering
\begin{subfigure}[b]{\linewidth}
\includegraphics[width=\linewidth]{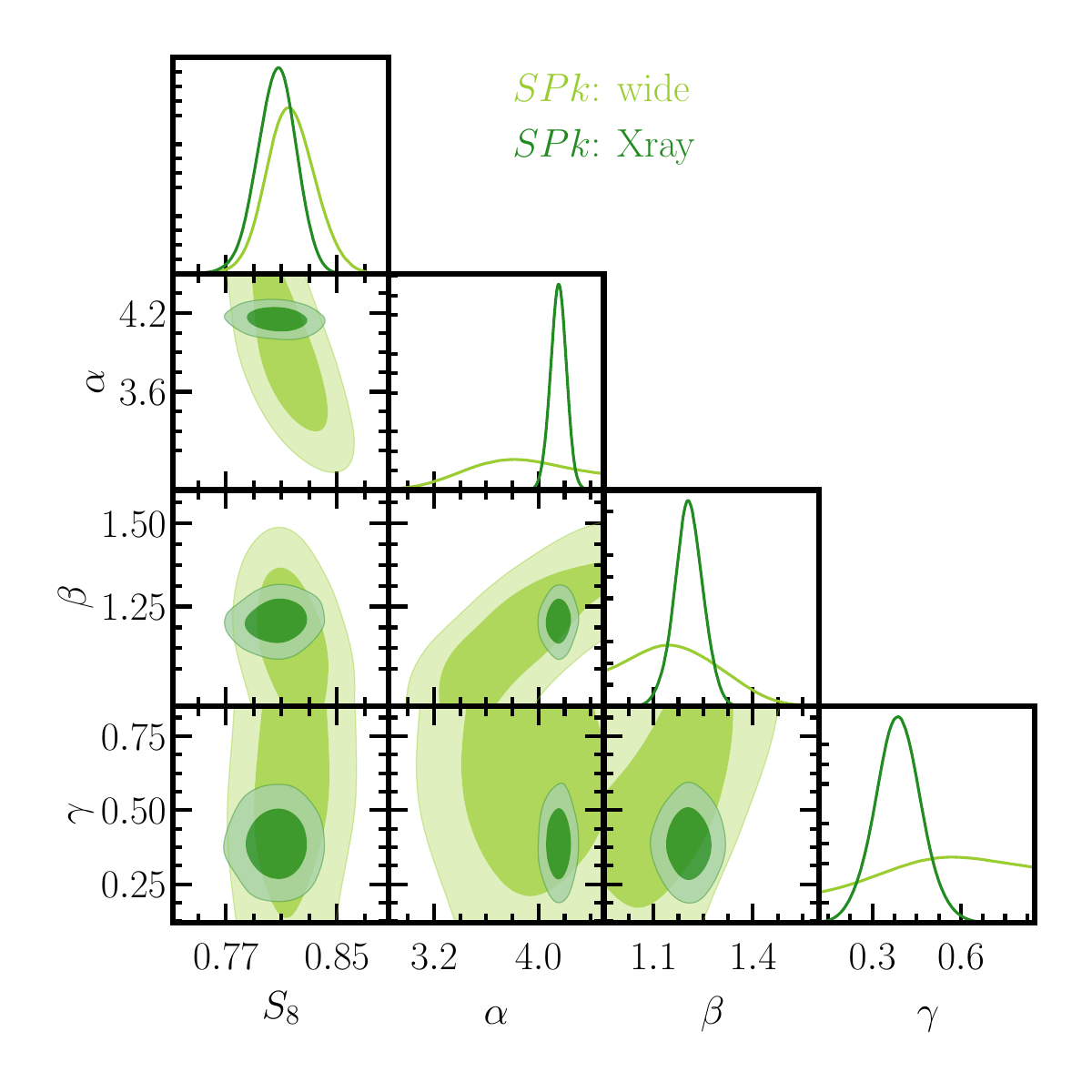}
\end{subfigure}
\begin{subfigure}[b]{\linewidth}
\includegraphics[width=\linewidth]{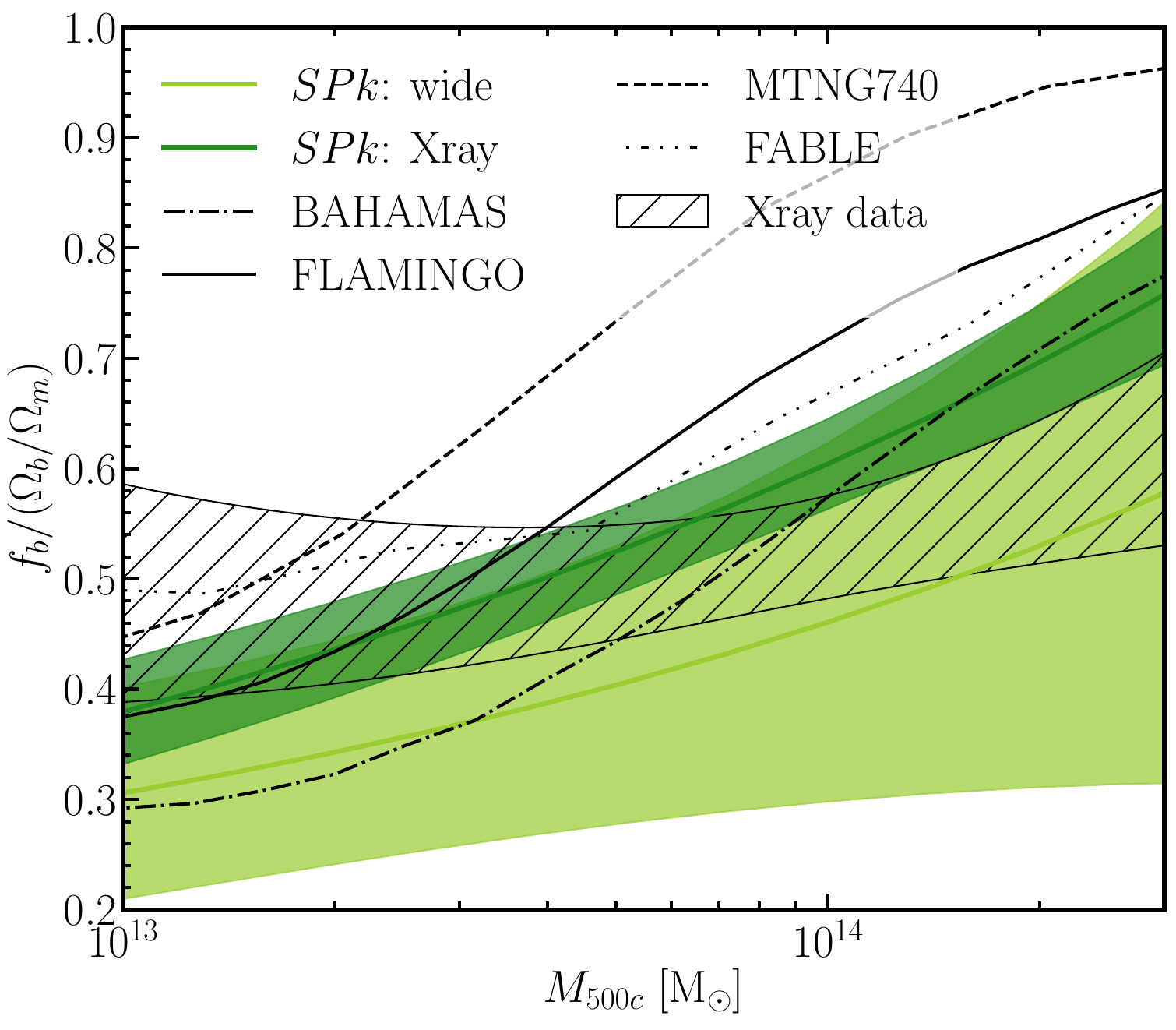}
\end{subfigure}
\caption{SP(k) marginalised posteriors when analysing the DES Y3 cosmic shear data with the wide prior on SP(k) (light green) and X-ray informed prior (dark green). We show the  68\% and 95\% confidence levels of the $S_8$ and the three model parameters (upper panel) and the mean total baryon fraction, $f_{\rm b}/\Omega_{\rm b}/\Omega_{\rm m}$, and $1\sigma$ uncertainty as a function of halo mass, $M_{500}$ (lower). The X-ray prior is derived from HSC-XXL $1\sigma$ constraints \citep[black hatched][]{Akino_2022}, in this plot scaled to the mean cosmology obtained from the $SPk$: X-ray analysis.}
\label{fig:spkparams}
\end{figure}

\begin{table}
\centering
\caption{The constraints on the SP(k) parameters attained by the WL-only SP(k) analyses with both the wide and X-ray informed prior choices.  We report the mean and 68\% confidence levels. }
\begin{tabular}{ccccc}
\hline
\hline
Parameter & Wide prior & X-ray prior \\
\hline
$\alpha$ &$3.830^{+0.409}_{-0.363}$ &$4.153^{+0.061}_{-0.058}$ \\
$\beta$ & $1.174^{+0.104}_{-0.158}$ &$1.206^{+0.043}_{-0.042}$ \\
$\gamma$ &  $0.519^{+0.265}_{-0.157}$ &$0.388^{+0.077}_{-0.076}$\\
\hline
\end{tabular}
\label{tab:spkparamconstraints}
\end{table}

The mean and 68\% confidence levels on the SP(k) parameters constrained by the WL-only SP(k) analyses with the wide and X-ray informed prior choices are reported in Table~\ref{tab:spkparamconstraints}.  The corresponding marginalised posteriors for these analyses are shown in the upper panel of Fig.~\ref{fig:spkparams}.  Since SP(k) directly maps between the baryon fraction in groups and clusters and the matter power spectrum suppression, we also show the constraints on the $f_{\rm b}/\Omega_{\rm b}/\Omega_{\rm m}$-$M_{500}$ relation predicted when analysing the DES cosmic shear with the two SP(k) analysis variants in the lower panel of Fig.~\ref{fig:spkparams}.

\subsection{Baryonification}\label{app:bcemu}
\begin{figure*}
	\centering
	\includegraphics[width=\linewidth]{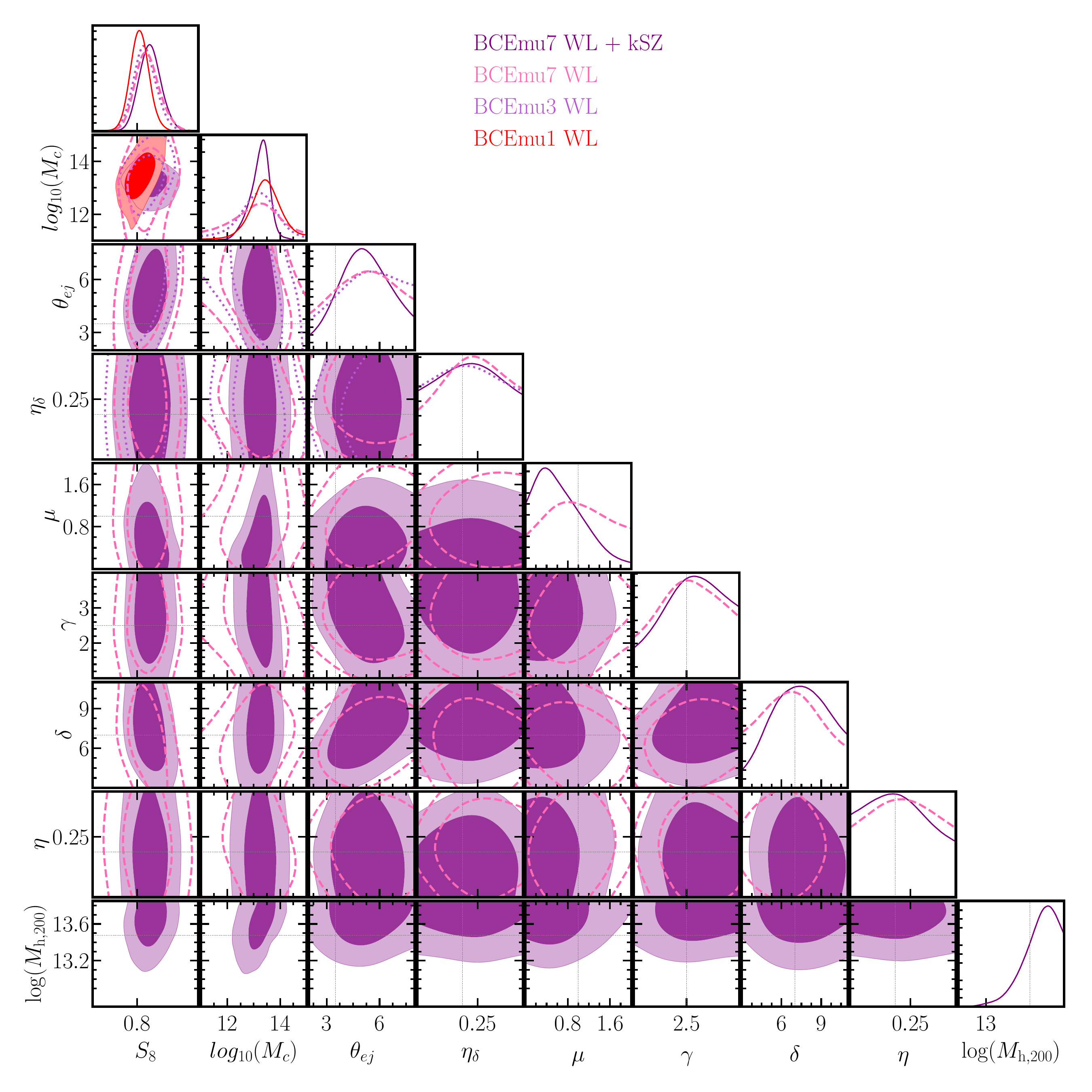} 
	\caption{The marginalised posteriors for $S_8$, the seven baryonification parameters, and the average halo mass og the kSZ sample $M_{\rm h, 200}$ attained  by the WL-only BCEmu1 (red solid line), BCEmu3 (light purple dotted line) and BCEmu7 (pink dashed line) analyses, as well as the WL + kSZ BCEmu7 (dark purple line) analysis.  The inner and outer contours show the 68\% and 95\% confidence levels respectively.  The grey dashed lines show the values the BCEmu parameters are fixed to in the case of using BCEmu1 and BCEmu3, and the mean halo mass of the CMASS sample reported by \citet{Schaan:2021}.}
	\label{fig:bcemuparams}
\end{figure*}

The mean and 68\% confidence levels on the baryonification parameters constrained by the WL-only BCEmu1, BCEmu3 and BCEmu7 analyses, in addition to the WL + kSZ BCEmu7 joint analysis (for which the average halo mass of the kSZ sample $M_{\rm h, 200}$ is also reported) are given in Table~\ref{tab:bcemuparamconstraints}. The corresponding marginalised posteriors for these analyses are shown in Fig.~\ref{fig:bcemuparams}.  

\begin{table}
\caption{The constraints on the baryonification parameters attained by the WL-only BCEmu1, BCEmu3 and BCEmu7 analyses, as well as the BCEmu7 WL + kSZ BCEmu7 analysis. For the WL + kSZ analysis we also report the constraint on the average halo mass of the kSZ sample  ($10^{13} M_{\odot}$).  We report the mean and 68\% confidence levels. }
\begin{tabular}{ccccc}
\hline
\hline
Param. & BCEmu1 & BCEmu3 & BCEmu7  & WL + kSZ  \\
\hline
$\log_{10}M_{\rm c}$ & $ 13.42^{+0.60}_{-0.54} $ & $13.13^{+0.79}_{-0.63} $& $ 13.06^{+1.01}_{-0.87}$ &  $13.22^{+0.42}_{-0.29}$ \\ 
$\theta_{\rm ej}$  & - & $5.31^{+2.17}_{-1.30} $& $ 5.14^{+1.75}_{-1.74}$ & $ 5.15^{+1.46}_{-1.48}$ \\  
 $\eta_{\delta}$ & - & $0.22^{+0.09}_{-0.12} $& $ 0.23^{+0.10}_{-0.10}$ & $ 0.22^{+0.10}_{-0.11}$  \\  
$\mu$  & - & - & $ 1.00^{+0.50}_{-0.61}$ & $ 0.68^{+0.25}_{-0.58}$  \\ 
$\gamma$  & - & -& $ 2.57^{+1.00}_{0.71}$ &$ 2.66^{+0.98}_{0.67}$ \\ 
$\delta$  & - & -& $ 6.83^{+2.09}_{-2.43}$ &  $ 7.35^{+2.16}_{-2.10}$  \\ 
$\eta$  & - & - & $ 0.22^{+0.10}_{-0.11}$ & $ 0.21^{+0.08}_{-0.13}$  \\ 
$M_{\rm h, 200} $ & - & - &- & $4.10^{+1.55}_{-1.64}$\\
\hline
\end{tabular}
\label{tab:bcemuparamconstraints}
\end{table}

\subsection{Prior choice for the halo mass of the kSZ sample}\label{app:mhalo}
Given the significant scatter in the literature, we choose to include an additional model parameter in the analysis, $M_{\rm h, 200}$, corresponding to the mean $M_{200}$ of the CMASS sample, with a prior range provided in Table~\ref{tab:modelparameters}.  For example, the stacked stellar mass--halo mass relation of \citet{Sonnenfeld2019} of CMASS galaxies derived from HSC galaxy--galaxy lensing measurements, when combined with the stellar mass distribution of CMASS, implies a mean $M_{200} \approx 0.5\times10^{13} {\rm M}_\odot$, whereas the abundance matching methods of \citet{Kravtsov2018}, \citet{Moster2018}, and \citet{Behroozi2019} imply mean $M_{200}$ values of $2.7\times10^{13} {\rm M}_\odot$, $4.4\times10^{13} {\rm M}_\odot$, and $6.6\times10^{13} {\rm M}_\odot$, respectively.  A HOD-based analysis of the clustering of BOSS CMASS galaxies by \citet{White2011} found a mean $M_{200}$ of $3.6\times10^{13} {\rm M}_\odot$.  Given this large study-to-study variance, we choose a flat prior range of $[0.5-7] \times10^{13} {\rm M}_\odot$ and marginalise over this parameter. 

The combined WL + kSZ analysis constrains $M_{\rm h,200}=4.098^{+1.548}_{-1.639} \times10^{13} {\rm M}_\odot$, indicating we are prior constrained at the $2\sigma$ level.  For comparison to the remainder of this work which generally quotes halo masses in $M_{500}$, this corresponds to $M_{h,500}=3.01\times10^{13} {\rm M}_\odot$, assuming a NFW profile and a concentration-mass relation from \citet{DuttonMaccio2014}.  

\begin{figure}
\centering
\begin{subfigure}[b]{\linewidth}
\includegraphics[width=0.9\columnwidth]{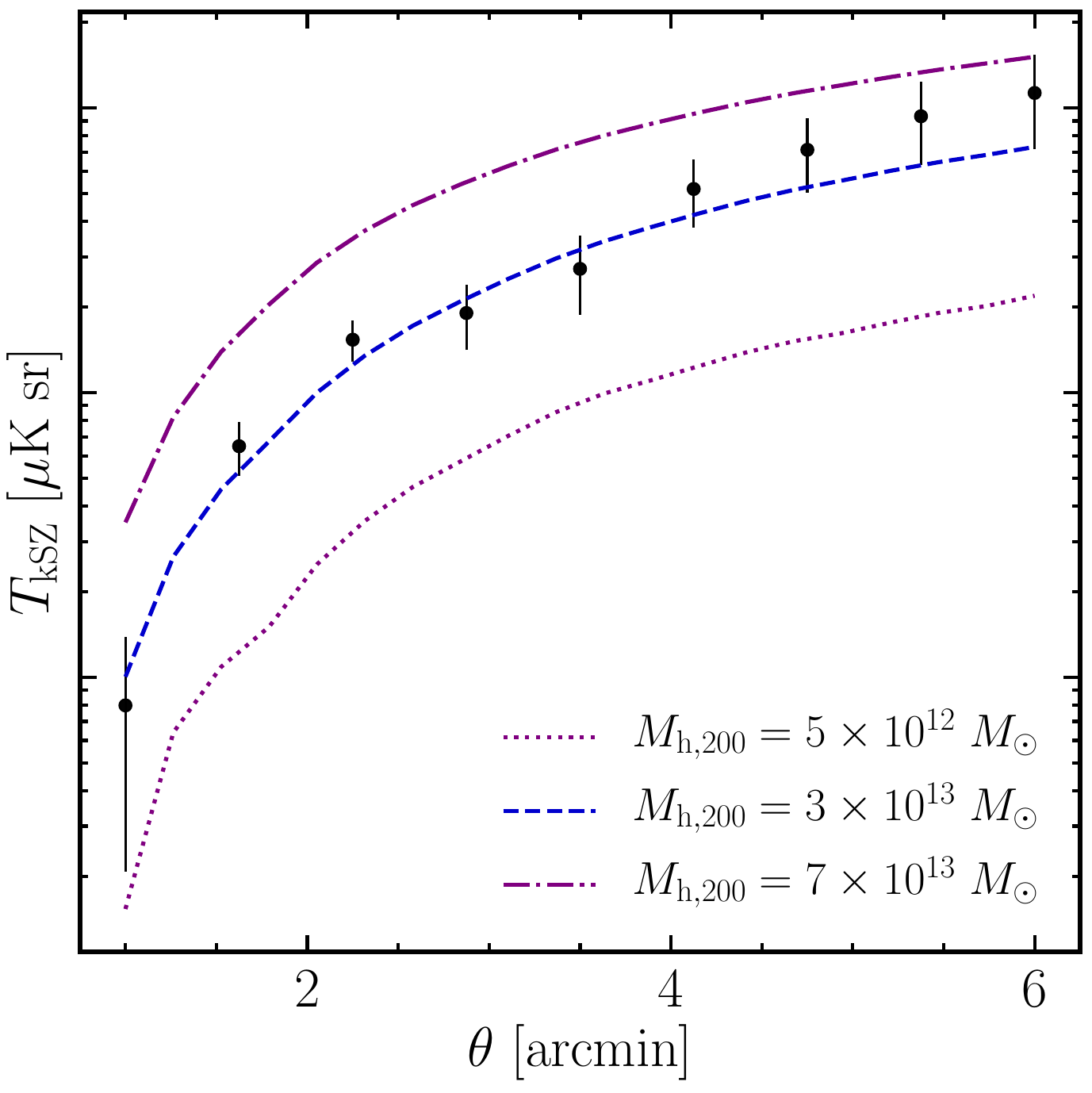}
\end{subfigure}
\begin{subfigure}[b]{\linewidth}
\includegraphics[width=\columnwidth]{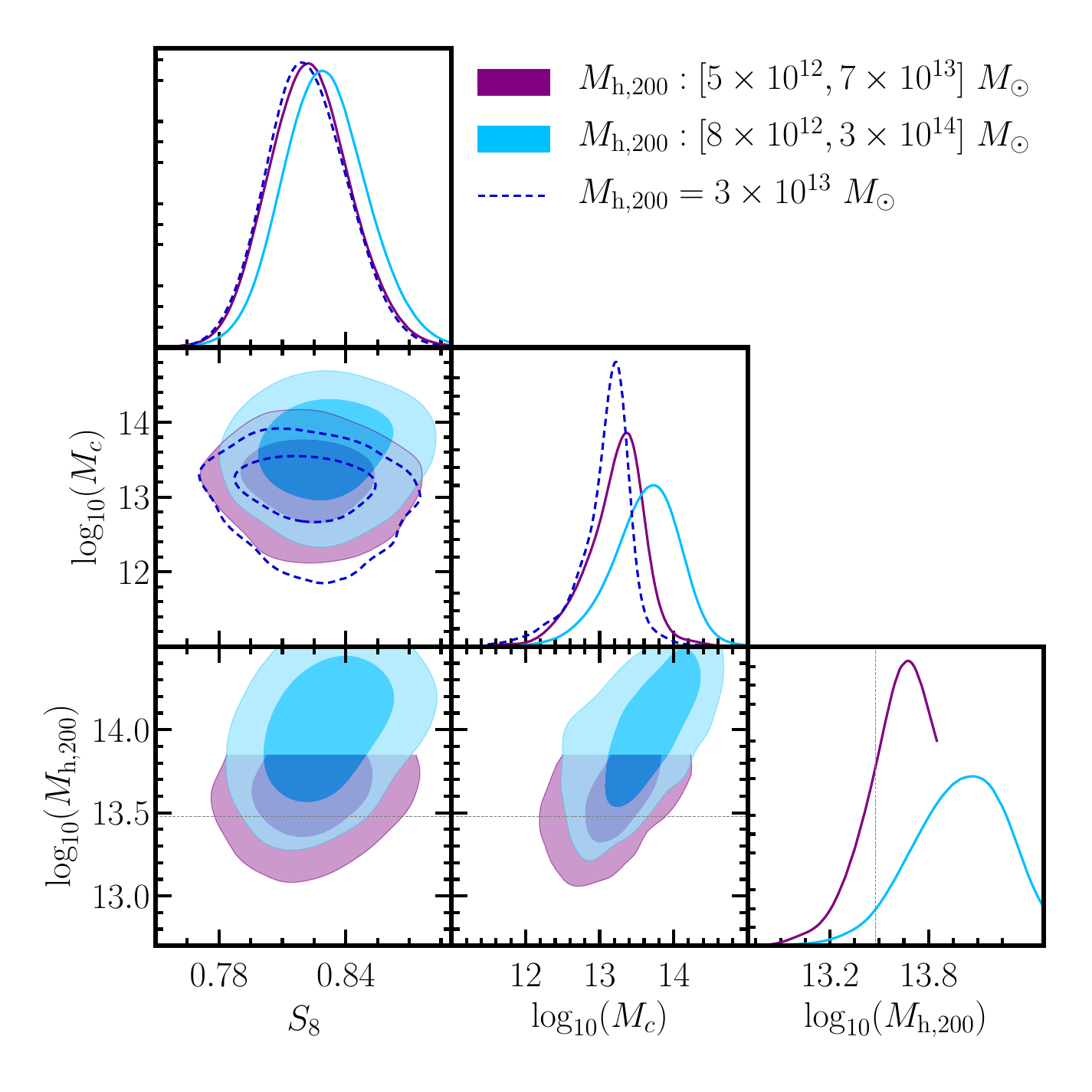}
\end{subfigure}
\caption{\textit{Upper:} The stacked kSZ temperature profile at 98GHz as a function of
angular radius, $\theta$, centred on the group or cluster (bottom) when varying the mean halo mass of CMASS galaxy sample modelled, $M_{\rm h, 200}$, within the limits of the fiducial prior choice  $M_{\rm h, 200}: [5\times10^{12},7\times10^{13}] M_{\odot}$. The ACT CMASS measurements at 98GHz are shown as the black data points in the bottom
panels and the model profiles are convolved with the f90 beam profile for comparison. \textit{Lower:} The marginalised posteriors for $S_8$, $\log_{10}(M_{\rm c})$ and $M_{\rm h, 200}$ in a WL + kSZ analysis with different prior choices on $M_{\rm h, 200}$.  We show the prior used in the fiducial analysis $M_{\rm h, 200}: [5\times10^{12},7\times10^{13}] M_{\odot}$ (purple), as well as a wide prior $M_{\rm h, 200}: [8\times10^{12},3\times10^{14}] M_{\odot}$ (light blue) and fixed $M_{\rm h, 200}=3\times10^{13}] M_{\odot}$ (dark blue, also shown as the dashed grey line).  The inner and outer contours show the 68\% and 95\% confidence levels respectively. }
\label{fig:massprior} 
\end{figure}

In this appendix we explore the impact on our cosmology and baryon model parameter constraints of choosing a wider prior $M_{\rm h, 200}:[0.8,30]\times10^{13} {\rm M}_\odot$.  We also consider a fixed $M_{\rm h, 200}$ analysis using the mean mass of the CMASS sample as determined by \citet{Schaan:2021} and used in \citep{Amodeo:2021}:  $M_{\rm h, 200}=3\times10^{13} {\rm M}_\odot$.   Fig.~\ref{fig:massprior} shows the marginalised posteriors for $\Omega_{\rm m}$, $S_8$, $\log_{10}(M_{\rm c})$, $\theta_{\rm ej}$ and $M_{\rm h, 200}$ in a WL + kSZ analysis with the different prior choices on $M_{\rm h, 200}$. The halo mass, $M_{\rm h, 200}$, is correlated with $\log_{10}(M_{\rm c})$ and in the case of a wider prior, both $M_{\rm h, 200}$ and $\log_{10}(M_{\rm c})$  prefer higher values. However, $S_8$ is relatively stable to the halo mass prior, shifting by 0.3$\sigma$. 

\subsection{The relationship between baryon fraction and matter power spectrum suppression}\label{sec:vandaalen}

\begin{figure}
	\centering
	\includegraphics[width=\linewidth]{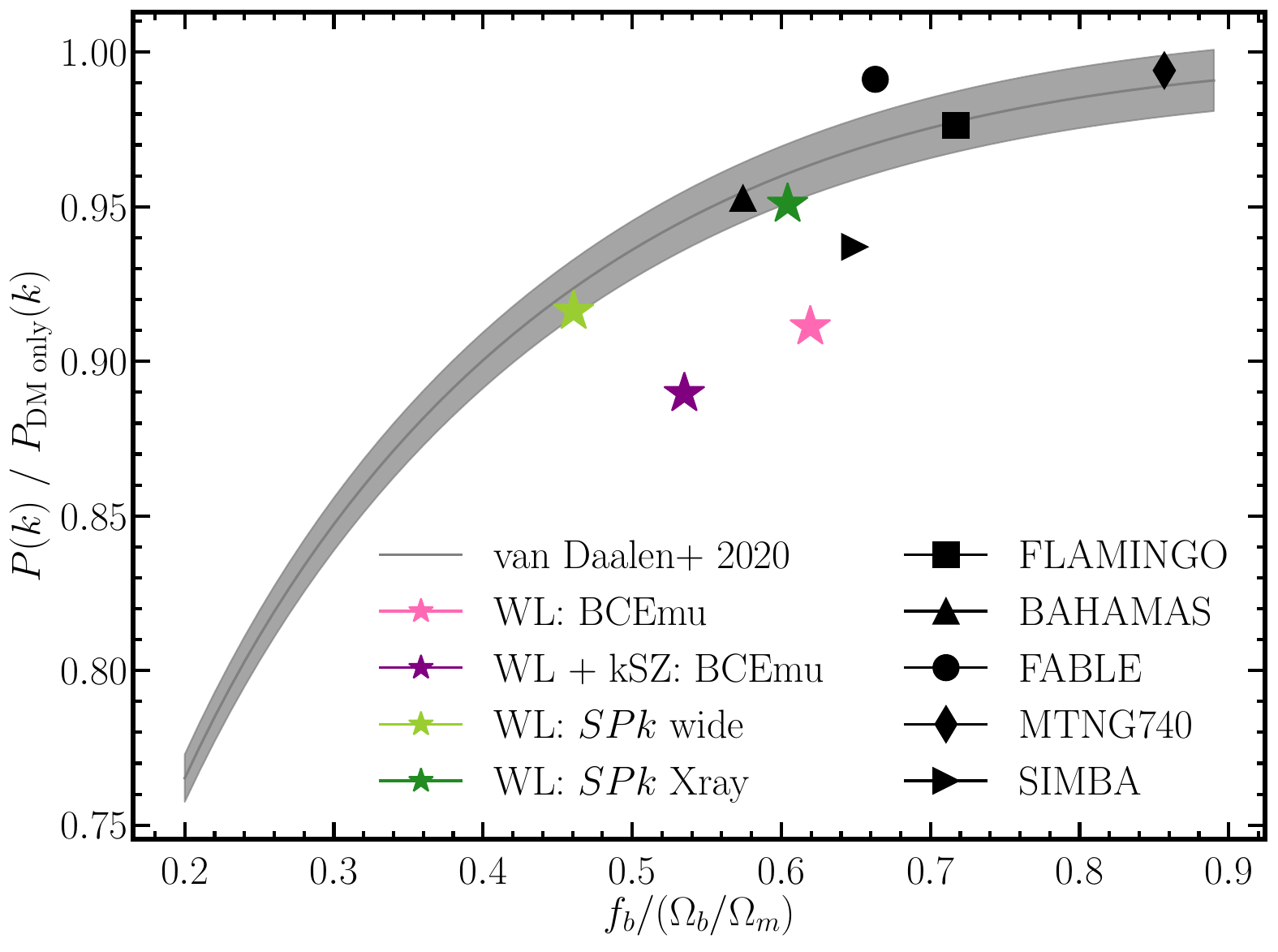} 
	\caption{The relation between the matter power spectrum suppression $P(k)/P_{\mathrm{DM only}}(k)$ at $k=1$ $h \mathrm{Mpc}^{-1}$ and the mean baryon fraction, $f_{\rm b}/\Omega_{\rm b}/\Omega_{\rm m}$, measured within $R_{500}$ for halos of mass $M_{500}=10^{14} M_{\odot}$.  We plot the empirical best-fit relation of \citet{vandaalen:2020} as the grey solid line, with the 1\% accuracy on the relation's ability to predict $P(k)/P_{\mathrm{DM only}}(k)$ shown as the shaded grey region.  We plot the constraints obtained in our analysis in the $P(k)/P_{\mathrm{DM only}}(k)$-$f_{\rm b}/\Omega_{\rm b}/\Omega_{\rm m}$ plane.  We show the WL-only (pink starred) and WL + kSZ (pruple starred) analyses with BCEmu7, and the WL-only results with $SPk$, showing both the wide (light green starred) and X-ray (dark green starred) prior choices.  We also plot the simulations FLAMINGO \citep{Schaye2023}; BAHAMAS \citep{McCarthy2017}; SIMBA \citep{Dave:2019}, MillenniumTNG \citep{Pakmor2023} and FABLE (\citealt{Henden2018}, Bigwood et al. in prep.) as the black datapoints.}
	\label{fig:vd}
\end{figure}

It has been demonstrated that in hydrodynamical simulations, the mean baryon fraction in halos of $M \sim 10^{14} \Msun$ can be predictive for the suppression of the matter power spectrum due to baryonic feedback effects, robust to a number of feedback prescriptions \citep{vandaalen:2020}.  This relationship can be described by an empirical fitting function relating the mean baryon fraction, $f_{\rm b}/\Omega_{\rm b}/\Omega_{\rm m}$, measured within $R_{500}$ for halos of mass $M_{500}=10^{14} M_{\odot}$ to the suppression of the matter power spectrum $P(k)/P_{\mathrm{DM only}}(k)$ at $k=1$ $h \mathrm{Mpc}^{-1}$ (equation 5 in their work).  The best-fit relation, shown as the solid grey line in Fig.~\ref{fig:vd}, was fit to the cosmo-OWLS \citep{LeBrun:2014} and BAHAMAS \citep{McCarthy2017} simulations and is accurate to 1\% for the simulations they test, shown as the grey shaded region in Fig.~\ref{fig:vd}.  In this section, we discuss where constraints of our analysis lie in the $P(k)/P_{\mathrm{DM only}}(k)$-$f_{\rm b}/\Omega_{\rm b}/\Omega_{\rm m}$ plane with respect to the \citet{vandaalen:2020} relation.  In Fig.~\ref{fig:vd} we plot the result for the WL-only and WL + kSZ analyses with BCEmu7 baryon modelling, as well as the WL-only results with SP(k)'s wide and X-ray informed prior choices.  For comparison, we also show the simulations discussed throughout this work: FLAMINGO \citep{Schaye2023}; BAHAMAS \citep{McCarthy2017}; SIMBA \citep{Dave:2019}, MillenniumTNG \citep{Pakmor2023} and FABLE (\citealt{Henden2018}, Bigwood et al. in prep.).  

This \citet{vandaalen:2020} relationship is at the core of the the SP(k) model \citep{Salcido2023}. Given that the ANTILLES simulations follow this relationship, the mean baryon fraction is used to calibrate their emulator.  Naturally, both of our WL-only analyses with SP(k) fall on the \citet{vandaalen:2020} relation to within 1\%. However, the BCEmu model does not enforce the \citet{vandaalen:2020} relationship between the  baryon fraction and matter power suppression, and allows for greater flexibility in the impact of feedback on the matter distribution, including scenarios which, according to the simulations, could be deemed unphysical.  Both the WL-only and WL + kSZ analyses with BCEmu give a mean constraint below the \citet{vandaalen:2020} relation. That is, they allow for a more extreme matter power spectrum suppression for their predicted mean baryon fraction predicted. In the future, it will be important to test this relationship with a range observations, including those probe lower halo masses \citep[see also][]{Pandey2023}.

\section{Intrinsic alignment (IA) model}\label{app:IA}

In this appendix we consider the impact of the IA model choice on our results.  Throughout this work, we use the NLA IA model, which is a sub-space of the TATT model.  A WL-only analysis using BCEmu7 with TATT constrains $S_8=0.802^{+0.028}_{-0.024}$, lying $\sim0.5\sigma$ lower than that constrained using NLA. This shift in $S_8$ between the two IA models is consistent when using BCEmu1 instead of BCEmu7, and similar to that found in previous work when using HM20 \cite{KiDSDES} and scale cuts \citep{Secco:2022, amon:2022}. As in the literature, both IA model choices give comparable $\chi^2_{\rm red}$ values (see Tab.\ref{tab:cosmologyresults}). In future work, it is important to determine the more accurate IA model. Here, we use this test as validation that our baryon model choice and the results of our baryon model comparison are independent of the choice of IA model.

\section{Sampling algorithm choice}\label{app:sampler}
In this appendix we compare the cosmological parameter estimates attained using the \texttt{multinest} and \texttt{polychord} samplers.  Throughout the work, we use the \texttt{multinest} settings $n_{\rm live} = 500$, efficiency = 0.3, tolerance = 0.1,
constant efficiency = False, max. iterations = 50,000 for computing efficiency.  For \texttt{polychord} we use $n_{\rm live} = 500$, tolerance = 0.01, $n_{\rm repeats}=60$, fast fraction$=0.1$.  
Considering the BCEmu7 WL-only analysis we find that sampling with \texttt{polychord} estimates a mean value of $S_8=0.817^{+0.025}_{0.025}$, consistent with that attained by \texttt{multinest} of  $S_8=0.818^{+0.017}_{0.024}$.  However, in agreement with the findings of \citet{KiDSDES}, we find that the 68\% confidence level for $S_8$ attained using \texttt{multinest} is $18\%$ smaller than that estimated with \texttt{polychord}.  Similarly we find a WL + kSZ analysis with \texttt{polychord} returns $S_8=0.821^{+0.020}_{0.023}$, consistent with the $S_8=0.823^{+0.019}_{0.020}$ using multinest but with a $9\%$ larger confidence region.

\begin{comment}
\begin{figure}
	\centering
	\includegraphics[width=\linewidth]{figs/sampler.pdf} 
	\caption{The marginalised posteriors for $\Omega_{\rm m}$, $\sigma_8$ and $S_8$ attained in a WL-only analysis with BCEmu7 when sampling with {\sc multinest} (pink) and {\sc polychord} (purple).  The inner and outer contours show the 68\% and 95\% confidence levels respectively.}
	\label{fig:sampler}
\end{figure} 
\end{comment}

\section*{Affiliations}
$^{1}$ Institute of Astronomy, University of Cambridge, Madingley Road, Cambridge CB3 0HA, UK\\
$^{2}$ Kavli Institute for Cosmology Cambridge, Madingley Road, Cambridge, CB3 OHA\\
$^{3}$ Department of Astrophysical Sciences, Princeton University, Peyton Hall, Princeton, NJ 08544, USA\\
$^{4}$ Institute for Computational Science, University of Zurich, Winterthurerstrasse 190, CH-8057 Zurich, Switzerland\\
$^{5}$ Astrophysics Research Institute, Liverpool John Moores University, 146 Brownlow Hill, Liverpool L3 5RF, UK\\
$^{6}$ Centre for Energy, Environmental and Technological Research, Av. Complutense, 40, Moncloa - Aravaca, 28040 Madrid, Spain\\
$^{7}$ Kavli Institute for Particle Astrophysics and Cosmology,
382 Via Pueblo Mall Stanford, CA 94305-4060, USA\\
$^{8}$ SLAC National Accelerator Laboratory 2575 Sand Hill Road Menlo Park, California 94025, USA\\
$^{9}$ Physics Division, Lawrence Berkeley National Laboratory, Berkeley, CA 94720, USA\\
$^{10}$ Berkeley Center for Cosmological Physics, Department of Physics,
University of California, Berkeley, CA 94720, USA\\
$^{11}$  Department of Astronomy, Cornell University, Ithaca, NY 14853, USA\\
$^{12}$ Department of Physics, University of Michigan, Ann Arbor, MI 48109, USA\\
$^{13}$ Kavli Institute for the Physics and Mathematics of the Universe (WPI), UTIAS, The University of Tokyo, Kashiwa, Chiba 277-8583, Japan\\
$^{14}$ Department of Physics, Carnegie Mellon University, Pittsburgh, Pennsylvania 15312, USA\\
$^{15}$ NSF AI Planning Institute for Physics of the Future, Carnegie Mellon University, Pittsburgh, PA 15213, USA\\
$^{16}$ Kavli Institute for Particle Astrophysics \& Cosmology, P. O. Box 2450, Stanford University, Stanford, CA 94305, USA\\
$^{17}$ SLAC National Accelerator Laboratory, Menlo Park, CA 94025, USA\\
$^{18}$ Laborat\'orio Interinstitucional de e-Astronomia - LIneA, Rua Gal. Jos\'e Cristino 77, Rio de Janeiro, RJ - 20921-400, Brazil\\
$^{19}$ Observat\'orio Nacional, Rua Gal. Jos\'e Cristino 77, Rio de Janeiro, RJ - 20921-400, Brazil\\
$^{20}$ Argonne National Laboratory, 9700 South Cass Avenue, Lemont, IL 60439, USA\\
$^{21}$ Institute of Space Sciences (ICE, CSIC),  Campus UAB, Carrer de Can Magrans, s/n,  08193 Barcelona, Spain\\
$^{22}$ Department of Astronomy and Astrophysics, University of Chicago, Chicago, IL 60637, USA\\
$^{23}$ Fermi National Accelerator Laboratory, P. O. Box 500, Batavia, IL 60510, USA\\
$^{24}$ Kavli Institute for Cosmological Physics, University of Chicago, Chicago, IL 60637, USA\\
$^{25}$ NASA Goddard Space Flight Center, 8800 Greenbelt Rd, Greenbelt, MD 20771, USA\\
$^{26}$ Instituto de F\'isica Gleb Wataghin, Universidade Estadual de Campinas, 13083-859, Campinas, SP, Brazil\\
$^{27}$ Center for Cosmology and Astro-Particle Physics, The Ohio State University, Columbus, OH 43210, USA\\
$^{28}$ Instituto de Astrofisica de Canarias, E-38205 La Laguna, Tenerife, Spain\\
$^{29}$ Department of Physics and Astronomy, University of Pennsylvania, Philadelphia, PA 19104, USA\\
$^{30}$ Universit\'e Grenoble Alpes, CNRS, LPSC-IN2P3, 38000 Grenoble, France\\
$^{31}$ University Observatory, Faculty of Physics, Ludwig-Maximilians-Universit\"at, Scheinerstr. 1, 81679 Munich, Germany\\
$^{32}$ Jet Propulsion Laboratory, California Institute of Technology, 4800 Oak Grove Dr., Pasadena, CA 91109, USA\\
$^{33}$ Brookhaven National Laboratory, Bldg 510, Upton, NY 11973, USA\\
$^{34}$ Department of Physics, ETH Zurich, Wolfgang-Pauli-Strasse 16, CH-8093 Zurich, Switzerland\\
$^{35}$ Institut de F\'{\i}sica d'Altes Energies (IFAE), The Barcelona Institute of Science and Technology, Campus UAB, 08193 Bellaterra (Barcelona) Spain\\
$^{36}$ Department of Astronomy/Steward Observatory, University of Arizona, 933 North Cherry Avenue, Tucson, AZ 85721-0065, USA\\
$^{37}$ Department of Physics, University of Arizona, Tucson, AZ 85721, USA\\
$^{38}$ School of Physics and Astronomy, Cardiff University, CF24 3AA, UK\\
$^{39}$ Centro de Investigaciones Energ\'eticas, Medioambientales y Tecnol\'ogicas (CIEMAT), Madrid, Spain\\
$^{40}$ Institut de Recherche en Astrophysique et Plan\'etologie (IRAP), Universit\'e de Toulouse, CNRS, UPS, CNES, 14 Av. Edouard Belin, 31400 Toulouse, France\\
$^{41}$ Department of Physics and Astronomy, University of Waterloo, 200 University Ave W, Waterloo, ON N2L 3G1, Canada\\
$^{42}$ Institute for Astronomy, University of Edinburgh, Edinburgh EH9 3HJ, UK\\
$^{43}$ Department of Physics, Northeastern University, Boston, MA 02115, USA\\
$^{44}$ Lawrence Berkeley National Laboratory, 1 Cyclotron Road, Berkeley, CA 94720, USA\\
$^{45}$ Jodrell Bank Center for Astrophysics, School of Physics and Astronomy, University of Manchester, Oxford Road, Manchester, M13 9PL, UK\\
$^{46}$ Nordita, KTH Royal Institute of Technology and Stockholm University, Hannes Alfv\'ens v\"ag 12, SE-10691 Stockholm, Sweden\\
$^{47}$ Physics Department, 2320 Chamberlin Hall, University of Wisconsin-Madison, 1150 University Avenue Madison, WI  53706-1390\\
$^{48}$ Department of Physics, University of Genova and INFN, Via Dodecaneso 33, 16146, Genova, Italy\\
$^{49}$ Center for Astrophysical Surveys, National Center for Supercomputing Applications, 1205 West Clark St., Urbana, IL 61801, USA\\
$^{50}$ Department of Astronomy, University of Illinois at Urbana-Champaign, 1002 W. Green Street, Urbana, IL 61801, USA\\
$^{51}$ Department of Physics, Duke University Durham, NC 27708, USA\\
$^{52}$ Department of Applied Mathematics and Theoretical Physics, University of Cambridge, Cambridge CB3 0WA, UK\\
$^{53}$ Kavli Institute for Cosmology, University of Cambridge, Madingley Road, Cambridge CB3 0HA, UK\\
$^{54}$ Department of Physics, Stanford University, 382 Via Pueblo Mall, Stanford, CA 94305, USA\\
$^{55}$ Physics Department, William Jewell College, Liberty, MO, 64068\\
$^{56}$ Department of Physics and Astronomy, Stony Brook University, Stony Brook, NY 11794, USA\\
$^{57}$ Department of Astronomy, University of Geneva, ch. d'\'Ecogia 16, CH-1290 Versoix, Switzerland\\
$^{58}$ Department of Astronomy, University of California, Berkeley,  501 Campbell Hall, Berkeley, CA 94720, USA\\
$^{59}$ Cerro Tololo Inter-American Observatory, NSFs National Optical-Infrared Astronomy Research Laboratory, Casilla 603, La Serena, Chile\\
$^{60}$ Institute of Cosmology and Gravitation, University of Portsmouth, Portsmouth, PO1 3FX, UK\\
$^{61}$ CNRS, UMR 7095, Institut d'Astrophysique de Paris, F-75014, Paris, France\\
$^{62}$ Sorbonne Universit\'es, UPMC Univ Paris 06, UMR 7095, Institut d'Astrophysique de Paris, F-75014, Paris, France\\
$^{63}$ Department of Physics \& Astronomy, University College London, Gower Street, London, WC1E 6BT, UK\\
$^{64}$ Institut d'Estudis Espacials de Catalunya (IEEC), 08034 Barcelona, Spain\\
$^{65}$ Hamburger Sternwarte, Universit\"{a}t Hamburg, Gojenbergsweg 112, 21029 Hamburg, Germany\\
$^{66}$ Department of Physics, IIT Hyderabad, Kandi, Telangana 502285, India\\
$^{67}$ Institute of Theoretical Astrophysics, University of Oslo. P.O. Box 1029 Blindern, NO-0315 Oslo, Norway\\
$^{68}$ Instituto de Fisica Teorica UAM/CSIC, Universidad Autonoma de Madrid, 28049 Madrid, Spain\\
$^{69}$ School of Mathematics and Physics, University of Queensland,  Brisbane, QLD 4072, Australia\\
$^{70}$ Santa Cruz Institute for Particle Physics, Santa Cruz, CA 95064, USA\\
$^{71}$ Department of Physics, The Ohio State University, Columbus, OH 43210, USA\\
$^{72}$ Center for Astrophysics $\vert$ Harvard \& Smithsonian, 60 Garden Street, Cambridge, MA 02138, USA\\
$^{73}$ Australian Astronomical Optics, Macquarie University, North Ryde, NSW 2113, Australia\\
$^{74}$ Lowell Observatory, 1400 Mars Hill Rd, Flagstaff, AZ 86001, USA\\
$^{75}$ George P. and Cynthia Woods Mitchell Institute for Fundamental Physics and Astronomy, and Department of Physics and Astronomy, Texas A\&M University, College Station, TX 77843,  USA\\
$^{76}$ LPSC Grenoble - 53, Avenue des Martyrs 38026 Grenoble, France\\
$^{77}$ Instituci\'o Catalana de Recerca i Estudis Avan\c{c}ats, E-08010 Barcelona, Spain\\
$^{78}$ Perimeter Institute for Theoretical Physics, 31 Caroline St. North, Waterloo, ON N2L 2Y5, Canada\\
$^{79}$ Ruhr University Bochum, Faculty of Physics and Astronomy, Astronomical Institute, German Centre for Cosmological Lensing, 44780 Bochum, Germany\\
$^{80}$ Department of Physics and Astronomy, Pevensey Building, University of Sussex, Brighton, BN1 9QH, UK\\
$^{81}$ School of Physics and Astronomy, University of Southampton,  Southampton, SO17 1BJ, UK\\
$^{82}$ \\
$^{83}$ Computer Science and Mathematics Division, Oak Ridge National Laboratory, Oak Ridge, TN 37831\\
$^{84}$ Max Planck Institute for Extraterrestrial Physics, Giessenbachstrasse, 85748 Garching, Germany\\
$^{85}$ Universit\"ats-Sternwarte, Fakult\"at f\"ur Physik, Ludwig-Maximilians Universit\"at M\"unchen, Scheinerstr. 1, 81679 M\"unchen, Germany\\

\end{document}